# Light selects non-equilibrium states of phase-separated droplets

Zi Lin,[1] Thomas Beneyton,[1] Suzanne Lafon,[1] Edison Rafael Jimenez Granda,[1] Liangfei Tian,[2] Alexandre Baron,[1,3]* Jean-Christophe Baret,[1,3]* Nicolas Martin[1]*

[1] Univ. Bordeaux, CNRS, Centre de Recherche Paul Pascal, UMR 5031, 115 avenue du Dr. Schweitzer; Pessac, 33600, France.

[2] Zhejiang Key Laboratory of Intelligent Sensing Technology and Advanced Medical Instrument and Key Laboratory for Biomedical Engineering of Ministry of Education, College of Biomedical Engineering & Instrument Science, Zhejiang University; Hangzhou, 310027, PR China.

[3] Institut Universitaire de France, 1 rue Descartes; Paris, 75231 Cedex 05, France.

*Corresponding author. Email: alexandre.baron@crpp.cnrs.fr; jean-christophe.baret@u-bordeaux.fr; nicolas.martin@crpp.cnrs.fr

**Abstract:** Driving phase-separated droplets out of equilibrium can oppose coarsening and trigger division. Selecting and sustaining these distinct states remains an important experimental challenge. Here, we use reversible photoswitching to tune molecular driving in DNA-azobenzene coacervates confined in microfluidic droplets without chemical fuel depletion or waste accumulation. Single-wavelength illumination sustaining bidirectional molecular switching arrests droplet coarsening at a finite size. Dual-wavelength illumination with wavelength-dependent penetration depths generates spatially asymmetric molecular driving within uniformly illuminated droplets, producing persistent interfacial instabilities. Under low-salt conditions, droplets divide and undergo recurrent growth-division cycles. Kinetic and thermodynamic measurements reveal how flux balance, phase stability and droplet size govern these distinct non-equilibrium states. These findings establish optically controlled molecular driving as a route to selecting sustained non-equilibrium droplet dynamics.

**Main Text:** A central frontier in soft matter is transforming passive systems into active assemblies capable of sustained dynamic evolution (*1*). Phase separation serves as a fundamental organizing principle of soft matter, governing structure formation in systems ranging from polymer blends (*2*) and viscoelastic fluids (*3*) to biomolecular condensates in living cells (*4–6*). At equilibrium, molecular interactions alone determine phase coexistence and transformation pathways (*7*). Phase-separated droplets relax toward equilibrium by reducing their interfacial area and coarsening, ultimately yielding a single domain. When phase separation is coupled to chemical reactions, however, sustained molecular fluxes continuously compete with thermodynamic relaxation (*8*). Theory predicts that reaction-driven fluxes can reshape phase behavior in ways inaccessible at equilibrium, giving rise to dissipative assembly, arrested coarsening, interfacial instabilities, and division-like dynamics (*9–12*), in which droplets create, rather than reduce, interfacial area. Selecting and sustaining these non-equilibrium states through molecular-driving remains a central challenge in soft matter science. Experiments on chemically fueled coacervates, where reactions produce and degrade phase-separating species, have revealed various non-equilibrium phenomena. A single reaction can drive transient phase separation (*13*, *14*), vacuolization (*15*, *16*) and non-spherical shaping (*17*), or asymmetric splitting (*18*) before ultimately relaxing. Competing reactions can produce transient coacervates (*19–22*), arrested growth (*23*) and repeated, yet limited, formation-dissolution cycles by successive substrate additions (*24–27*). Open microreactors that continuously supply fuel and remove waste can sustain non-equilibrium intra-droplet vacuoles (*28*) and self-generated droplet oscillations (*29*) for hours. These studies establish that chemical driving can push phase-separated matter beyond its equilibrium fate, but also reveal a major limitation: changing the reaction flux necessarily entails consuming reactants and generating waste products. The resulting dynamics are therefore tied to the reaction network that sustains them, rather than being a tunable property of the phase-separated material itself. A fundamentally different approach would be to sustain molecular driving by feeding a chemically reversible reaction with a physical energy reservoir that cannot be depleted, thus enabling selection of non-equilibrium droplet states that can survive indefinitely.

Here photons serve as an inexhaustible source of energy. We use molecular photoswitching to impose molecular driving in phase-separated droplets in a closed chemical system. Photoswitches interconvert between molecular states with different intermolecular interactions, enabling reversible control over phase separation (*30–35*). The Illumination wavelength and intensity tune the rates of the forward and reverse pathways without consuming a chemical fuel or accumulating waste products. Using a minimal light-responsive coacervate system formed by a cationic azobenzene photoswitch and DNA (*30*) confined within monodisperse microfluidic droplets, we quantitatively link reaction kinetics, phase boundaries, and droplet dynamics under continuous, spatially uniform irradiation (Fig. 1). Sustained bidirectional photoisomerization under single-wavelength illumination arrests coarsening and stabilizes finite-size droplets. With two wavelengths preferentially exciting the two molecular states, differences in isomer absorption and concentration profiles generate spatially asymmetric molecular driving within otherwise uniformly illuminated droplets. Tuning the relative irradiation intensities then selects distinct dynamical states, from stable droplets and interfacial undulations to budding and, at higher phase stability, complete droplet division and recurrent growth-division cycles. These results establish molecular photoswitching as a route to select distinct non-equilibrium states in phase-separated matter, from arrested coarsening to recurrent division.

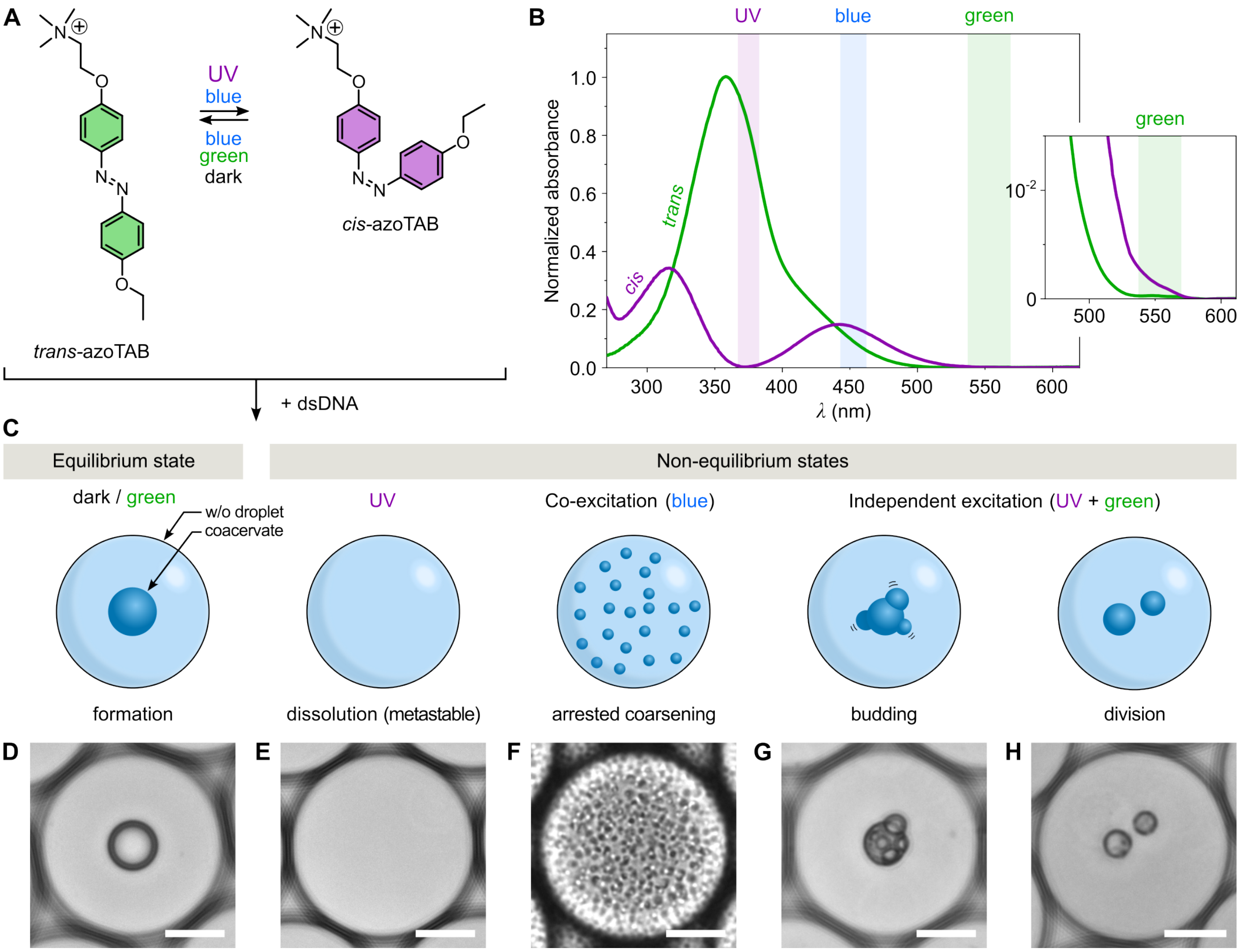

**Figure 1. Light-controlled molecular switching selects distinct coacervate states.** (**A**) Schematic of *trans*-*cis* photoisomerization of azoTAB. (**B**) UV/vis absorption spectra of *trans*- (green) and *cis*-azoTAB (purple). Shaded regions indicate the excitation wavelengths used to drive isomerization (purple: 375 nm, selective *trans*→*cis* conversion; blue: 460 nm, bidirectional *trans*↔*cis* switching; green: 560 nm, selective *cis*→*trans* conversion). Inset shows a zoom of the green wavelength range. (**C**) Schematic of phase behavior of dsDNA/azoTAB coacervates confined in water-in-oil (w/o) droplets under different illumination conditions. (**D-H**) Representative bright-field microscopy images of the observed regimes ($c_0$ = 20 mM): stable droplet (D), dissolution (E), arrested coarsening (F), budding (G), and division (H). Scale bars, 20 μm.

## Programming phase separation with molecular switching

We used an established photoswitchable coacervate system formed by complexing short double-stranded DNA (dsDNA) with the cationic azobenzene surfactant *trans*-azobenzenetrimethylammonium bromide (*trans*-azoTAB) (*30*), and confined it in monodisperse water-in-oil (w/o) microfluidic droplets at matched stoichiometric charge concentrations $c_0$ in 100 mM NaCl (Fig. 2A and fig. S1). In the dark, *trans*-azoTAB drives phase separation into a single coacervate droplet; UV illumination (375 nm) converts the surfactant predominantly to its *cis* form and dissolves the coacervate within seconds, while green light (560 nm) restores the *trans* state and re-induces phase separation within minutes (Figs. 2A,C, movies S1 and S2). These reversible transitions are governed by wavelength-dependent photostationary states of azoTAB (fig. S2, table S1 and Note S1).

To quantify this photoswitchable behavior, we mapped the phase diagram as a function of *trans* fraction $x_{trans}$. In the dark ($x_{trans}$ = 100%), the coacervate volume fraction $\phi$ increases linearly with $c_0$ (fig. S3), consistent with mass conservation between a dilute and a dense phase of fixed composition (*36*, *37*); the resulting coexistence concentrations determined from droplet size measurements are in quantitative agreement with independent bulk UV-visible measurements (Note S2). Extending this analysis across the full range of $x_{trans}$ using thermal relaxation experiments, in which coacervates reform in the dark following UV pre-irradiation as *cis* relaxes to *trans*, resolves the dependence of $\phi$ on $x_{trans}$, quantitatively captured by a diffusion-limited growth model (Note S3 and fig. S4). This reconstructs the binodal, independently validated by bulk spin-down and turbidity assays (Fig. 2B). As $x_{trans}$ decreases from 100% to 10%, the dilute-phase concentration increases more than 20-fold while the dense-phase concentration remains nearly constant, indicating that compositional changes are absorbed almost entirely by the dilute phase. A coarse-grained two-component Flory-Huggins model captures the overall phase boundary (Note S4 and fig. S5).

Light also sets the kinetics of these transitions. Under UV illumination, strong absorption confines photoisomerization to a thin interfacial layer (fig. S6), producing progressive dissolution of the coacervate (Fig. 2C). Photothermal contributions are negligible under these conditions owing to rapid thermal diffusion at the droplet scale. The dissolution time $\tau_{\mathrm{dis}}$ shows a reaction-limited regime at low UV intensity, matching independently measured *trans→cis* isomerization kinetics (fig. S7 and Note S5), and approaches a diffusion-limited regime at high intensity for all $c_0$ tested (Fig. 2D). In the crossover regime, a reaction-diffusion model coupling photoisomerization to diffusive transport quantitatively predicts the dissolution dynamics, and collapses individual coacervate radius trajectories onto a master curve governed by a single dimensionless ratio between photochemical and diffusive rates (fig. S8 and Note S6). Time-interrupted illumination further shows that dissolution becomes light-independent once the composition crosses the binodal, confirming that UV drives the system across the phase boundary, after which the coacervate dissolves spontaneously in the thermodynamic sense (Fig. 2F, purple triangles, and fig. S9).

Following UV-induced dissolution, green illumination restores phase separation through nucleation and growth (Fig. 2C). Because green light is only weakly absorbed, photoisomerization proceeds uniformly throughout the droplet (fig. S6), and nucleation occurs only after the *trans* fraction reaches the phase boundary. Consistent with this picture, the nucleation time $\tau_{\mathrm{nuc}}$ scales inversely with green-light intensity (Fig. 2E), matching the *cis→trans* photoisomerization kinetics (fig. S7), and converts to a critical *trans* fraction coinciding with the equilibrium binodal (Fig. 2F, green triangles, Note S7). Independently generating defined *trans* fractions under controlled green illumination reconstructs the same phase boundary (fig. S10). Classical nucleation theory accurately describes the nucleation kinetics (fig. S11 and Note S8). Subsequent droplet growth by coalescence proceed independently of illumination, indicating that light controls the onset of phase separation whereas post-nucleation evolution remains governed by equilibrium transport (fig. S12).

Together, these measurements identify $x_{trans}$ as the sole thermodynamic control parameter governing phase separation, recovered consistently across equilibrium, kinetic, and illumination-controlled measurements. They establish molecular switching as a quantitative handle over both the thermodynamics and kinetics of phase separation.

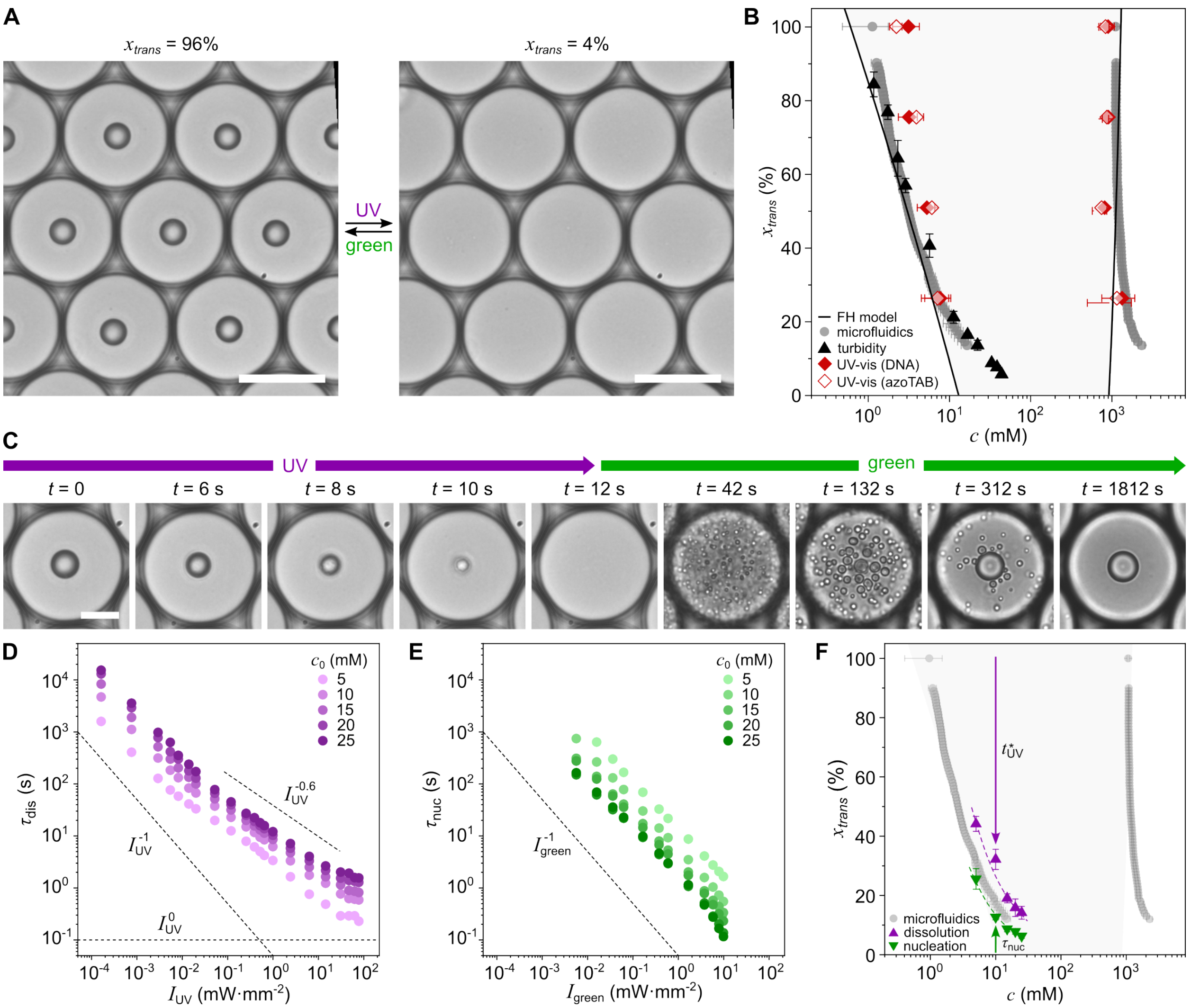

**Figure 2. Photoswitchable coacervation in microfluidic droplets.** (**A**) Bright-field microscopy images of w/o emulsion droplets containing azoTAB and dsDNA ($c_0$ = 15 mM) under green (left) and UV (right) illumination. Scale bars, 50 µm. (**B**) Phase diagram of the azoTAB-dsDNA coacervate system at varying $x_{trans}$. Grey symbols show values derived from relaxation experiments in microfluidics. Error bars on concentration represent the standard error obtained from linear fits of $\phi$ versus $c_0$ at fixed $x_{trans}$. Phase boundaries are also independently determined using bulk UV/vis spin-down assays at pre-defined *trans*-azoTAB fractions (solid red diamonds: dsDNA concentration; open red diamonds: azoTAB concentration) and turbidity recovery assays at varying $c_0$ (solid black triangles). Error bars for UV/vis measurements represent mean ± s.d. (*n* = 3 independent samples prepared at three different $c_0$ values); turbidity data are shown as mean ± s.d. (*n* = 3 independent samples per concentration). Black solid lines indicate the binodal curve obtained from a Flory-Huggins (FH)–based model (Note S4). (**C**) Time-lapse bright-field microscopy images of a coacervate ($c_0$ = 15 mM) irradiated continuously with UV light ($I_{UV}$ = 1.3 mW·mm$^{-2}$), followed by continuous green light illumination ($I_{green}$ = 16.4 mW·mm$^{-2}$), showing coacervate dissolution followed by reformation. Scale bar, 20 µm. (**D**) Dissolution time, $\tau_{dis}$, of coacervate droplets prepared at varying $c_0$ as a function of UV light intensity, $I_{UV}$. Data are shown as mean ± s.d. (*n* = 7 droplets). Scaling exponents for reaction-limited, diffusion-limited, and crossover regimes are indicated by dotted lines. (**E**) Nucleation time, $\tau_{nuc}$, of coacervate droplets prepared at varying $c_0$ as a function of green light intensity, $I_{green}$. Data are shown as mean ± s.d. (*n* = 4 droplets). The scaling exponent of the isomerization-driven process is highlighted as a dotted line. (**F**) Critical fraction of *trans*-azoTAB for coacervate dissolution and nucleation reported on the phase diagram. Upward purple triangles correspond to values estimated from the coacervate

volume fraction measured at the critical UV irradiation time $t^*_{\mathrm{UV}}$, beyond which dissolution becomes light-independent (fig. S9). Downward green triangles indicate values obtained from the nucleation time $\tau_{\mathrm{nuc}}$ under green illumination, by converting $\tau_{\mathrm{nuc}}$ into $x_{trans}$ (Note S7). Data are shown as mean ± s.d. ($n$ = 6 light intensities for UV; $n$ > 10 light intensities for green light), and overlaid to values derived from relaxation experiments in microfluidics (grey symbols). The dotted lines serve as guides to the eye.

### Continuous molecular cycling selects a finite droplet size

We next tested whether continuous molecular switching could alter the dynamical evolution of coacervates at a fixed average molecular composition. Unlike UV or green illumination alone, which drive the system toward nearly pure *cis* or *trans* endpoints, respectively, blue light continuously excites both photoisomerization pathways, establishing a photostationary state with sustained *cis*⇌*trans* turnover at constant average composition (~64% *trans*; table S1). Phase separation therefore remains thermodynamically favorable while molecules continuously switch between strongly and weakly interacting states.

Following UV-induced dissolution, blue illumination triggers nucleation with isomerization-controlled kinetics (Figs. 3A,B), consistent with the relatively weak absorption of blue light. However, unlike green illumination, growth is arrested before a single condensate droplet is obtained, yielding a persistent population of micron-scale droplets that do not coarsen (Fig. 3A and movie S3). The steady-state droplet size can be tuned from approximately 0.5 to 2 μm across the investigated concentrations (Figs. 3C,D and fig. S13). Conversely, pre-existing coacervates fragment under blue illumination into an indistinguishable population of micron-sized droplets (Fig. 3A and movie S4). Therefore, the same steady state is reached from opposite initial conditions, identifying it as a non-equilibrium attractor rather than a kinetically arrested intermediate.

When illumination ceases, droplets immediately resume coarsening and relax toward a single equilibrium coacervate (Fig. 3A), with a size-derived *trans* fraction of 57 ± 6% (mean ± s.d., $n$ = 180 droplets across 5 different $c_0$ and both nucleation and fragmentation protocols, fig. S14), close to the photostationary *trans* fraction of 64%. Likewise, equilibrium samples prepared at the same average *trans*/*cis* ratio, but without continuous switching, coarsen normally to a single droplet (fig. S4). Thus, the persistence of finite-sized droplets cannot be explained by the average molecular composition alone: it requires continuous molecular switching through sustained energy input. This arrested coarsening is consistent with the mechanism predicted for chemically active phase separation, in which reversible conversion between phase-favoring and phase-disfavoring species counterbalances diffusive coarsening (*9*, *38*). Here, photoswitching provides this reversible molecular turnover without net chemical consumption, with light supplying the required energy input. The forward and reverse isomerization fluxes remain yet coupled under blue illumination, with both rates determined by the same light intensity.

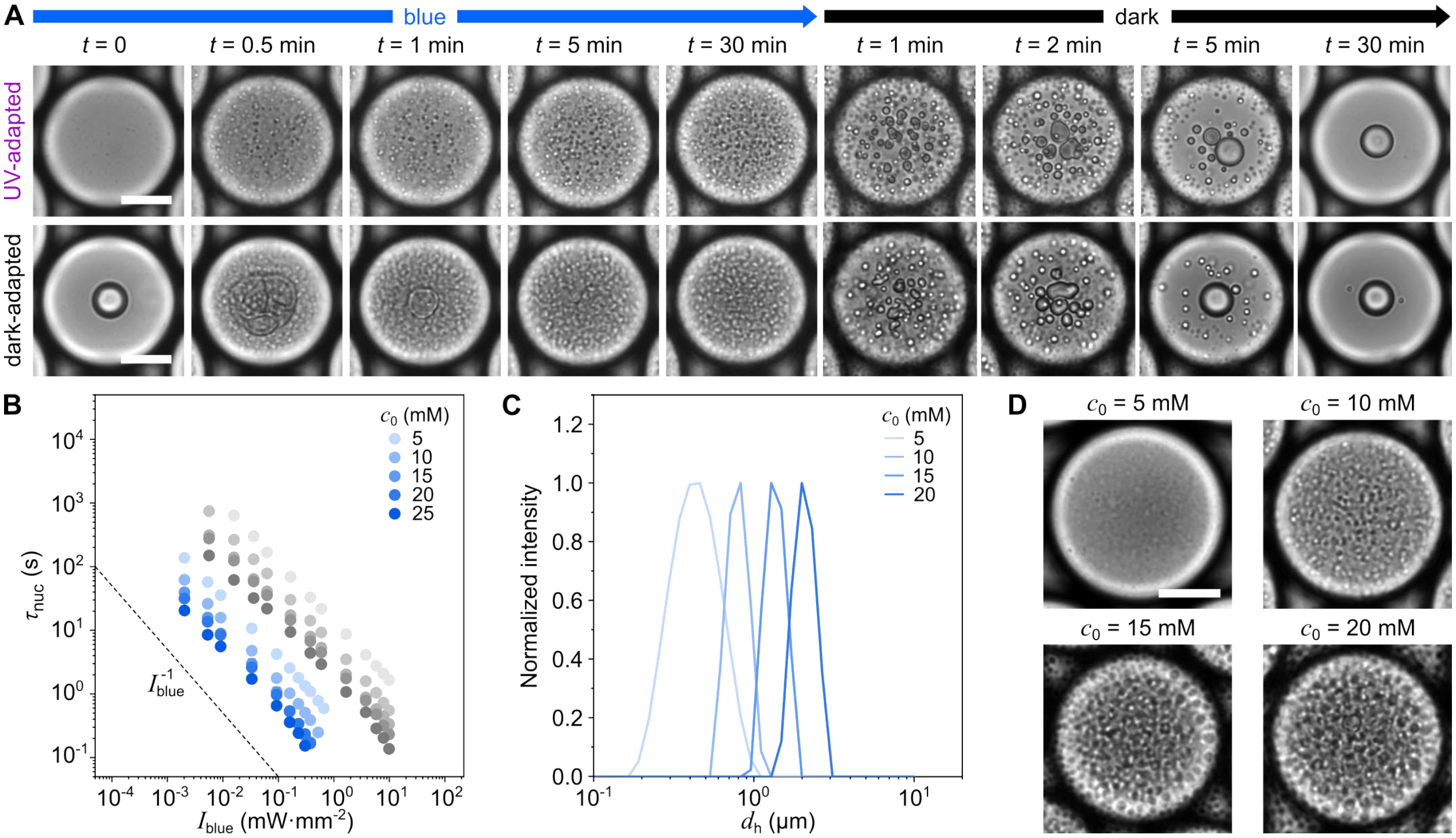

**Figure 3. Arrested coarsening under photostationary switching.** (**A**) Time-lapse bright-field microscopy images of a coacervate ($c_0$ = 15 mM) initially equilibrated either under UV illumination or in the dark, then exposed to continuous blue light ($I_{\text{blue}}$ = 60 mW·mm$^{-2}$) for 30 min, followed by incubation in the dark for 30 min. UV-pretreated samples undergo blue-light-induced nucleation of micron-sized droplets, whereas pre-formed coacervates fragment into a similar persistent droplet population. Upon cessation of illumination, droplets resume coarsening and relax to a single condensate. Scale bars, 20 μm. (**B**) Nucleation time, $\tau_{\text{nuc}}$, of coacervate droplets prepared at varying $c_0$ as a function of blue light intensity, $I_{\text{blue}}$ (blue symbols). Grey symbols show to the nucleation time determined under green light irradiation. Data are shown as mean ± s.d. ($n$ = 4 droplets). (**C**) Intensity-weighted hydrodynamic diameter distribution ($d_h$) for bulk dsDNA/azoTAB solutions prepared at varying $c_0$ and irradiated for 5 min with blue light. (**D**) Bright-field microscopy images of the steady-state coacervate population of samples prepared at varying $c_0$ and exposed to continuous blue light ($I_{\text{blue}}$ = 60 mW·mm$^{-2}$) for 30 min.

## Independent reaction fluxes destabilize droplet interfaces

Simultaneous UV and green illumination provides a means to independently tune the magnitude of the opposing reaction fluxes, thereby extending control beyond the coupled switching achieved under blue illumination. Because UV light is strongly absorbed, *trans*→*cis* conversion occurs predominantly near the droplet surface, whereas green light penetrates throughout the droplet and drives *cis*→*trans* conversion in the bulk (fig. S6), imposing distinct spatial photoisomerization profiles. This combination of independently tunable and spatially distinct reaction fluxes provides a route to drive interfacial instabilities, as predicted theoretically (*10*).

The resulting competing molecular fluxes indeed produce an interfacial instability absent under single wavelength excitation. Increasing the UV-to-green intensity ratio $I_{\text{UV}}/I_{\text{green}}$ drives a progression from stable droplets to small-amplitude surface undulations (movie S5), persistent budding, and finally dissolution (Fig. 4A). During budding, protrusions repeatedly form and remerge with the parent droplet without detaching (Fig. 4B and movie S6). This interfacial instability is independent of preparation history, arising equally from pre-formed coacervates and from UV-dissolved states (fig. S15), and exists only while both reaction pathways remain active: removing either or both wavelengths restores equilibrium behavior, whereas reinstating dual illumination restores budding (Fig. 4C and fig. S16).

The instability selects a characteristic geometric scale. Buds form with a nearly constant radius (~3 μm), largely independent of illumination conditions and $c_0$, whereas the parent droplet size decreases continuously with increasing $I_{\mathrm{UV}}/I_{\mathrm{green}}$ ratio until reaching $R/R_0 \approx 0.65$, beyond which dissolution occurs (Figs. 4D–F and fig. S17). The region of the $(I_{\mathrm{UV}}, I_{\mathrm{green}})$ parameter space supporting instability expands with increasing $c_0$, although the characteristic bud size remains essentially unchanged (Fig. 4G and fig. S18).

Together, these observations identify independently programmable opposing reaction fluxes as a sufficient condition for destabilizing coacervate interfaces. The spatial asymmetry in photoisomerization is expected to generate chemical potential gradients between the droplet surface and bulk, providing a plausible driving force for the interfacial instability. Whether this instability originates directly from reaction-flux gradients, from changes in the effective interfacial tension induced by sustained molecular turnover as in active interfaces (*8*, *39*, *40*), or from a combination of both remains an open theoretical question. Nevertheless, our results identify spatially asymmetric reaction flux as an essential ingredient for destabilizing the droplet interface, in agreement with theoretical paradigms (*10*, *12*). The resulting instability exhibits a characteristic protrusion size that remains largely invariant across conditions, whereas the illumination ratio quantitatively controls the onset of shape-symmetry breaking.

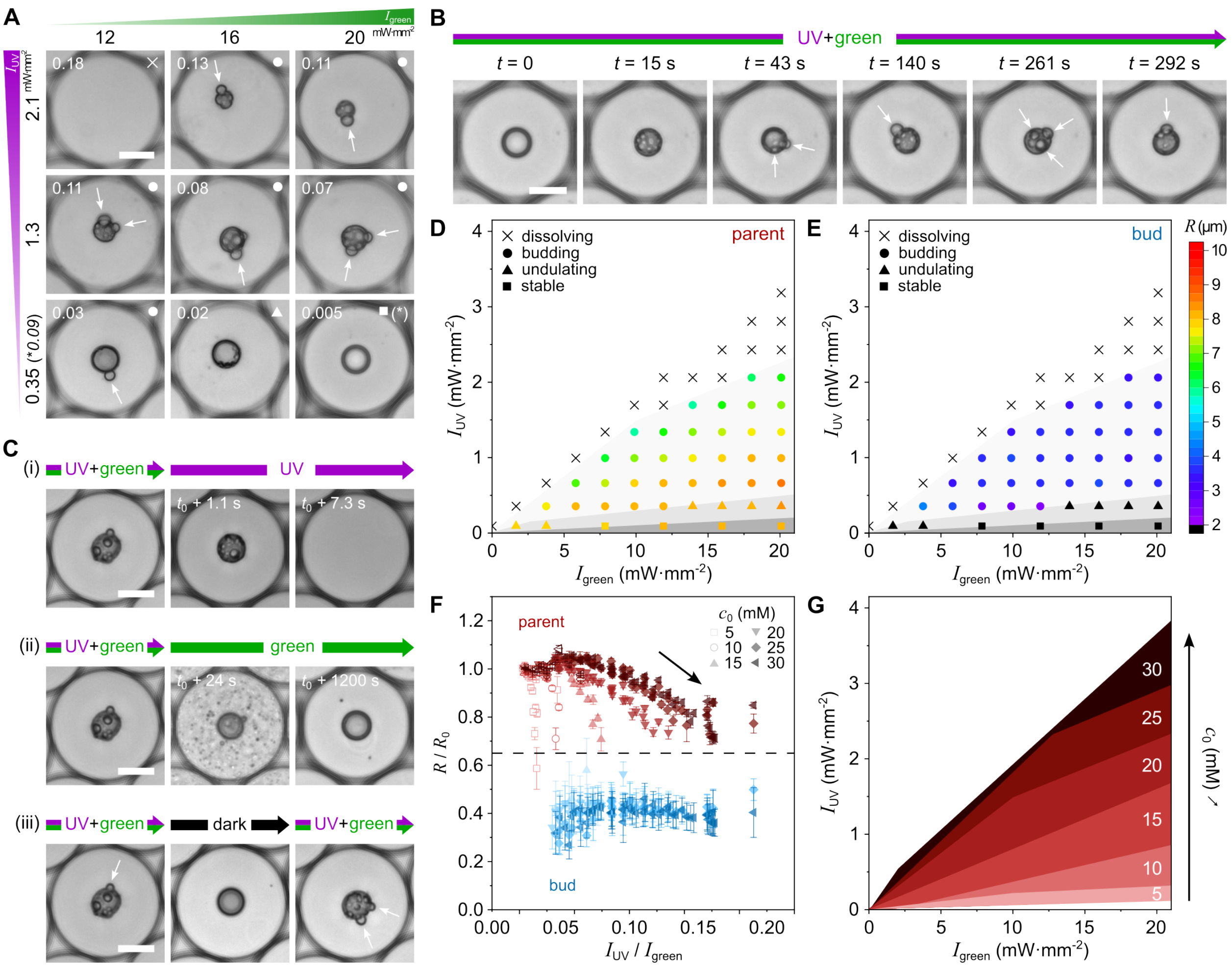

**Figure 4. Competing photochemical fluxes drive interfacial instabilities.** (**A**) Representative bright-field microscopy images of a coacervate ($c_0$ = 20 mM) at varying $I_{\mathrm{UV}}$ and $I_{\mathrm{green}}$, showing different regimes identified by white symbols on images: stable (square), undulating (triangle), budding (circle) and dissolving (cross) droplets. White arrows indicate buds; numbers on images indicate the corresponding $I_{\mathrm{UV}}/I_{\mathrm{green}}$ ratio. Scale bar, 20 µm. (**B**) Time-lapse bright-field microscopy images of a coacervate ($c_0$ = 20 mM) co-irradiated continuously with UV ($I_{\mathrm{UV}}$ = 1.3 mW·mm$^{-2}$) and green ($I_{\mathrm{green}}$ = 20 mW·mm$^{-2}$) light, showing representative budding events (see also movie S6). White arrows indicate buds. Additional buds emerge along the optical axis and appear with different contrast due to defocusing. Scale bar, 20 µm. (**C**) Representative bright-field microscopy images of a coacervate ($c_0$ = 20 mM) co-irradiated with UV ($I_{\mathrm{UV}}$ = 1.0 mW·mm$^{-2}$) and green ($I_{\mathrm{green}}$ = 20 mW·mm$^{-2}$) light, followed by either single irradiation with UV (i) or green (ii), or incubation in the dark before re-instating dual UV and green light illumination. White arrows indicate buds. Scale bar, 20 µm. (**D**,**E**) Nonequilibrium phase response in the ($I_{\mathrm{UV}}$, $I_{\mathrm{green}}$) parameter space showing dissolving (crosses), budding (circles), undulating (triangles) and stable (squares) regimes for $c_0$ = 20 mM. Symbol color denotes the steady-state radius measured ∼1 min after dual illumination for the parent droplet (D) and bud (E) (rainbow color bar, 2-10 µm). Grey shading indicates regime boundaries: dark grey, stable; grey, undulating; light grey, budding. In E, black symbols correspond to conditions where no budding occurs. (**F**) Normalized parent droplet (red) and bud (blue) radius as a function of the illumination ratio $I_{\mathrm{UV}}/I_{\mathrm{green}}$ for varying $c_0$: 5 mM (squares), 10 mM (circles), 15 mM (upward triangles), 20 mM (downward triangles), 25 mM (diamonds), 30 mM (leftward triangles). Darker hues correspond to higher $c_0$. Filled symbols indicate droplets exhibiting budding; open symbols correspond to droplets not developing buds. The dotted line mark $R/R_0$ = 0.65, indicating the approximate onset of coacervate

dissolution. Data are shown as mean ± s.d. ($n \geq 10$ parent droplets and buds). (**G**) Nonequilibrium phase boundary in the in the ($I_{\text{UV}}$, $I_{\text{green}}$) parameter space at and varying $c_0$ (full datasets in fig. S18).

## Phase stability converts interfacial instability into active division

Lowering the ionic strength transforms the same interfacial instability into complete droplet division. Whereas droplets at 100 mM NaCl dissolve after reaching the instability threshold, droplets at 10 mM NaCl instead undergo a reproducible sequence of shrinking, protrusion, necking and splitting into a remnant and daughter droplet (Fig. 5A and movie S7). Both droplets subsequently continue shrinking until the daughter dissolves, consistent with division emerging at the boundary between budding and dissolution (Fig. 5B and fig. S19). This division-decay pathway dominates at all illumination ratios in the division regime (fig. S20). Division timing is stochastic across droplets within a sample, though its average onset decreases with illumination ratio, consistent with a faster approach to the instability threshold (fig. S21).

Division does not arise from a different interfacial instability, but from modifying the thermodynamic landscape in which it develops. Reducing the ionic strength shifts the phase boundary toward lower $x_{\text{trans}}$ (i.e. increases the phase separation propensity) and slows molecular diffusion in the dense phase, as independently measured by FRAP (fig. S22), consistent with stronger molecular interactions that suppress complete dissolution under UV light (fig. S23). As a result, droplets evolve beyond the instability threshold identified during budding and proceed to complete splitting rather than dissolving (Fig. 5C).

Once division becomes accessible, droplet size emerges as a second key parameter governing its dynamical outcome. Smaller droplets enter the division regime at lower UV-to-green intensity ratios than larger droplets (fig. S24), consistent with the increasing contribution of surface-localized UV reactions at higher surface-to-volume ratio. Reducing the initial droplet size therefore narrows the overall instability region in ($I_{\text{UV}}$, $I_{\text{green}}$) space but shifts its dominant outcome from budding to division (fig. S25).

Under weak driving (low $I_{\text{UV}}/I_{\text{green}}$), daughter droplets coexist for extended periods without shrinking or recoalescing, producing a long-lived multi-droplet state (fig. S26). Under stronger driving, the daughter gradually shrinks while the remnant droplet regrows, eventually re-entering the instability regime, and dividing again, producing sustained growth-division cycles under continuous illumination (Fig. 5D,E, fig. S27 and movie S8). We observed repeated growth-division cycles over multiple successive events, with individual droplets undergoing up to ten divisions under continuous illumination (fig. S28). The emergence of these cyclic dynamics at lower driving coincides with higher photostationary *trans* fractions under these conditions (fig. S29), which favor phase separation and likely enable regrowth of the remnant between successive division events.

Division of DNA condensates has recently been reported following spatially localized irradiation, which triggers division at a selected location within the droplet (*33*). Here, by contrast, division emerges and recurs under spatially uniform illumination. Although the incident light is spatially uniform, the two wavelengths indeed generate different internal photoisomerization profiles because of their distinct absorption and the spatial distribution of the two isomers. This internal asymmetry provides the spatially heterogeneous driving required for the interfacial instability without externally imposed illumination patterns.

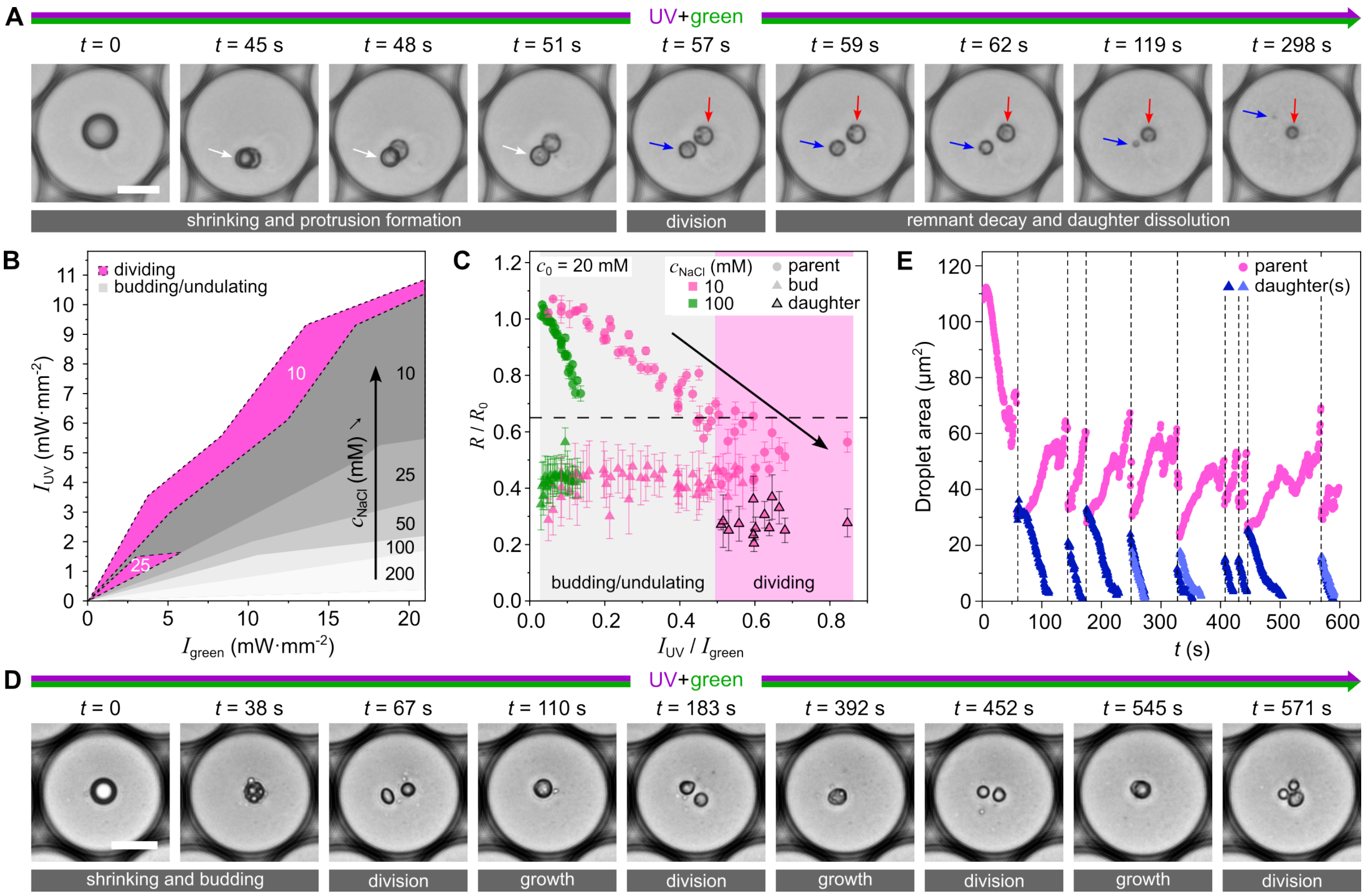

**Figure 5. Phase stability controls the transition from interfacial instability to active droplet division.** (**A**) Time-lapse bright-field microscopy images of a w/o emulsion droplet prepared at $c_0$ = 20 mM in 10 mM NaCl and co-irradiated continuously with UV ($I_{UV}$ = 2.4 mW·mm$^{-2}$) and green ($I_{green}$ = 3.8 mW·mm$^{-2}$) light, showing dividing droplets (see also movie S7). White arrows indicate a bud protruding from the coacervate surface, while red and blue arrows point to the remnant and daughter droplet, respectively. Scale bars, 20 μm. (**B**) Nonequilibrium phase boundaries in the ($I_{UV}$, $I_{green}$) parameter space showing undulating/budding regimes (grey) and dividing regimes (pink, dotted boundary) at $c_0$ = 20 mM and varying NaCl concentrations $c_{NaCl}$, as indicated (full datasets in fig. S19). (**C**) Normalized parent droplet (circles) and bud (triangles) or daughter droplet (edged triangles) radius as a function of the illumination ratio $I_{UV}/I_{green}$ for $c_0$ = 20 mM in $c_{NaCl}$ = 100 mM (green) or 10 mM (pink). The dotted line mark $R/R_0$ = 0.65. The undulating/budding (resp. dividing) regime is identified as a grey (resp. pink) shaded area. Data are shown as mean ± s.d. ($n \geq 4$ parent droplets, buds and daughter droplets). (**D**) Representative time-lapse bright-field microscopy images of division/growth cycles ($c_0$ = 10 mM, $c_{NaCl}$ = 10 mM, $I_{UV}$ = 1.0 mW·mm$^{-2}$, $I_{green}$ = 9.9 mW·mm$^{-2}$; see also movie S8). Scale bar, 20 μm. (**E**) Time-dependent evolution of the remnant (pink) and daughter (blue) droplets area under continuous UV ($I_{UV}$ = 1.0 mW·mm$^{-2}$) and green ($I_{green}$ = 9.9 mW·mm$^{-2}$) light illumination. Vertical dotted lines highlight division events, characterized by an abrupt decrease of the remnant droplet size and the formation of one or two daughter droplets, followed by the gradual dissolution of the former and re-growth of the remnant droplet.

Our results show that reversible molecular switching turns molecular driving into a tunable determinant of non-equilibrium phase behavior, without irreversible chemical turnover. The balance of opposing photoisomerization pathways and the stability of the condensed phase together determine whether droplets show arrested coarsening, interfacial destabilization or recurrent division. Strikingly, division emerges under continuous, spatially uniform illumination: the spatial asymmetry required for interface destabilization arises internally from the different penetration depths of the two wavelengths and the resulting asymmetric photoisomerization fluxes. Spatially heterogeneous molecular driving therefore need not be imposed externally; it can emerge from the coupling between light transport, molecular switching, and phase separation. These findings provide a physical framework for converting external energy input into selectable non-equilibrium dynamics of phase-separated matter through reversible molecular switching. The emergence of recurrent growth and division from such minimal molecular components also illustrates how dynamic compartmentalization could provide a physical basis for protocell proliferation and, ultimately, the evolution of compartmentalized molecular systems.

**Acknowledgments:** We thank Alexis Maillard for help with preliminary investigations, Xavier Brilland for help with azoTAB synthesis, Lionel Buisson for help with optical set-ups, Cristian Villalobos for help with image analysis, and Frédéric Nallet, Jean-François Joanny, Jacques Prost David Zwicker and Guido Kusters for stimulating discussions. ChatGPT (OpenAI) was used for language editing. The authors reviewed and approved the final text and take full responsibility for its content.

**Funding:**

Chinese Scholarship Council (ZL)

French National Research Agency grant ANR-21-CE06-0022-01 (NM)

IdEx Bordeaux (NM)

Région Nouvelle-Aquitaine grant AAPR 2020-2019-8330510 (NM)

Univ. Bordeaux Research Network Frontiers of Life (NM, JCB)

Institut Universitaire de France (AB, JCB)

**Author contributions:**

Conceptualization: ZL, LT, AB, JCB, NM

Methodology: ZL, TB, SL, ERJG, NM

Investigation: ZL, SL, ERJG

Visualization: ZL, NM

Funding acquisition: ZL, AB, JCB, NM

Project administration: NM

Supervision: JCB, NM

Writing – original draft: ZL, NM

Writing – review & editing: ZL, LT, AB, JCB, NM

**Competing interests:** The authors declare no competing interests.

**Data, code, and materials availability:** All data are available in the main text or the supplementary materials.

## Supplementary Materials

Materials and Methods

Supplementary Notes

Figs. S1 to S29

Tables S1 to S3

References (*41–55*)

Movies S1 to S8

# Supplementary Materials for

## Light selects non-equilibrium states of phase-separated droplets

Zi Lin, Thomas Beneyton, Suzanne Lafon, Edison Rafael Jimenez Granda, Liangfei Tian, Alexandre Baron, Jean-Christophe Baret, Nicolas Martin

Corresponding author: alexandre.baron@crpp.cnrs.fr; jean-christophe.baret@u-bordeaux.fr; nicolas.martin@crpp.cnrs.fr

**The PDF file includes:**

**Other Supplementary Materials for this manuscript include the following:**

Movies S1 to S8

## Materials and Methods

Materials

Low molecular weight double-stranded DNA (dsDNA) from salmon sperm (ref. 31149) and sodium chloride were purchased from Sigma-Aldrich. Tetramethylrhodamine-labelled single-stranded DNA (TAMRA-ssDNA) and Cyanine 5-labelled single-stranded DNA (Cy5-ssDNA) were purchased from Integrated DNA Technologies (sequences are given in table S3). AzoTAB was synthesized using a three-step reaction by azocoupling *p*-ethoxyaniline with phenol, followed by alkylation with dibromoethane then quaternisation with trimethylamine (see below).

Milli-Q water was used to prepare stock solutions of dsDNA (100 mM nucleobase concentration, $M_w = 303.7$ g mol$^{-1}$, pH 8), azoTAB (100 mM, $M_w = 408.5$ g mol$^{-1}$, pH 8) and NaCl (2 M). The pH of the dsDNA and azoTAB solutions was adjusted to 8 using 1 M or 0.1 M NaOH. The dsDNA stock solution was aliquoted and stored at -20°C. The azoTAB stock solution was heated at 80°C for 1 hour before use to ensure complete isomerisation to the *trans* state, and stored at room temperature in the dark (wrapped in aluminium foil).

Synthesis of azoTAB

The synthesis of azoTAB was performed following previously reported protocols (see refs. (*30*, *41*)) as described below.

***Synthesis of 4-ethoxy-4'-hydroxy-azobenzene (azoH):*** Concentrated HCl (17 mL) and ice (80 g) were added to a 1:1 v/v ethanol : water solution (160 mL) containing p-ethoxyaniline (10.3 mL, 80 mmol, 1 equiv.) and sodium nitrite (5.5 g, 80 mmol, 1 equiv.) in an ice bath (T = 0°C) in a fume hood. The mixture was stirred for 1h. Cold water (42 mL) containing phenol (7.5 g, 80 mmol, 1 equiv.) and NaOH (6.4 g, 160 mmol, 2 equiv.) was then carefully added to the solution and the mixture was stirred for another 90 min by keeping the temperature below 5°C. The pH of the solution was then adjusted to 1 with concentrated HCl and left to stand for 30 min. The resulting precipitate was filtered, thoroughly washed with water and dried under vacuum overnight to give 4-ethoxy-4'-hydroxyazobenzene (azoH) as a dark brown powder (76% yield). $^1$H NMR (400 MHz, $CDCl_3$): δ = 7.86 (d, $^3$J(H-H) = 8 Hz, 2H; Ar-H), 7.82 (d, $^3$J(H-H) = 8 Hz, 2H; Ar-H), 6.99 (d, $^3$J(H-H) = 8 Hz, 2H; Ar-H), 6.94 (d, $^3$J(H-H) = 8 Hz, 2H; Ar-H), 4.12 (q, $^3$J(H-H) = 6 Hz, 2H; $CH_2$), 1.46 ppm (t, $^3$J(H-H) = 4 Hz, 3H; $CH_3$); $^{13}$C NMR (400 MHz, $CDCl_3$): δ = 161.0 (Ar-C), 157.9 (Ar-C), 147.1 (Ar-C), 146.8 (Ar-C), 124.6 (Ar-C), 124.4 (Ar-C), 115.8 (Ar-C), 114.7 (Ar-C), 63.8 ($CH_2O$), 14.8 ppm ($CH_3$).

***Synthesis of 4-ethoxy-(4'-(2-bromoethyloxy)phenyl)azobenzene (azoBr):*** A mixture of 4-ethoxy-4'-hydroxy-azobenzene (2.4 g, 10 mmol, 1 equiv.), 1,2-dibromoethane (5.6 g, 3 equiv.), potassium carbonate (2.07 g, 1.5 equiv.) and potassium iodide (0.083 g, 0.05 equiv.) were refluxed in 50 mL of butanone for 48 h in the dark in a fume hood. The reaction mixture was filtered hot to remove solid impurities and salt, and the residue was washed with butanone. The filtrate was collected and the solvent was removed under reduced pressure. The obtained solid was dissolved in dichloromethane (20 mL) and extractions were performed with NaOH solution (1M, 2 × 8 mL) then pure water (2 × 8 mL). The organic phase was dried with $MgSO_4$ and concentrated. The crude product was recrystallized with hot filtration from ethanol and dried under vacuum to give 4-ethoxy-(4'-(2-bromoethyloxy)phenyl)azobenzene (azoBr) as an orange powder (54% yield). $^1$H NMR (400 MHz, $CDCl_3$): δ = 7.92 (q, $^3J$(H-H) = 8.2 Hz, 4H; Ar-H), 7.00 (dd, $^3J$(H-H) = 8.6 Hz, 4H; Ar-H), 4.37 (t, $^3J$(H-H) = 8 Hz, 2H; $CH_2O$), 4.12 (q, $^3J$(H-H) = 6 Hz, 2H; $CH_2O$), 3.67 (t, $^3J$(H-H) = 7 Hz, 2H; $CH_2Br$), 1.46 ppm (t, $^3J$(H-H) = 6 Hz, 3H; $CH_3$); $^{13}$C NMR (400 MHz,

$CDCl_3$): δ = 161.5 (Ar-C), 160.2 (Ar-C), 146.8 (Ar-C), 146.3 (Ar-C), 124.8 (Ar-C), 124.6 (Ar-C), 114.9 (Ar-C), 114.8 (Ar-C), 68.0 ($CH_2O$), 63.9 ($CH_2O$), 28.8 ($CH_2Br$), 14.8 ppm ($CH_3$).

***Synthesis of azobenzene trimethylammonium bromide (azoTAB):*** 1 g of 4-ethoxy-(4'-(2-bromoethyloxy)phenyl)azobenzene (4.4 mmol, 1 equiv.) was dissolved in 80 mL of dry THF, followed by the addition of a 33% solution of trimethylamine in ethanol (4.2 mL, 11.5 mmol, 4 equiv.). The mixture was stirred for 6 days in the dark in a fume hood. The resulting precipitate was filtered, washed with THF, and dried under vacuum. The crude product was recrystallized twice from ethanol and dried under vacuum overnight to give azobenzene trimethylammonium bromide (azoTAB) as an orange powder (36% yield). $^{1}$H NMR (400 MHz, DMSO): δ = 7.86 (d, $^{3}J$(H-H) = 8 Hz, 2H; Ar-H), 7.82 (d, $^{3}J$(H-H) = 8 Hz, 2H; Ar-H), 7.17 (d, $^{3}J$(H-H) = 8 Hz, 2H; Ar-H), 7.08 (d, $^{3}J$(H-H) = 8 Hz, 2H; Ar-H), 4.56 (m, 2H; $CH_2O$), 4.11 (q, $^{3}J$(H-H) = 6 Hz, 2H; $CH_2O$), 3.82 (m, 2H; $CH_2N$), 3.18 (s, 9H; $CH_3N$), 1.35 ppm (t, $^{3}J$(H-H) = 4 Hz, 3H; $CH_3$); $^{13}$C NMR (400 MHz, DMSO): δ = 161.3 (Ar-C), 159.9 (Ar-C), 147.1 (Ar-C), 146.4 (Ar-C), 124.7 (Ar-C), 124.5 (Ar-C), 115.8 (Ar-C), 115.4 (Ar-C), 64.5 ($CH_2O$), 64.1 ($CH_2N$), 62.5 ($CH_2O$), 53.6 ($CH_3N$), 15.0 ppm ($CH_3$).

Photophysical properties of azoTAB

***Equilibrium fraction of trans- and cis-azoTAB under irradiation.*** AzoTAB solutions stored in the dark contain 100% *trans*-azoTAB, which is the thermodynamically more stable isomer. The fraction of *trans*- and *cis*-azoTAB at the photostationary state (PSS) after irradiation at different wavelengths was determined by UV/vis spectroscopy. 2 mL of a 30 µM solution of azoTAB were prepared by diluting the stock *trans*-azoTAB solution in Milli-Q water in a quartz cuvette (10 × 10 × 45 mm). The cuvette was wrapped in aluminum foil to limit the influence of ambient light, and the solution stirred using a magnetic stirring bar (8 × ø3 mm, FisherBrand) at 800 rpm on a magnetic stirring plate (IKAMAG). The solution was irradiated at different wavelengths (UV, 375 nm; blue, 460 nm; green, 530-580 nm) from the top of the cuvette for at least 5 minutes to reach the photostationary state using a LED (CoolLED, pE-300) equipped with an optical fiber. Baseline-corrected UV/vis spectra were acquired on a spectrophotometer (Varian, Cary 100 Bio) immediately after irradiation. Changes in the absorbance at 380 nm were used to determine the *trans*- and *cis*-azoTAB fractions at each photostationary state as listed in table S1 and detailed in Note S1.

***Extinction coefficients.*** Pure *trans*-azoTAB and UV-adapted *cis*-azoTAB solutions were prepared at varying concentrations (5-40 µM) in a quartz cuvette (10 × 10 × 45 mm), and their UV/vis spectra were measured on a spectrophotometer (Varian, Cary 100 Bio). The spectra of pure *cis*-azoTAB were deduced from that of the UV-adapted sample using the residual *trans*-azoTAB fraction determined above. The extinction coefficients of both azoTAB isomers were extracted from the slope of the linear fit of absorbance at a specific wavelength as a function of concentration, with the intercept set to 0, and are listed in table S2.

***Thermal relaxation kinetics.*** A quartz cuvette (10 × 10 × 45 mm) containing 2 mL of 30 µM azoTAB was first irradiated with UV light for ca. 5 minutes to reach the photostationary state. The cuvette was then incubated in the dark inside a UV/vis spectrophotometer (Varian, Cary 100 Bio) equipped with a Peltier temperature controller. Changes in the absorbance at 380 nm were monitored over time at varying temperatures (T = 20, 25, 30, 35, 40, 45 °C) until it reached a plateau. Data were fitted with a mono-exponential function to recover the fraction of *trans*-azoTAB during thermal relaxation (see Note S3.5).

***Photoisomerisation kinetics.*** The azoTAB isomerization kinetics under different light irradiation conditions was followed by UV/vis spectroscopy using a fiber optic spectrometer (Ocean Optics, USB 400) to allow simultaneous side light irradiation with a LED and absorption measurements (pump-probe experimental setup). 2 mL of a 50 µM *trans*-azoTAB solution were prepared in a quartz cuvette (10 × 10 × 45 mm) placed inside a cuvette holder (Thorlabs, CVH100) and stirred using a magnetic stirring bar (8 × ø3 mm, FisherBrand) at 800 rpm on a magnetic stirring plate (IKAMAG). The solution was irradiated at different wavelengths (UV, 375 nm; blue, 460 nm; green, 530-580 nm) from the side of the cuvette (orthogonally to the optical path) using a LED (CoolLED, pE-300) equipped with an optical fiber. A halogen white light source (Ocean Optics, HL-2000-FHSA equipped with a band-pass blue filter, 340-480 nm) was used for absorbance measurements. The absorbance was recorded through the solution on the fiber optic spectrometer as $A = -\log I/I_0$, where $I_0$ was the reference intensity measured on pure Milli-Q water. Time-dependent changes of the absorbance at 380 nm were recorded using the Spectra Suite software (Ocean Optics) until the photostationary state was reached, and used to extract a characteristic isomerization time (see Note S5). Measurements were performed at different irradiation intensities. The light intensity at a specific wavelength was measured at the sample position with a power meter (Thorlabs, PM100) coupled to a photodiode power sensor (Thorlabs, S120VC). After every measurement, the sample was irradiated with green or UV light to regenerate 100% *trans*-azoTAB or 96% *cis*-azoTAB isomers, respectively, before the next kinetic measurement was acquired.

Microfluidics experiments

***Fabrication of microfluidics chips.*** Microfluidic circuits (fig. S1) designed on AutoCAD were transferred onto poly(dimethylsiloxane) (PDMS, Sylgard 184) chips from SU8-3000 negative photoresist (MicroChem Corp) molds, fabricated using standard soft-lithography procedures. Microfluidic chips were assembled by bonding the engraved PDMS chips onto glass slides using a plasma cleaner (Diener electronic, Pico, 80 % power with air at <0.4 mbar for 30 s) to activate both surfaces. The microfluidic channels were rinsed with fluorosilane (Aquapel) to achieve hydrophobic coating, followed by fluorinated oil (Novec 7500) to remove any residues before use.

***Microfluidic experiment set up.*** The microfluidic chip was placed on an inverted optical microscope (Olympus, IX71) and connected to 1 mL syringes (Braun) controlled by Nemesys syringe pumps (Cetoni) via PTFE tubing (Fisher Scientific; inner diameter, 0.3 mm; outer diameter, 0.76 mm). The flows inside channels and sizes of droplets formed were checked during production using a high-speed camera (Phantom v210) coupled to the microscope.

***Production of coacervates in water-in-oil droplets.*** Unless stated otherwise, all coacervates were prepared in the presence of 100 mM NaCl. To ensure this controlled saline environment for coacervation within water-in-oil (w/o) droplets, aqueous starting solutions of dsDNA and azoTAB were prepared in 100 mM NaCl prior to microfluidic encapsulation. Three aqueous solutions – dsDNA and azoTAB, and a central saline stream (100 mM NaCl) – were injected into separate microfluidic channels, merged at a flow-focusing junction, and emulsified into monodisperse w/o droplets using fluorinated oil (Novec 7500) containing 2 wt% perfluoropolyether-polyethyleneglycol diblock copolymer surfactant (PFPE-PEG) (*42*). Flow rates of 140 $\mu L \cdot h^{-1}$ were used for both the dsDNA and azoTAB streams, while the central saline stream was set at 80 $\mu L \cdot h^{-1}$, resulting in a 2.57-fold dilution of both dsDNA and azoTAB in the final w/o droplets. Accordingly, the concentrations of the starting solutions were adjusted to obtain final concentrations of dsDNA and azoTAB of 5, 10, 15, 20, 25, 30 mM within w/o droplets, while maintaining an equimolar charge ratio of both components. Emulsion droplet diameter was adjusted at approximately 60 µm by varying the oil stream flow rate between 180 and 280 µL·h$^{-1}$, yielding a droplet generation frequency of approximately 5 Hz.

***Observation chamber.*** For long-term imaging, w/o droplets were loaded into a custom observation chamber (fig. S1) designed to confine droplets into a quasi-two-dimensional array. The chamber was assembled from two glass microscopy slides (76 × 25 × 1 mm), separated by a 60 µm-thick double-sided adhesive spacer (1375 SDAG, Adhesif) patterned using a cutting plotter (Graphtec CE6000-40). Two inlet holes were created in the top slide by micro-sandblasting, and NanoPortT M assembled with NanoTightT M fittings (IDEX Health and Science) were fixed using UV-curing adhesive (Loctite 3529, Henkel). The assembled chamber had internal dimensions of approximately 1 × 3 cm and 60 µm in depth, corresponding to a volume of 18 µL, allowing storage of more than $10^4$ w/o droplets. Prior to use, the chamber was treated with Aquapel and dried under argon. Droplets containing *trans*-azoTAB and dsDNA were transferred directly from the microfluidic device via external tubing, after which the chamber was sealed and disconnected. Samples were stored in the dark for several hours (or irradiated with green light as specified) to induce coacervate formation. The chamber was reused multiple times after cleaning between each experiment by flushing fluorinated oil.

Coacervation phase diagram

***Coacervate volume fraction measurement.*** Bright-field images were acquired by focusing on the equator plane of the w/o emulsion droplets or the coacervates on an inverted microscope (Leica DMI4000B) equipped with a 20×/0.40 NA air objective (Leica HC PL FLUOTAR L). Two different approaches were then used to determine the w/o droplet and coacervate volumes, and to calculate the coacervate volume fraction.

*For w/o emulsion droplets:* we targeted the production of emulsion droplets closer in diameter to the height of the observation chamber ($h = 60$ µm) by adjusting the flow rates of the oil stream. In this regime, non-confined droplets adopt a spherical shape considering the limit of small capillary number. However, in some cases, the emulsion droplets exhibited a diameter slightly higher than the height of the observation chamber, resulting in a flattened shape departing from a pure spherical shape. To account for this distortion, $h_0 = h/(2L)$ is defined where $L$ is the droplet radius calculated from the projected 2D surface area of an emulsion droplet, measured using FIJI (ImageJ).

- If $h_0 \geq 1$, the droplet was non-confined and its volume was thus calculated as $(4/3)\pi L^3$.
- If $h_0 < 1$, the droplet shape was approximated as a filled torus, and its volume was calculated using the following simplified expression (*43*):

$$V_{droplet} = \frac{\pi(2L)^3 h_0}{4}\left[1 - 2h_0 + \frac{5}{3}{h_0}^2 + \frac{\pi}{2}h_0(1 - h_0)\right] \qquad \text{(eq. S1)}$$

*For coacervates:* given their smaller size compared to w/o droplets, coacervates systematically adopted a spherical shape (as confirmed by confocal fluorescence imaging). Segmentation by intensity thresholding was applied to target the dark outer rim resulting from internal light reflection within the coacervate (*44*), using an optimal cut-off value in the grayscale histogram (automated "Huang" method). The area of the segmented coacervate region was then measured using particle analysis, from which the radius $R$ was calculated, and the volume using $V = (4/3)\pi R^3$.

*For both*: measurements were conducted on over 250 droplets from 5 independent samples for each coacervate component concentration, and the average value and standard deviation determined.

***Relaxation-induced coacervate reformation in microfluidic droplets.*** The observation chamber containing coacervates-in-w/o droplets was placed on an inverted microscope (Zeiss, LSM 980) equipped with a 10×/0.45 NA air objective (Zeiss, Plan-Apochromat). Coacervates were first dissolved by irradiation with UV light (385 nm) for ca. 1 minute using an epifluorescence source (Colibri 5, Zeiss) and an appropriate band-pass filter. The chamber was then kept in the dark at a fixed temperature (~25 ± 1 °C), and bright-field images were acquired every 30 minutes. A long-pass filter (> 600 nm) was placed in the illumination path to prevent unintended azoTAB photoisomerization during imaging. For each condition, the volume of the re-forming coacervate and that of the surrounding w/o droplet were measured from bright-field images, allowing determination of the coacervate volume fraction. Measurements were performed on more than 150 droplets per sample, from which mean values and standard deviations were calculated. Experiments were conducted across a range of azoTAB and dsDNA concentrations (5, 10, 15, 20, 25, 30 mM, equimolar charge ratio). These relaxation-induced reformation data were used to construct the phase diagram of the system (Note S3).

***Production of coacervates at predefined trans-azoTAB fractions.*** To validate the relaxation-based phase diagram against equilibrium preparations, coacervates were independently generated at predefined *trans*-azoTAB fractions under otherwise identical microfluidic conditions. These experiments were performed at a fixed final concentration of 5 mM dsDNA and 5 mM azoTAB in the w/o droplets. AzoTAB solutions were prepared by mixing 100 mM stock solutions of *trans*-azoTAB (dark-adapted at 65 °C for 30 min) and *cis*-azoTAB (UV-irradiated for 30 min) at nominal *trans* fractions of 30%, 40%, 50%, 60%, or 80%. After mixing, the solution was immediately injected into the microfluidic device and emulsified as described above. Because *cis*-azoTAB undergoes thermal relaxation toward the *trans* state after mixing, the elapsed time between preparation of the azoTAB solution and the first observation of coacervates in the droplets was recorded. The corresponding increase in *trans*-azoTAB fraction during this interval was estimated from independently measured thermal relaxation kinetics at the same temperature (Note S3.5) and added to the nominal value to obtain the effective *trans*-azoTAB fraction at the time of observation. Uncertainty in this effective *trans*-azoTAB fraction was calculated by propagating a ± 1 °C temperature uncertainty during coacervate formation.

***Concentration measurements via spin-down assays.*** The concentrations of dsDNA and azoTAB in the coacervate and supernatant phases were measured by UV/vis spectroscopy after centrifugation ("spin-down" assay), to validate the phase diagram obtained via microfluidics.
Coacervate suspensions (100 µL) were prepared by mixing dsDNA and azoTAB at equimolar charge ratio (10, 15 or 20 mM each) in 100 mM NaCl. Mixtures of *trans*- and UV-adapted *cis*-azoTAB were used at four defined *trans* fractions (100%, 75%, 50%, 25%). Sample were equilibrated in the dark at 4 °C for 30 minutes, then centrifuged at 20,000 ×g at 4 °C for 10 minutes to separate the dense and dilute phases. The supernatant (90 µL) was collected, while the remaining 10 µL containing the coacervate phase were diluted with 90 µL of 1 M NaCl to induce complete coacervate dissolution. Both supernatant and dissolved coacervate samples were sonicated, vortexed and irradiated with UV to ensure full dissociation of dsDNA and azoTAB. Two parallel quantification protocols were then applied.

*For dsDNA*: Samples were applied to a pre-equilibrated G-25 Sephadex column (2.1 mL, Cytiva) using 1M NaCl as eluent to separate azoTAB and dsDNA by size exclusion. 10 fractions of ca. 200 µL were collected sequentially; the first fractions contained dsDNA only, as confirmed by UV/vis spectra showing a characteristic 260 nm peak and no azoTAB absorbance in the 350-450 nm range. The exact volume of each fraction was measured by pipetting, and *ds*DNA concentrations were determined at 260 nm using an extinction coefficient of 7,950 ± 65 $L \cdot mol^{-1} \cdot cm^{-1}$ obtained from calibration. Column recovery was also calibrated by loading pure dsDNA solutions of known concentration (1, 5, 10, 15, 20, 25, 30 mM), yielding a linear relation between injected and recovered dsDNA amounts and allowing correction for material loss in the resin. Total dsDNA amounts in the coacervate and supernatant phases were calculated by summing concentrations across dsDNA-containing fractions, correcting for collected volumes and column recovery calibration.

*For azoTAB*: Size exclusion separation was unnecessary since azoTAB has a distinct absorption band far from dsDNA absorption. Samples were analyzed directly by UV/vis spectroscopy after heating to 75 °C to ensure complete conversion of all azoTAB to the *trans* form. AzoTAB concentrations were quantified at 380 nm using an extinction coefficient of 18,030 $L \cdot mol^{-1} \cdot cm^{-1}$ from calibration. Each condition was prepared and measured in triplicate, and the reported data represent mean values with pooled standard deviations.

Finally, the measured dsDNA and azoTAB concentrations were converted to concentrations in the coacervate and diluted phase phases using the coacervate volume fractions independently determined from w/o microfluidic droplet measurements at the corresponding *trans*-azoTAB fractions.

***Turbidity assays.*** UV-adapted *cis*-azoTAB and dsDNA were mixed at equimolar charge ratios (40-80 µL, 1-40 mM each) in 100 mM NaCl. Each mixture was prepared in triplicate to obtain mean values and standard deviations. Samples were irradiated with UV light for ca. 5 minutes to ensure complete conversion of azoTAB to the *cis* form, then transferred to a 384-well microplate (Greiner Bio-One). The plate was placed in a microplate reader (SpectraMax iD3, Molecular Devices) maintained at 30 °C. Turbidity was monitored by measuring the absorbance at 700 nm every 5-10 min for up to 18 h. Two reference wells containing (i) pure UV-adapted *cis*-azoTAB (30 µM) and (ii) pure *trans*-azoTAB (30 µM) were simultaneously monitored at 380 nm to track the gradual photochemical relaxation of the *cis* sample toward the *trans* state and to define the 100% *trans* reference, respectively. The onset of turbidity (sharp increase in absorbance at 700 nm) was taken as the starting point of coacervate formation, marking the transition from the homogeneous phase to the two-phase region (i.e., crossing the bimodal). The corresponding *trans*-azoTAB fraction was estimated from the parallel absorbance relaxation kinetics at 380 nm and assigned as the threshold composition at the dilute branch of the binodal, where the bulk concentrations in the well equal the saturation concentration for coacervate formation.

Light-activated phase transitions

***Coacervate dissolution under UV light.*** Coacervate dissolution under UV light was monitored via bright-field imaging at the equatorial plane of coacervates using a long-pass filter (> 600 nm) to prevent photo-isomerization by the white light source. UV light irradiation was performed through the objective of an inverted epifluorescence microscope (details below). The light power at the sample plane was measured at different output settings using a power meter (Thorlabs, PM100) fitted with a photodiode sensor (Thorlabs, S120VC).

*Determination of the dissolution time:* UV light irradiation was performed on an inverted epifluorescence microscope (Zeiss, LSM 980) equipped with a 63×/0.75NA air objective (Zeiss, LD Plan-Neofluar) and a LED light source (Colibri 5, Zeiss) with an appropriate band-pass filter (385 nm). The light power was controlled via ZEN3.3 software (Zeiss) together with an attenuator (neutral density filter, Zeiss). The dissolution time ($\tau_{dis}$) was determined from the temporal evolution of the mean grey value within entire individual w/o droplets using FIJI (ImageJ). The intensity ratio between each frame and the final image was computed; the first frame for which this ratio exceeded 0.99, indicating complete coacervate dissolution into a homogeneous one-phase solution, was defined as the dissolution time. 7 droplets per sample, located at the center of the field of view, were analyzed to obtain mean values and standard deviations.

*Time-interrupted illumination experiments:* UV light irradiation was performed on an inverted microscope (Olympus, IX71) equipped with a 20×/0.50NA air objective (Olympus, UPlanSApo) and a UV LED (365 nm, Thorlabs, M365LP1) connected to a driver (Thorlabs, LEDD1B) and an automatic shutter with controller (Thorlabs, SHB1T). Data acquisition and shutter control were synchronized via a DAQ card and LabVIEW program (National Instruments). Samples were exposed to UV light for varying durations controlled by automated shutter timing, and then monitored for at least an additional minute after the shutter was closed. The shutter-open period was recorded as the irradiation time. The shortest irradiation time that resulted in complete dissolution within the subsequent one-minute observation window was defined as the critical irradiation time $t^*$. The subsequent dissolution time in the dark $t_{dark}$ was defined as the interval between shutter closure and the frame corresponding to complete coacervate dissolution. 6 droplets per condition were analyzed to obtain mean values and standard deviations.

*Measurement of coacervate radius:* The coacervate radius was determined from time-lapse bright-field images using FIJI (ImageJ). In most datasets, droplets were segmented by intensity thresholding to detect their contours, from which the projected area was measured and converted to radius assuming a spherical geometry. In some recordings, thresholding accuracy decreased toward the end of dissolution due to changes in focal plane and image contrast; in these cases, the image contrast was inverted before applying the same segmentation procedure. For samples displaying internal reorganization or vacuolization (where the dark outer rim was less distinguishable), segmentation was instead performed using the Trainable Weka Segmentation plugin in FIJI (ImageJ), a machine learning-based approach (*45*). Regions corresponding to coacervates and continuous phase were manually annotated in a subset of representative images (typically 5-10 frames spanning early, mid, and late stages), and the trained classifier was then applied to the full time series for automated contour detection, followed by area-based size measurement.

***Coacervate reformation under green light.*** Unless stated otherwise, green light irradiation was performed through the objective of on an inverted epifluorescence microscope (Zeiss, LSM 980) equipped with a 63×/0.75NA air objective (Zeiss, LD Plan-Neofluar) and a LED light source (Colibri 5, Zeiss) with an appropriate band-pass filter (555 nm). The light power was controlled via ZEN3.3 software (Zeiss) together with an attenuator (neutral density filter, Zeiss), and measured at the sample plane using a power meter (Thorlabs, PM100) fitted with a photodiode sensor (Thorlabs, S120VC). In all conditions, the focus was adjusted to the equatorial plane of coacervates formed in the dark, after which samples were briefly irradiated with UV light to dissolve the encapsulated coacervates prior to green-light exposure.

*Determination of the nucleation time:* Bright-field imaging was performed using a long-pass filter (> 600 nm) to prevent photo-isomerization by the white light source. Image acquisition began immediately upon green light illumination. The mean grey value within each w/o droplet was extracted from time-lapse sequences using FIJI (ImageJ) and normalized to the initial value. This normalized intensity was plotted as a function of time, and the onset of coacervate nucleation was determined as the intersection between the initial plateau and the tangent to the first decreasing segment of the curve. The corresponding time point was defined as the characteristic nucleation time ($\tau_{\mathrm{nuc}}$). Each experiment was performed on four independent droplets in the same sample to obtain mean values and standard deviations.

*Determination of the sedimentation time:* Coacervate sedimentation was visualized by fluorescence microscopy. TAMRA-labelled ssDNA (TAMRA-ssDNA, ex/em = 546/579 nm) was added to the microfluidic starting solution of dsDNA at a molar ratio of ssDNA : dsDNA = 1 : 2.5 × $10^4$. Fluorescence images were recorded for 10 minutes under continuous green-light irradiation, and the resulting sequences were analyzed in FIJI (ImageJ). For each droplet, the fluorescence signal within the region corresponding to the final single coacervate was normalized to the total fluorescence within the whole w/o droplet, defining a central fluorescence intensity fraction (fig. S12). This fraction was plotted as a function of time and further normalized between 0 and 1 using the mean values from the first and last frames of the sequence as the minimum and maximum, respectively. The onset of sedimentation was determined from the point where the tangent to the initial rising portion of the normalized curve intersected the time axis. The characteristic sedimentation time ($\tau_{\mathrm{sed}}$) was defined as the interval between this onset time and the previously determined nucleation time obtained from bright-field measurements under the same conditions. Measurements were repeated on three independent droplets in the same sample to obtain mean values and standard deviations.

*Determination of the nucleation rate:* Fluorescent coacervate nuclei were detected in TAMRA-ssDNA-doped samples within a fixed observation volume using a laser-scanning confocal microscope (Zeiss LSM 980) equipped with a 63×/0.75NA air objective (Zeiss, LD Plan-Neofluar). A 561 nm laser was used both to trigger coacervate nucleation and to image emerging nuclei sequestering TAMRA-ssDNA. Laser intensity was varied between 5% and 30% (5% increments) using ZEN 3.3 software (Zeiss). The focal depth was adjusted to approximately 1.1 μm using a 1 Airy Unit pinhole aperture. Nuclei were identified using FIJI (ImageJ) within a region of interest (1400 $\mu m^2$) corresponding to the interior of a single w/o droplet. Automated segmentation was performed using a fixed global threshold determined from the bimodal fluorescence histogram of TAMRA-ssDNA, followed by particle analysis to count distinct nuclei. The same segmentation parameters were applied to all datasets to ensure comparability. Only the initial 20 s of each recording were analyzed to avoid the influence of subsequent coalescence. Nuclei density was calculated as the number of detected nuclei per unit volume (1400 × 1.1 $\mu m^3$) and plotted as a function of time. The nucleation rate $J$ was obtained from the slope of a linear fit to the initial rise in nuclei density, corresponding to the number of nuclei formed per unit time and volume. Measurements were repeated for three independent w/o droplets in the same sample, and reported values represent mean values and standard deviation.

*Coacervate reformation under limited green light irradiation.* Reformation experiments under temporally controlled green-light exposure were performed on an inverted microscope (Leica DMI4000 B) equipped with a 20×/0.40 NA air objective (Leica HC PL FLUOTAR L). Green light irradiation was delivered through the objective using a LED light source (CoolLED pE-300) and an appropriate band-pass filter (> 600 nm) for defined durations, controlled by Micro-Manager software. Following irradiation, samples were incubated in the dark for 30 min before measuring coacervate volume fractions from more than 30 droplets in the same sample to obtain mean values and standard deviations. Experiments were repeated at initial concentrations of 5, 10, 15, 20, 25 and 30 mM.

***Coacervate size arrest under blue light.*** Blue light irradiation was performed through the objective of on an inverted epifluorescence microscope (Zeiss, LSM 980) equipped with a 63×/0.75NA air objective (Zeiss, LD Plan-Neofluar) and a LED light source (Colibri 5, Zeiss) with an appropriate band-pass filter (475 nm). The light power was controlled via ZEN3.3 software (Zeiss) and measured at the sample plane using a power meter (Thorlabs, PM100) fitted with a photodiode sensor (Thorlabs, S120VC). In all conditions, the focus was adjusted to the equatorial plane of coacervates formed in the dark, after which samples were briefly irradiated with UV light to dissolve the encapsulated coacervates prior to blue-light exposure.

*Determination of the nucleation time:* Bright-field imaging was performed using a long-pass filter (> 600 nm) to prevent photo-isomerization by the white light source. Image acquisition began immediately upon blue light illumination. The mean grey value within each w/o droplet was extracted from time-lapse sequences using FIJI (ImageJ) and normalized to the initial value. This normalized intensity was plotted as a function of time, and the onset of coacervate nucleation was determined as the intersection between the initial plateau and the tangent to the first decreasing segment of the curve. The corresponding time point was defined as the characteristic nucleation time ($\tau_{\text{nuc}}$). Each experiment was performed on four independent droplets in the same sample to obtain mean values and standard deviations.

***Non-equilibrium behavior under co-irradiation with UV and green lights.*** Experiments were performed on an inverted epifluorescence microscope (Zeiss, LSM 980) through a 63×/0.75NA air objective (Zeiss, LD Plan-Neofluar), using a LED light source (Colibri 5, Zeiss) together with an appropriate multiband-pass filter (UV, 385 nm; green, 555 nm). The UV and green light intensities were controlled independently via ZEN3.3 software (Zeiss) and measured at the sample plane using a power meter (Thorlabs, PM100) fitted with a photodiode sensor (Thorlabs, S120VC). Samples were continuously irradiated with combined UV and green light at varying intensities and imaged for 5 min (560 ms per frame) to reach a steady state. Time-lapse microscopy images were analyzed using a custom MATLAB pipeline code.

*Budding regime:* images were normalized and contrast-enhanced using adaptive intensity stretching, after which contours of budding coacervates were extracted using Canny edge detection. Coacervates were initially localized in the first frame using a circular Hough transform within a predefined radius range, and subsequently tracked within a local search regions centered on their previous positions. The largest connected edge was taken as the contour. From these contours, the centroid and mean radial distance (defined as the mean radius) were computed and propagated across frames.

Each contour was resampled into 200 evenly spaced points and circularly smoothed to reduce noise. The radial profile relative to the centroid, together with the corresponding curvature and circularity, were computed based on the reconstructed contour. The equilibrium size of the parent coacervate was defined as the mean ± s.d. radius of the non-budding circular contour, considering only the frames ($n > 3$) with circularity > 0.99 in the 5$^{\text{th}}$ minute.

Bud candidates were identified as local maxima in the radial profile, requiring a minimum valley drop of 1 μm on both sides of the peak. Their boundaries were refined by locating curvature minima flanking each peak, corresponding to neck regions. Each segmented region was fitted with a circle to obtain the bud center and radius, while the remaining contour was also fitted with a circle to define the core droplet. These detected bud candidates were classified and tracked independently for each coacervate using a nearest-neighbor criterion: a bud was linked to a trajectory from the previous frame if the center-to-center distance was below 0.7 × the radius of the previously detected bud; otherwise, a new bud trajectory was initiated. The characteristic bud size was defined as the fitted radius at the frame where the bud center reached its maximal distance from the fitting center of the core droplet, and the equilibrium bud sizes (mean ± s.d.) were obtained by averaging over the 5$^{\text{th}}$ minute ($n > 3$ frames).

The final reported mean ± s.d. values of equilibrium size of parent coacervate and bud were computed across droplets ($n > 3$ droplets) using weighted statistics: the overall mean was computed as a weighted average of individual droplet means, using the sample count as weights; the overall standard deviation was calculated using a pooled formulation that accounts for both within-droplet variance and between-droplet variability.

*Division regime:* a condition was classified as a division regime when more than 50% of coacervates within the field of view underwent at least one division event (typically n ≥ 5 coacervates analyzed per condition). The onset of division was defined as the first frame in which the daughter and remnant coacervates became fully separated. The sizes of the daughter (smaller) and remnant (larger) coacervates were measured from this frame. The size of the parent coacervate immediately before division was estimated from the combined volume of the daughter and remnant coacervates.

<u>Modeling light attenuation within droplets</u>

The spatial distribution of light intensity within a coacervate (radius $R$ = 4 μm) or w/o droplet (radius $L$ = 30 μm) was estimated using a Beer-Lambert attenuation model under collimated illumination. Each spherical domain was discretized on a 2D Cartesian grid (500 × 500 points) representing a cross-sectional plane of the sphere. The absorption coefficient $\mu$ was obtained from the molar extinction coefficient $\varepsilon$ of the considered species (table S2) and its concentration $c$ via $\mu = \ln(10)\varepsilon c$.

Illumination was modelled as a uniform plane wave incident from the bottom of the sphere, propagating along the positive $z$ direction. For each point inside the sphere, the optical path length $l(x, z)$ was defined as the vertical distance from the entry point at the lower spherical boundary to the point of interest: $l(x, z) = z + \sqrt{R^2 - x^2}$. The local light intensity was then evaluated using the Beer-Lambert law: $I(x, z) = I_0 e^{-\mu l(x,z)}$.

The resulting 2D intensity field was visualized using a colormap, where intensity values were normalized to the maximum intensity ($I/I_{\max}$) and mapped to a continuous scale from 0 to 1. Points outside the spherical domain were masked out.

<u>Experiments on bulk coacervate suspensions</u>

***Imaging of bulk coacervate suspension.*** A custom-made observation chamber was assembled using glass coverslips covalently functionalized with poly(ethylene glycol) (PEG) brushes to avoid wetting of coacervates. PEGylation was performed by incubation of ethanol-rinsed glass coverslips (0.13-0.17 mm thick) for 48 hours into a 10 mL toluene solution containing 500 μL of 3-[methoxy(polyethyleneoxy)propyl]tri-methoxysilane (90%, 6-9 PE units, abcr GmbH). Coverslips where subsequently rinsed with ethanol and water, and dried with compressed air before being assembled into an observation capillary chamber.

Bulk coacervate solutions were prepared at a total concentration of $c_0$ = 15 mM in 100 mM NaCl by diluting the stock solutions of NaCl (2 M), *trans*-azoTAB (100 mM, pH 8) and dsDNA (100 mM nucleobase concentration, pH 8) in this order in Milli-Q water in an Eppendorf tube. The sample was kept in the dark for ca. 5 minutes, then an aliquot of the suspension was transferred into a custom-made PEG-functionalized capillary chamber. Coacervates were left to settle on the PEGylated coverslip for ca. 5 minutes before observation. Optical microscopy imaging was performed on an inverted microscope (Leica, DMI4000B) equipped with a 63×/1.4 NA oil immersion objective (HC PL APO). The halogen light source was filtered using a long-pass filter (> 600 nm) to avoid unwanted azoTAB isomerization in the sample during imaging. A LED light source (CoolLED, pE-300) was used for *in situ* light irradiation of the coacervates through the objective, using a manual shutter and band-pass filters to select specific wavelength ranges (UV, 370-380 nm; blue: 450-460 nm; green: 555-565 nm).

***Dynamic light scattering (DLS).*** DLS was performed to determine the hydrodynamic radius of coacervates formed under blue light. Bulk coacervate solutions were prepared at a total concentration of $c_0$ = 5-20 mM in 100 mM NaCl. Samples (100 µL) were loaded into a low-volume quartz cuvette (Hellma Analytics) and irradiated with blue light (CoolLED pE-400, 450 nm) before measurement. DLS measurements were performed at 25°C using a Malvern Zetasizer. For each measurement, 5 consecutive acquisitions of 10 s were collected, ensuring a short total acquisition time to minimize sample evolution during DLS acquisition (performed in the dark). After each set, samples were re-irradiated with blue light, and measurements were repeated three times.

***Coacervate characterization at varying salt concentration.***

*Fusion dynamics:* A freshly prepared *trans*-azoTAB/dsDNA droplet suspension ($c_0$ = 10 mM, $c_{\mathrm{NaCl}}$ = 0-200 mM) was rapidly loaded onto a custom-made PEG-functionalized capillary slide. Several coalescence events between contacting droplets could then be observed in real time by optical microscopy imaging (same set-up as above). The aspect ratio of 10 individual droplets undergoing coalescence was measured as a function of time using FIJI (ImageJ) as the ratio of the long and short axes, respectively, of the droplets. The data was fitted to a mono-exponential decay function to determine the relaxation time $\tau_{\mathrm{r}}$, and the characteristic length scale, $d$, of the droplets was the measured radius after coalescence. $\tau_{\mathrm{r}}$ is expected to be directly proportional to $d$: $\tau_{\mathrm{r}} \approx (\eta/\gamma)d$, which gives the inverse capillary velocity as the ratio of the viscosity of the droplets, $\eta$, to surface tension, $\gamma$.

*Fluorescence recovery after photobleaching (FRAP):* A freshly prepared *trans*-azoTAB/dsDNA droplet suspension ($c_0$ = 20 mM, $c_{\mathrm{NaCl}}$ = 0-200 mM) was labelled with Cy5-ssDNA (23 bases, 5 µM) in an Eppendorf tube. The sample was kept in the dark for ca. 5 minutes, then an aliquot of the suspension was transferred into a custom-made PEG-functionalized capillary chamber. Coacervates were left to settle on the PEGylated coverslip for ca. 5 minutes before observation. FRAP studies were conducted with excitation at 639 nm on a laser-scanning confocal microscope (Zeiss LSM 980) equipped with a 63×/1.4NA oil immersion objective (Zeiss, LD Plan-Apochromat). The photobleached area was chosen as a circular spot of 1 µm diameter in the center of a single droplet of at least 8 µm in diameter. Measurements were performed on six different droplets from the same sample. A 5-frame pre-bleach sequence was used, followed by a 1-frame bleach at 30% 639 nm. Fluorescence recovery was monitored by measuring the fluorescence intensity during the post-bleach sequence on FIJI (ImageJ). Data was normalized through the double normalization method, then fit to a mono-exponential function to give $\tau_{\mathrm{f}}$ as the fluorescence recovery time constant. The apparent diffusion coefficient $D_{\mathrm{FRAP}}$ was obtained by the formula used for 2D diffusion: $D_{\mathrm{FRAP}} = (0.88r^2)/(4\tau_{\mathrm{f}} \ln 2)$ where $r$ is the radius of the bleached area.

## Supplementary Text

Note S1. Determination of photostationary states of azoTAB.

The UV/vis spectra shown in fig. S2 were used to determine the *trans*- and *cis*-azoTAB fractions at each photostationary state (PSS), following the procedure described in ref. (*46*). Briefly, the absorbance spectrum of an azoTAB solution at the PSS under irradiation at the excitation wavelength $\lambda_{\text{ex}}$ (denoted as $A_{\text{ex}}(\lambda)$) varies with the *trans*:*cis* composition according to the Beer-Lambert law:

$$A_{\text{ex}}(\lambda) = [\varepsilon_{trans}(\lambda)x_{trans}^{\text{ex}} + \varepsilon_{cis}(\lambda)x_{cis}^{\text{ex}}]c_0 l = A_{trans}(\lambda)x_{trans}^{\text{ex}} + A_{cis}(\lambda)x_{cis}^{\text{ex}} \qquad (\text{eq. S2})$$

where $c_0$ is the total molar concentration of azoTAB and $l$ is the optical path length; $x_{trans}^{\text{ex}}$ and $x_{cis}^{\text{ex}}$ are the photostationary fractions of *trans*- and *cis*-azoTAB, respectively, after irradiation of the excitation wavelength $\lambda_{\text{ex}}$, where $x_{trans}^{\text{ex}} + x_{cis}^{\text{ex}} = 1$; $\varepsilon_{trans}(\lambda)$ and $\varepsilon_{cis}(\lambda)$ are the wavelength-dependent molar extinction coefficients of the pure *trans* and *cis* isomers, respectively; $A_{trans}(\lambda)$ and $A_{cis}(\lambda)$ their corresponding absorption spectra.

The dark-adapted solution gives the reference spectrum of 100% *trans*-azoTAB ($x_{trans}^{\text{dark}} = 1$): $A_{\text{dark}}(\lambda) = A_{trans}(\lambda)$. By contrast, a purely *cis*-azoTAB solution cannot be obtained experimentally, since a small fraction of the *trans* isomer remains even under prolonged UV irradiation. The spectrum of the pure *cis* isomer, $A_{cis}(\lambda)$, can yet be deduced from the measured spectra under UV and blue light irradiation using:

$$A_{\text{ex}}(\lambda) = x_{trans}^{\text{ex}}A_{trans}(\lambda) + (1 - x_{trans}^{\text{ex}})A_{cis}(\lambda) \qquad (\text{eq. S3})$$

which gives:

$$A_{cis}(\lambda) = \frac{A_{\text{ex}}(\lambda) - x_{trans}^{\text{ex}}A_{trans}(\lambda)}{(1 - x_{trans}^{\text{ex}})} \qquad (\text{eq. S4})$$

To estimate the initial values of $x_{trans}^{\text{ex}}$, we note that the absorbance of the *cis* isomer is negligible compared to that of the *trans* isomer in the wavelength range 360-390 nm. At the wavelength of minimum absorbance in the UV-adapted sample ($\lambda_{\text{UV,min}}$ = 380 nm), the measured absorbance thus arises almost entirely from the *trans* species. This provides an upper bound for the *trans* fraction, given by:

$$x_{trans}^{\text{ex,max}} = \frac{A_{\text{ex}}(\lambda_{\text{UV,min}})}{A_{trans}(\lambda_{\text{UV,min}})} \qquad (\text{eq. S5})$$

These initial estimates are then refined through an iterative procedure. The spectrum of the pure *cis* isomer is recalculated independently from the spectra acquired under UV and blue light using eq. S4. At each iteration, $x_{trans}^{\text{ex}}$ is updated for both spectra until the two calculated $A_{cis}(\lambda)$ curves converge into a single one shown in fig. S2. This spectrum of pure *cis*-azoTAB is subsequently used to compute the *trans*:*cis* fractions under green light and to determine the extinction coefficient of the *cis* isomer. The obtained *trans*-/*cis*-azoTAB compositions at each photostationary state are listed in table S1.

<u>Note S2. Size-based determination of equilibrium compositions.</u>

Given that coacervates were assembled at equimolar charge concentrations of azoTAB and dsDNA, the system was treated as an effective single-component system characterized by a dilute-phase concentration $c_{\mathrm{dil}}$ and a dense-phase concentration $c_{\mathrm{coa}}$. At equilibrium, mass conservation within a droplet of volume $V_{\mathrm{drop}}$

$$V_{\mathrm{dil}}c_{\mathrm{dil}} + V_{\mathrm{coa}}c_{\mathrm{coa}} = V_{\mathrm{drop}}c_0 \qquad \text{(eq. S6)}$$

where $V_{\mathrm{dil}}$ and $V_{\mathrm{coa}}$ represent the volume of dilute- and dense phases, respectively, with $V_{\mathrm{dil}} + V_{\mathrm{coa}} = V_{\mathrm{drop}}$; and $c_0$ is the total concentration of coacervate components, with $c_0 = c_{\mathrm{DNA}} = c_{\mathrm{azo}}$ (where $c_{\mathrm{DNA}}$ is the DNA concentration expressed as nucleobase concentration, and $c_{\mathrm{azo}}$ is the concentration of azoTAB).

Defining the coacervate volume fraction as $\phi = V_{\mathrm{coa}}/V_{\mathrm{drop}}$, eq. S6 can be written as:

$$(1-\phi)c_{\mathrm{dil}} + \phi c_{\mathrm{coa}} = c_0 \qquad \text{(eq. S7)}$$

which yields a linear dependence:

$$\phi = \frac{1}{c_{\mathrm{coa}} - c_{\mathrm{dil}}}c_0 - \frac{c_{\mathrm{dil}}}{c_{\mathrm{coa}} - c_{\mathrm{dil}}} \qquad \text{(eq. S8)}$$

Thus, linear regression of $\phi$ as a function of $c_0$ provides direct access to $c_{\mathrm{dil}}$ (from the $x$-intercept) and $c_{\mathrm{coa}} - c_{\mathrm{dil}}$ (from the slope), eventually yielding $c_{\mathrm{coa}}$.

*Validation*

The above volume fraction measurements give $c_{\mathrm{dil}}$ = 1.0 ± 0.6 mM (mean ± s.e.) and $c_{\mathrm{coa}}$ = 1.07 ± 0.04 × $10^3$ mM (mean ± s.e.). These compositions are in good agreement with the *trans*-azoTAB and DNA concentrations determined via independent UV/vis measurements in bulk suspensions: $c_{\mathrm{dil}}^{\mathrm{azo}}$ = 1.9 ± 0.3 mM and $c_{\mathrm{coa}}^{\mathrm{azo}}$ = 0.8 ± 0.1 × $10^3$ mM (mean ± s.d., $n$ = 9); $c_{\mathrm{dil}}^{\mathrm{DNA}}$ = 2.7 ± 1 mM and $c_{\mathrm{coa}}^{\mathrm{DNA}}$ = 0.9 ± 0.2 × $10^3$ mM (mean ± s.d., $n$ = 3).

Note S3. Thermal relaxation-based determination of the equilibrium phase diagram.

To determine the phase boundaries of the photoswitchable coacervate system, we combined microfluidic measurements with controlled variation of the *trans*-azoTAB fraction.

*3.1 – Coacervate growth during thermal relaxation*

To relate phase behavior to molecular composition, we exploited the $cis \rightarrow trans$ thermal relaxation of azoTAB in the dark. Samples prepared at varying $c_0$ were first converted to the predominant *cis* state under UV illumination, leading to coacervate dissolution, then incubated at constant temperature in the dark, leading to gradual recovery of the *trans* fraction $x_{trans}(t)$ and spontaneous reformation of coacervates (fig. S4A).

The coacervate volume fraction $\phi(t)$ was measured by microscopy (fig. S4B), while $x_{trans}(t)$ was independently determined from UV/vis spectroscopy at the same temperature (fig. S4C, see details in section 3.5). Assuming that phase separation proceeds faster than molecular relaxation (see validation in section 3.4), the instantaneous coacervate volume fraction directly reflects the current *trans* fraction. This approach provides $\phi(x_{trans})$ without requiring pre-equilibrated mixtures (fig. S4D). The resulting curve agrees well with measurements performed on samples prepared at predefined *trans*/*cis* ratios (fig. S4E), confirming that coacervate size is governed by the average *trans* fraction.

*3.2 – Reconstruction of the binodal*

To enable quantitative comparison across datasets acquired at different total concentrations $c_0$ (fig. S4F), the experimental $\phi(x_{trans})$ curves were interpolated onto a common grid with 1% increments in $x_{trans}$ (fig. S4G). At each value of $x_{trans}$, $\phi$ was regressed as a function of $c_0$ (fig. S4H), yielding $c_{\mathrm{dil}}$ and $c_{\mathrm{coa}}$ as described in Note S2.

Uncertainty in $x_{trans}$, arising from temperature fluctuations (± 1 °C), was propagated using Monte Carlo sampling based on the independently measured relaxation kinetics. Measurement uncertainty in $\phi$ was estimated from droplet-to-droplet variability (n > 120). The total uncertainty was obtained by combining both contributions in quadrature.

This procedure yields the phase boundaries over the range $x_{trans} \in [0.12,0.9]$ for which sufficient data points are available; the extracted phase diagram was independently validated by bulk measurements using spin-down assays and turbidity recovery experiments, showing quantitative agreement (Fig. 2B).

*3.3 – Interpretation*

To corroborate these data, we developed a model for coacervate growth driven by spontaneous $cis \rightarrow trans$ conversion. We consider the growth of a single coacervate droplet of radius $R(t)$ inside a closed microfluidic droplet of radius $L$ containing DNA and azoTAB initially in a predominantly *cis* state (no coacervate). Thermal relaxation continuously converts *cis*-azoTAB into the *trans* isomer, which alone participates in phase separation. The instantaneous *trans* fraction $x_{trans}$ therefore acts as the control parameter driving condensation.

We seek to relate the coacervate volume fraction $\phi = (R/L)^3$ to $x_{trans}$. The temporal evolution of the latter follows first-order kinetics with thermal rate constant $k_{\mathrm{T}}$ (see section 3.5):

$$\frac{dx_{trans}}{dt} = k_{\mathrm{T}}(1 - x_{trans}) \qquad \text{(eq. S9)}$$

Phase separation occurs only when $x_{trans}(t)$ exceeds a threshold value $x^*_{trans}$, corresponding to the binodal of the system. Above this threshold, a single coacervate droplet nucleates and grows by diffusive capture of *trans* molecules from the surrounding dilute phase. Because the system is closed, growth is ultimately limited by finite mass conservation, leading to saturation at long times.

In the post-nucleation regime, the coupled effects of (i) continuous production of condensable material via *cis*→*trans* conversion and (ii) depletion of the dilute reservoir can be coarse-grained into an effective first-order relaxation of the coacervate volume fraction toward a finite asymptotic value:

$$\frac{d\phi}{dt} = k_{\text{eff}}\,(\phi_{\max} - \phi), \qquad x_{trans} > x^{*}_{trans} \qquad \text{(eq. S10)}$$

with $\phi$ = 0 below the threshold. Here $k_{\text{eff}}$ is an effective growth rate incorporating diffusive transport and partitioning between phases, and $\phi_{\max}$ is the saturation volume fraction set by total material conservation in the confined droplet.

The key assumption underlying this reduction is a separation of timescales: condensate relaxation is fast compared to the *cis*→*trans* conversion, such that the droplet volume adiabatically follows the amount of available condensable material. This allows the explicit dependence on microscopic transport to be absorbed into $k_{\text{eff}}$ and $\phi_{\max}$, yielding a single effective relaxation law for the macroscopic variable $\phi(t)$.

To obtain the dependence on the *trans* fraction, we combine eq. S9 and eq. S10:

$$\frac{d\phi}{dx_{trans}} = \frac{k_{\text{eff}}}{k_{\text{T}}}\frac{\phi_{\max} - \phi}{1 - x_{trans}} \qquad \text{(eq. S11)}$$

Separation of variables and integration from the nucleation point ($x^{*}_{trans}$, $\phi$ = 0) to ($x_{trans}$, $\phi$) yield the closed-form solution:

$$\phi(x_{trans}) = \phi_{\max}\left[1 - \left(\frac{1 - x_{trans}}{1 - x^{*}_{trans}}\right)^{\alpha}\right], \qquad x_{trans} > x^{*}_{trans} \qquad \text{(eq. S12)}$$

where $\alpha = k_{\text{eff}}/k_{\text{T}}$ is an effective exponent comparing the rate of droplet growth to the rate of *cis*→*trans* conversion. This expression captures the experimentally observed behavior: vanishing volume fraction below the critical *trans* fraction, rapid increase immediately after nucleation, and progressive saturation at large $x_{trans}$.

*3.4 – Validation*

Using the experimentally determined relation between $\phi$ and $x_{trans}$, simultaneous fits of all six $c_0$ concentrations with a single shared exponent (fig. 4G) yield $\alpha$ = 3.86 ± 0.03, confirming that condensate growth is significantly faster than the underlying chemical conversion.

In addition, conversion of the dimensionless threshold $x^{*}_{trans}$ for each $c_0$ gives a fixed absolute *trans* concentration at nucleation of $c_{\text{sat}}$ = 1.65 ± 0.09 mM for all datasets. This observation is consistent with classical phase separation theory: nucleation is controlled by the supersaturation in the dilute phase rather than by the relative fraction of *trans* isomer. Consequently, plotting $\phi$ versus absolute *trans* concentration produces a universal onset for all $c_0$ (fig. 4I), confirming the robustness of the extracted $x^{*}_{trans}$ values and providing a direct physical interpretation in concentration units. Additionally, rescaling the coacervate volume fraction by $\phi_{\max}$ and the *trans* concentration by $u = (x_{trans} - x^{*}_{trans})/(1 - x^{*}_{trans})$ collapses all datasets onto a single master curve (fig. 4J), consistent with the universal growth law predicted by:

$$\frac{\phi}{\phi_{\max}} = 1 - (1 - u)^{\alpha} \qquad \text{(eq. S13)}$$

*3.5 – Modelling the thermal relaxation kinetics of azoTAB.*

To model the *cis*→*trans* relaxation dynamics of azoTAB in the dark, we defined $k_\mathrm{T}$ as the relaxation rate constant at temperature $T$. Since there is no photon influx during thermal relaxation in the dark, the relaxation rate is described as:

$$\frac{dc_{trans}}{dt} = -\frac{dc_{cis}}{dt} = k_\mathrm{T} c_{cis}(t) \qquad \text{(eq. S14)}$$

where $c_{trans}$ and $c_{cis}$ are the molar concentration of *trans*- and *cis*-azoTAB, respectively, with $c_{trans} + c_{cis} = c_0$. Integration of eq. S14 yields:

$$c_{cis}(t) = c_{cis}(0)e^{-k_\mathrm{T} t} \qquad \text{(eq. S15)}$$

where $c_{cis}(0)$ is the initial molar concentration of *cis*-azoTAB. According to Beer-Lambert's Law, the measured absorbance $A(t, \lambda)$ is related to $c_{cis}(t)$ as follows:

$$\begin{aligned} A(t,\lambda) &= \varepsilon_{trans}(\lambda)[c_0 - c_{cis}(t)]l + \varepsilon_{cis}(\lambda)c_{cis}(t)l \\ &= \varepsilon_{trans}(\lambda)c_0 l + \left(\varepsilon_{cis}(\lambda) - \varepsilon_{trans}(\lambda)\right)c_{cis}(t)l \end{aligned} \qquad \text{(eq. S16)}$$

where $c_0 = c_{trans}(t) + c_{cis}(t)$, $l$ is the optical path-length and $\varepsilon_{trans}(\lambda)$, $\varepsilon_{cis}(\lambda)$ are the extinction coefficients of *trans*-, *cis*-azoTAB at a given wavelength, respectively. Substituting eq. S15 into eq. S16, $A(t, \lambda)$ is cast as:

$$A(t,\lambda) = a_1 + b_1 e^{-k_T t} \qquad \text{(eq. S17)}$$

where $a_1$ and $b_1$ depend on extinction coefficients and optical path-lengths.

Experimentally, the absorbance was monitored at $\lambda$ = 380 nm in a quartz cuvette of $l$ = 1 cm path-length containing a UV-adapted aqueous azoTAB solution prepared at $c_0$ = 30 µM. Eq. S17 was fitted to the experimental data collected at different temperatures $T$, yielding an Arrhenius plot of $k_\mathrm{T}$ vs. $1/T$. This allowed us to determine $k_\mathrm{T}$ at any given temperature. The obtained $k_\mathrm{T}$ value was further used to estimate the time-dependent fractions of *trans*-azoTAB during thermal relaxation $x_{trans}(t)$ by reinjecting eq. S15 as follows:

$$x_{trans}(t) = \frac{c_{trans}(t)}{c_0} = 1 - \frac{c_{cis}(0)e^{-k_\mathrm{T} t}}{c_0} = 1 - x_{cis}(0)e^{-k_\mathrm{T} t} \qquad \text{(eq. S18)}$$

where the initial *cis*-azoTAB fraction $x_{cis}(0)$ was 96%, as determined from UV/vis spectra (table S1).

Note S4. Theory of photoswitchable DNA coacervation.

We developed a simple model for photoswitchable DNA coacervation by examining the phase behavior of dsDNA chains in solution. Based on the Flory-Huggins mean-field theory for polymer solutions (*47, 48*) the free energy of mixing per unit volume is given as:

$$f = \frac{\varphi}{N}\ln\varphi + (1-\varphi)\ln(1-\varphi) + \chi\varphi(1-\varphi) \qquad \text{(eq. S19)}$$

where $N$ is the ratio of the effective DNA segment volume to that of the solvent, $\varphi$ is the DNA volume fraction, and $\chi$ is the Flory-Huggins interaction parameter.

In practice, experimental concentrations expressed in nucleobase units ($c_0$ = $c_{\mathrm{DNA}}$) were converted to the effective volume fraction entering the Flory-Huggins model using $\varphi = c_{\mathrm{DNA}}M_{\mathrm{nb}}/\rho_{\mathrm{DNA}}$ where $M_{\mathrm{nb}}$ is the molar mass per nucleobase (≈ 307.3 $g \cdot mol^{-1}$) and $\rho_{\mathrm{DNA}}$ is the density of dsDNA (≈ 1.7 $g \cdot mL^{-1}$).

The Flory parameter quantifies the balance between DNA-DNA and DNA-solvent interactions and can be written as:

$$\chi = \frac{z}{k_{\mathrm{B}}T}\left[u_{\mathrm{ds}} - \frac{1}{2}(u_{\mathrm{dd}} + u_{\mathrm{ss}})\right] \qquad \text{(eq. S20)}$$

where $u_{\mathrm{ij}}$ are the mean-field energies per site that quantify the interactions between species $i$ and $j$ ($'\mathrm{d}'$ standing for DNA and $'\mathrm{s}'$ standing for the solvent) and $z$ is the coordination number (namely, the number of nearest neighbors for a lattice site, each one occupied either by one chain segment or a solvent molecule). When $\chi$ exceeds a critical value $\chi_{\mathrm{c}}$ (typically ~ 1/2 for long polymers), the energy term outweighs the mixing entropy term, leading to a region of negative curvature in the free energy of mixing. Under these conditions, the system becomes unstable, resulting in phase separation into a DNA-rich phase (coacervate) and a DNA-poor phase.

In our model, we consider that azoTAB modulates the mean field DNA-DNA interaction. Specifically, the neutralization of the phosphate charges of the DNA backbone by azoTAB molecules reduces the electrostatic repulsion between DNA segments, effectively lowering the DNA-DNA repulsive interaction parameter $u_{\mathrm{dd}}$. Previous studies on cationic azobenzenes with longer alkyl tails and extended linkers between the azobenzene unit and the trimethylammonium head group have shown that both *trans* and *cis* isomers can bind to DNA, even in the presence of 0.1 M NaCl, albeit through different binding modes (*49, 50*). The *cis* isomer tends to form disordered pseudo-micelles interacting with DNA, while the *trans* isomer assembles into antiparallel (head-to-tail) aggregates around 0.1 M NaCl, forming intermolecular ion bridges between DNA chains. Within this coarse-grained description, azoTAB binding modifies both the effective DNA-DNA interaction parameter and the effective segment volume entering the Flory-Huggins free energy.

Given that both isomers can, in principle, bind to DNA, we assume that $u_{\mathrm{dd}}$ is proportional to the total number of *trans*-azoTAB ($N_{trans}$) and *cis*-azoTAB molecules ($N_{cis}$) in solution, with distinct proportionality factors for each isomer. Thus, we express $u_{\mathrm{dd}}$ as:

$$u_{\mathrm{dd}} = -2\left(\kappa_{trans}\frac{N_{trans}}{N_0} + \kappa_{cis}\frac{N_{cis}}{N_0}\right) \qquad \text{(eq. S21)}$$

where $N_0 = N_{trans} + N_{cis}$, $\kappa_{trans}$ and $\kappa_{cis}$ represent the proportionality factors for the *trans* and *cis* isomer, respectively. Using this expression of $u_{\mathrm{dd}}$, we can recast the formula for $\chi$ as:

$$\chi = \chi_0 + \theta x_{trans} \quad \text{(eq. S22)}$$

where $\chi_0 = \frac{z}{k_{\mathrm{B}}T}\left(u_{\mathrm{ds}} - \frac{1}{2}u_{\mathrm{ss}} + \kappa_{cis}\right)$, $\theta = \frac{z}{k_{\mathrm{B}}T}(\kappa_{trans} - \kappa_{cis})$, and $x_{trans} = N_{trans}/N_0$.

The linear dependence of $\chi$ on $x_{trans}$ is consistent with the mean-field interpretation of $\chi$ as an effective interaction parameter. Similar linear relations are commonly observed in polymer and biomolecular condensates, where $\chi$ is expressed as $\chi = A + B/T$ (see, e.g., ref. (*51*)), reflecting the temperature dependence of interaction energies. In the present system, $x_{trans}$ plays an analogous role by modulating the effective DNA-DNA interaction strength through azoTAB binding.

The volume fractions of DNA in the dilute and coacervate phases, denoted as $\varphi_{\mathrm{dil}}$ and $\varphi_{\mathrm{coa}}$, respectively, correspond to the points where the chemical potentials $\mu$ and osmotic pressures $\Pi$ are equal, namely $\mu(\varphi_{\mathrm{dil}}) = \mu(\varphi_{\mathrm{coa}})$ and $\Pi(\varphi_{\mathrm{dil}}) = \Pi(\varphi_{\mathrm{coa}})$. Given that $\mu(\varphi) = f'(\varphi)$ and $\Pi(\varphi) = \varphi f'(\varphi) - f(\varphi)$, we use eq. S19 to express the chemical potential and osmotic pressure as:

$$\mu(\varphi) = \frac{1}{N} - 1 + \frac{\ln\varphi}{N} - \ln(1-\varphi) + \chi(1-2\varphi) \quad \text{(eq. S23)}$$

$$\Pi(\varphi) = \left(\frac{1}{N} - 1\right)\varphi - \ln(1-\varphi) - \chi\varphi^2 \quad \text{(eq. S24)}$$

Using eq. S23 and eq. S24, and applying the conditions of equal chemical potential and osmotic pressure, we obtain:

$$\chi = -\frac{1}{N}\frac{1}{(\varphi_{\mathrm{coa}} - \varphi_{\mathrm{dil}})(2 - \varphi_{\mathrm{coa}} - \varphi_{\mathrm{dil}})}\ln\frac{\varphi_{\mathrm{dil}}}{\varphi_{\mathrm{coa}}} + \frac{1 - 1/N}{2 - \varphi_{\mathrm{coa}} - \varphi_{\mathrm{dil}}} \quad \text{(eq. S25)}$$

For any value of $N$, χ can be determined from the experimental data of $\varphi_{\mathrm{dil}}$ and $\varphi_{\mathrm{coa}}$ as a function of $x_{trans}$ using eq. S25.

Because the experimental coacervate concentrations at very low *trans*-azoTAB fractions display a non-monotonic evolution inconsistent with a Flory-Huggins binodal, we restricted the analysis to $x_{trans} > 0.2$. For each trial value of $N$ in the range 1-200, χ was calculated from eq. S25 using the experimental values of $\varphi_{\mathrm{dil}}$ and $\varphi_{\mathrm{coa}}$ as varying $x_{trans}$. The optimal value of $N$ was determined by maximizing the linearity of $\chi(x_{trans})$, quantified by minimizing the ratio between the standard deviation of $\mathrm{d}\chi/\mathrm{d}x_{trans}$ and its mean value. This procedure yielded $N = 117$. This value should be interpreted as an effective degree of polymerization within the coarse-grained Flory-Huggins description. It is consistent with the expected size of the polydisperse DNA fragments used here (< 200 base pairs, centered around ~50 base pairs (*30*)), noting that N represents an effective parameter that also reflects chain length polydispersity and coarse-graining of the lattice model.

The corresponding $\chi(x_{trans})$ data are shown in fig. S5A. Linear fitting using eq. S22 then provided $\kappa = 0.069$ and $\chi_0 = 0.66$.

Next, the binodal curve is theoretically computed for $N = 117$ using the algorithm proposed in ref. (*52*), which consists in considering the following vector after $i$ iterations:

$$\mathbf{X}^{(i)} = \begin{bmatrix} \varphi_{\text{dil}}^{(i)} \\ \varphi_{\text{coa}}^{(i)} \end{bmatrix} = \mathbf{G}^i X^{(0)} \qquad (\text{eq. } S26)$$

where $X^{(0)}$ is initialized with the Ginzburg-Landau solution:

$$\varphi_{\text{dil,coa}} = \varphi_{\text{c}} \pm \sqrt{\frac{3(\chi - \chi_{\text{c}})}{2\chi_{\text{c}}^2 \sqrt{N}}} \qquad (\text{eq. } S27)$$

with:

$$\varphi_{\text{c}} = \frac{1}{1 + \sqrt{N}} \qquad (\text{eq. } S28)$$

$$\chi_{\text{c}} = \frac{1}{2}\left(1 + \frac{1}{\sqrt{N}}\right)^2 \qquad (\text{eq. } S29)$$

The **G** operator is defined as:

$$\mathbf{G}\mathrm{X}^{(i)} = \begin{bmatrix} \dfrac{1 - \exp\left\{g_1(\varphi_{\text{dil}}^{(i-1)}, \varphi_{\text{coa}}^{(i-1)})\right\}}{1 - \exp\left\{-N\left[g_1\left(\varphi_{\text{dil}}^{(i-1)}, \varphi_{\text{coa}}^{(i-1)}\right) - g_2(\varphi_{\text{dil}}^{(i-1)}, \varphi_{\text{coa}}^{(i-1)})\right]\right\} \exp\left\{g_1(\varphi_{\text{dil}}^{(i-1)}, \varphi_{\text{coa}}^{(i-1)})\right\}} \\ \dfrac{1 - \exp\left\{-g_1(\varphi_{\text{dil}}^{(i-1)}, \varphi_{\text{coa}}^{(i-1)})\right\}}{1 - \exp\left\{-N\left[g_1\left(\varphi_{\text{dil}}^{(i-1)}, \varphi_{\text{coa}}^{(i-1)}\right) - g_2(\varphi_{\text{dil}}^{(i-1)}, \varphi_{\text{coa}}^{(i-1)})\right]\right\} \exp\left\{-g_1(\varphi_{\text{dil}}^{(i-1)}, \varphi_{\text{coa}}^{(i-1)})\right\}} \end{bmatrix} (\text{eq. } S30)$$

where the $g_1$ and $g_2$ functions are defined as follows

$$g_1(\varphi_{\text{dil}}, \varphi_{\text{coa}}) = \left(1 - \frac{1}{N}\right)(\varphi_{\text{coa}} - \varphi_{\text{dil}}) + \chi(\varphi_{\text{coa}}^2 - \varphi_{\text{dil}}^2) \qquad (\text{eq. } S31)$$

$$g_2(\varphi_{\text{dil}}, \varphi_{\text{coa}}) = 2\chi(\varphi_{\text{coa}} - \varphi_{\text{dil}}) \qquad (\text{eq. } S32)$$

For each value of $x_{trans}$, the experimentally determined $\chi$ was used as input for the iterative algorithm, which converged after $i_{\text{max}}$ = 95 iterations to a precision below 1%. The procedure yields theoretical volume fractions $\varphi_{\text{dil}}^*$ and $\varphi_{\text{coa}}^*$ defining the binodal curve. When plotted directly against the experimental data, the theoretical binodal exhibited a systematic offset, with $\varphi_{\text{dil}}^*$ and $\varphi_{\text{coa}}^*$ larger than $\varphi_{\text{dil}}$ and $\varphi_{\text{coa}}$ by an approximately constant factor over the full range of $x_{trans}$. This indicates a mismatch between the effective lattice segment volume used in the Flory-Huggins description and the experimental conversion from concentration to volume fraction.

To account for this difference, we introduced a single multiplicative rescaling factor $m$ between theoretical and experimental volume fractions, determined from the average ratio $\varphi_{\text{dil}}^*/\varphi_{\text{dil}}$ across all $x_{trans}$ values. This yielded $m$ = 1.8. After applying this rescaling, the theoretical binodal reproduces the experimental phase boundary over the explored range of $x_{trans}$ (fig. S5B). For completeness, fig. S5C shows the binodal in the ($\chi$, $\varphi$) representation, combining experimental $\chi$ values and the theoretical curve. This representation highlights the characteristic minimum of the Flory-Huggins binodal, which is not directly visible in the ($x_{trans}$, $\varphi$) plots due to the limited accessible $\chi$ range.

The applied rescaling reflects the coarse grained nature of the Flory-Huggins lattice model, in which the segment volume is not known a priori. In the present system, binding of azoTAB to DNA lead to an effective segment volume larger than that estimated from DNA alone, resulting in a uniform shift between theoretical and experimental volume fractions. More specifically, under the equimolar charge conditions used here ($c_0$ = $c_{\mathrm{DNA}}$ = $c_{\mathrm{azo}}$), each nucleobase is associated on average with one azoTAB molecule. Using the partial specific volume of dsDNA (≈ 0.55 $cm^3 \cdot g^{-1}$, corresponding to a molecular volume of ≈0.3 $nm^3$ per nucleobase) and the molecular volume of azoTAB (≈ 0.64 $nm^3$ from ref. (*53*)), a simple additive estimate would yield a total occupied volume approximately threefold larger than that of DNA alone. This value represents an upper bound, because binding leads to partial interdigitation of azobenzene molecules (e.g. via π-π stacking), release of counterions, and overlap of hydration shells, which reduce the effective excluded volume. The obtained factor $m$ = 1.8 is therefore consistent with the expected increase in effective segment volume upon azoTAB binding based on molecular volume considerations.

Note S5. Modelling the photoisomerisation kinetics under illumination.

The azoTAB photoconversion kinetics are determined in a continuously stirred azoTAB solution of total concentration $c_0$ under pump–probe illumination. The absorbance at 380 nm is monitored while irradiating the sample at various wavelengths and intensities. Because thermal *cis*→*trans* relaxation is much slower than photoisomerization under illumination, the system is treated as a two-level *trans*↔*cis* system.

The probe absorbance $A_\mathrm{p}$ is related to the *trans* concentration $c_{trans}$ via the Beer-Lambert law:

$$A_\mathrm{p}(t) = \left[\varepsilon_{trans}^\mathrm{p} c_{trans}(t) + \varepsilon_{cis}^\mathrm{p} c_{cis}(t)\right] l_\mathrm{p} = \left[(\varepsilon_{trans}^\mathrm{p} - \varepsilon_{cis}^\mathrm{p}) c_{trans}(t) + \varepsilon_{cis}^\mathrm{p} c_0\right] l_\mathrm{p} \quad (\text{eq. S33})$$

where $c_{cis}$ is the concentration of *cis* isomer, $\varepsilon_{trans}^\mathrm{p}$ and $\varepsilon_{cis}^\mathrm{p}$ are the molar extinction coefficients of the *trans* and *cis* isomers, respectively, at the probe wavelength, and $l_\mathrm{p}$ is the optical path-length of the probe.

The pump beam is attenuated as it propagates through the sample, so that the absorbed intensity depends on the instantaneous absorbance:

$$I = I_\mathrm{ex}\left(1 - 10^{-A_\mathrm{ex}(t)}\right) \quad (\text{eq. S34})$$

where $A_\mathrm{ex}$ is the pump absorbance and $I_\mathrm{ex}$ is the pump intensity. The rate of photoisomerization is proportional to the absorbed photon flux. Neglecting thermal relaxation and probe-induced conversion (given that $I_\mathrm{ex} \sim 10^{-1} - 10^{3}$ mW·cm$^{-2}$ $\gg I_\mathrm{p} \sim 0.02$ mW·cm$^{-2}$ as validated experimentally), the *trans* concentration therefore obeys:

$$\frac{dc_{trans}}{dt} = -kI_\mathrm{ex}\left(1 - 10^{-A_\mathrm{ex}(t)}\right) \quad (\text{eq. S35})$$

where $k$ is an effective rate constant incorporating quantum yields and photon energy. Using the Beer-Lambert relation for the pump absorbance:

$$A_\mathrm{ex}(t) = \left[(\varepsilon_{trans}^\mathrm{ex} - \varepsilon_{cis}^\mathrm{ex}) c_{trans}(t) + \varepsilon_{cis}^\mathrm{ex} c_0\right] l_\mathrm{ex} \quad (\text{eq. S36})$$

where $\varepsilon_{trans}^\mathrm{ex}$ and $\varepsilon_{cis}^\mathrm{ex}$ are the molar extinction coefficients of the *trans* and *cis* isomers, respectively, at the pump wavelength, and $l_\mathrm{ex}$ is the optical path-length of the pump, one obtains a closed differential equation for $A_\mathrm{ex}$:

$$\frac{dA_\mathrm{ex}}{dt} = -\frac{1 - 10^{-A_\mathrm{ex}}}{\tau_\mathrm{iso}} \quad (\text{eq. S37})$$

with $\tau_\mathrm{iso} \propto I_\mathrm{ex}^{-1}$ is the characteristic isomerization time, expressed as:

$$\tau_\mathrm{iso} = \frac{1}{kI_\mathrm{ex}(\varepsilon_{trans}^\mathrm{ex} - \varepsilon_{cis}^\mathrm{ex}) l_\mathrm{ex}} \quad (\text{eq. S38})$$

Integration of eq. S37 yields:

$$A_{\mathrm{ex}}(t) = \log\left(1 - (1 - 10^{A_{\mathrm{ex}}(0)})10^{-t/\tau_{\mathrm{iso}}}\right) \quad \text{(eq. S39)}$$

By combining eq. S33, eq. S36 and eq. S39, the measured probe absorbance can be written in the form:

$$A_{\mathrm{p}}(t) = a_2 \log\left(1 - (1 - 10^{A_{\mathrm{ex}}(0)})10^{-t/\tau_{\mathrm{iso}}}\right) + b_2 \quad \text{(eq. S40)}$$

where $a_2$ and $b_2$ depend on extinction coefficients and optical path-lengths. After normalization, the temporal evolution of the *trans* fraction $x_{trans} = c_{trans}/c_0$ becomes:

$$x_{trans}(t) = a_2^* \log\left(1 - (1 - 10^{A_{\mathrm{ex}}(0)})10^{-t/\tau_{\mathrm{iso}}}\right) + b_2^* \quad \text{(eq. S41)}$$

This expression captures the non-exponential kinetics arising from attenuation of the pump beam in optically thick conditions. Eq. S41 is used to globally fit the photoisomerization data shown in fig. S6A-C with $a_2^*$ and $b_2^*$ fixed by the initial and photostationary *trans* fractions. The extracted values of $\tau_{\mathrm{iso}}$ show a $I_{\mathrm{ex}}^{-1}$ scaling (fig. S6D), as predicted by the model.

Note S6. Modelling coacervate dissolution under UV light.

The photochemical process under UV irradiation corresponds to the *trans*→*cis* isomerization induced by photon absorption. Owing to the high concentration of azoTAB in the coacervate phase, the system is optically thick, leading in general to spatially dependent light attenuation and non-linear reaction-diffusion kinetics discussed in Note S5. A full treatment of this regime results in non-elementary expressions involving dilogarithmic functions, which hinder direct comparison with experimental data

We therefore introduce a coarse-grained description in which spatial variations of the light field are absorbed into an effective homogeneous *trans*→*cis* isomerization rate within the coacervate, $k_{\mathrm{eff}}$. The relevant dynamical variable is the local *trans* concentration in the coacervate, $c_{trans}^{\mathrm{coa}}(t)$, which governs phase stability. Its evolution is approximated by a first-order rate equation:

$$\frac{dc_{trans}^{\mathrm{coa}}}{dt} = -k_{\mathrm{eff}} I_{\mathrm{UV}} c_{trans}^{\mathrm{coa}}(t) \qquad \text{(eq. S42)}$$

where $I_{\mathrm{UV}}$ is the incident UV intensity, yielding:

$$c_{trans}^{\mathrm{coa}}(t) = c_{trans}^{\mathrm{coa}}(0) e^{-k_{\mathrm{eff}} I_{\mathrm{UV}} t} \qquad \text{(eq. S43)}$$

The conversion to the *cis* isomer reduces the affinity of azoTAB for the coacervate phase, resulting in a net outward flux of material toward the dilute phase. In a diffusion-limited description, the droplet radius obeys:

$$\frac{dR}{dt} = -\frac{D_{\mathrm{eff}}}{\rho R}\left(c_{cis}^{\mathrm{coa}}(t) - c_{cis}^{\mathrm{dil}}\right) \qquad \text{(eq. S44)}$$

where $\rho$ is the coacervate density, and the dilute-phase *cis* concentration $c_{cis}^{\mathrm{dil}}$ is set by the binodal under UV irradiation. Under UV irradiation, the dilute phase acts as an effective sink due to its low concentration and continuous photoconversion, such that $c_{cis}^{\mathrm{coa}} \gg c_{cis}^{\mathrm{dil}}$. Using $c_{cis}^{\mathrm{coa}}(t) = c_{trans}^{\mathrm{coa}}(0) - c_{trans}^{\mathrm{coa}}(t) = c_{trans}^{\mathrm{coa}}(0)(1 - e^{-k_{\mathrm{eff}} I_{\mathrm{UV}} t})$, this reduces to:

$$\frac{dR}{dt} = \frac{D_{\mathrm{eff}}}{\rho R} c_{trans}^{\mathrm{coa}}(0)(1 - e^{-k_{\mathrm{eff}} I_{\mathrm{UV}} t}) \qquad \text{(eq. S45)}$$

Integration yields:

$$R(t) = R_0 \sqrt{1 + \frac{2 c_{trans}^{\mathrm{coa}}(0) D_{\mathrm{eff}}}{k_{\mathrm{eff}} I_{\mathrm{UV}} \rho R_0^2}(1 - e^{-k_{\mathrm{eff}} I_{\mathrm{UV}} t} - k_{\mathrm{eff}} I_{\mathrm{UV}} t)} \qquad \text{(eq. S46)}$$

This expression predicts an initially slow evolution with a vanishing slope at short times, reflecting the progressive build-up of *cis*-azoTAB, followed by accelerated dissolution once a sufficient fraction of *trans* is converted. The experimental coacervate radius trajectories during dissolution $R(t)$ are well fitted with this expression (fig. S8).

The dynamics are governed by two characteristic rates: a photochemical rate $\kappa = k_{\text{eff}} I_{\text{UV}}$ and an effective diffusion-limited transport rate $\Gamma = 2c_{trans}^{\text{coa}}(0)D_{\text{eff}}/(\rho R_0^2)$. Introducing these quantities, the solution can be written in dimensionless form as:

$$\frac{R(t)}{R_0} = \sqrt{1 + \frac{\Gamma}{\kappa}(1 - e^{-\kappa t} - \kappa t)} \qquad \text{(eq. S47)}$$

This formulation shows that the dissolution dynamics is controlled by a single dimensionless parameter $\Gamma/\kappa$, which compares the rate of diffusive transport to that of photoinduced conversion.

The dissolution time $\tau_{\text{dis}}$, defined by $R(\tau_{\text{dis}}) = 0$, satisfies:

$$\Gamma\tau_{\text{dis}} + \frac{\Gamma}{\kappa}(e^{-\kappa\tau_{\text{dis}}} - 1) = 1 \qquad \text{(eq. S48)}$$

This equation captures the crossover between regimes controlled by photochemical conversion and diffusive transport. In this intermediate regime, neither process is rate-limiting, and dissolution results from their coupled action. By fitting the experimental trajectories of $R(t)$, we find that $\kappa\tau_{\text{dis}}$ = 0.03-1 $\approx \mathcal{O}(1)$. Expanding the exponential in eq. S48 to second order yields:

$$\tau_{\text{dis}} \sim \sqrt{\frac{2}{\kappa\Gamma}} \propto \frac{1}{\sqrt{k_{\text{eff}} I_{\text{UV}}}} \qquad \text{(eq. S49)}$$

Hence, the model predicts a $I_{\text{UV}}^{-0.5}$ dependency in the crossover regime, in good agreement with the –0.6 exponent observed experimentally (Fig. 2D).

In comparison, in the low-intensity limit, the rate of *cis* production is slow compared to diffusive transport, so that dissolution is limited by the photochemical conversion, yielding:

$$\tau_{\text{dis}} \sim \frac{1}{\kappa} = \frac{1}{k_{\text{eff}} I_{\text{UV}}} \qquad \text{(eq. S50)}$$

In contrast, at high intensity, conversion is effectively instantaneous and dissolution is controlled by diffusive transport, giving:

$$\tau_{\text{dis}} \sim \frac{1}{\Gamma} = \frac{\rho R_0^2}{2c_{trans}^{\text{coa}}(0)D_{\text{eff}}} \qquad \text{(eq. S51)}$$

Note S7. Determination of photoconversion in w/o droplets under green light.

To estimate the photoconversion kinetics inside w/o droplets, we apply the pump-probe model to a spherical geometry by taking the droplet diameter (60 µm) as the optical path-length. Under our experimental conditions ($c_0$ = 5-30 mM), the initial absorbance at 555 nm for this path-length is small (0.8 - 4.8 × $10^{-3}$), indicating that droplets are optically thin.

Building on Note S5 in the limit $A_{\mathrm{ex}}(0) \ll 1$, eq. S41 can be expanded to first order, yielding:

$$x_{trans}(t) \simeq a_2^* \, A_{\mathrm{ex}}(0) 10^{-t/\tau_{\mathrm{iso}}} + b_2^* \qquad (\mathrm{eq.\,S52})$$

This expression predicts an exponential kinetics independent of $c_0$ in the optically thin regime. The dependence of $\tau_{\mathrm{iso}}$ on the excitation intensity is extracted from linear fits of $1/\tau_{\mathrm{iso}}$ versus $I_{\mathrm{ex}}$. These values are then used to predict $x_{trans}(t, I_{\mathrm{green}})$ inside droplets, which allows locating the corresponding binodal concentration pair ($c_{\mathrm{dil}}$ and $c_{\mathrm{coa}}$) in the phase diagram for a given green irradiation duration and intensity. These concentrations are extracted from linear fitting of the coacervate volume fraction $\phi$ as a function of the initial concentration $c_0$ under the same green light dose (as described in Note S2).

Note S8. Coacervate nucleation rate under green light.

To interpret the experimentally measured nucleation rates, we adopt the framework of classical nucleation theory (CNT). Although complex coacervates are multicomponent systems and may not strictly satisfy the assumptions of classical nucleation theory, CNT provides a useful phenomenological framework for relating nucleation rates to an effective thermodynamic driving force for phase separation.

In our system, coacervate formation is triggered by light-induced *cis*→*trans* photoisomerization of azoTAB. The total concentrations of DNA and azoTAB are fixed and equal, while the fraction of *trans*-azoTAB increases progressively during green illumination. Nucleation occurs once the concentration of *trans*-azoTAB exceeds a threshold value corresponding to the onset of phase separation.

Following previous work on light-responsive biomolecular condensates (*54*), we define the supersaturation as:

$$S = \log\left(\frac{c_0}{c_{\text{sat}}}\right) \qquad \text{(eq. S53)}$$

where $c_0$ is the total concentration of the coacervate components and $c_{\text{sat}}$ is an effective saturation concentration associated with the nucleation threshold.

Prior to nucleation, the system remains in a single metastable phase, such that the total concentration equals the dilute-phase concentration. In this multicomponent system, the supersaturation defined above should be interpreted as a phenomenological measure of the thermodynamic driving force associated with the accumulation of *trans*-azoTAB and its interaction with DNA.

Within CNT, the steady-state nucleation rate is given by:

$$J = K\ \exp\left(-\frac{\Delta G^*}{k_B T}\right) \qquad \text{(eq. S54)}$$

where $J$ is the nucleation rate, $k_B T$ is the thermal energy, $K$ is a kinetic prefactor, and $\Delta G^*$ is the free-energy barrier for formation of a critical nucleus.

For a liquid droplet nucleating from a supersaturated solution, the barrier height can be written as:

$$\Delta G^* = \frac{16\pi}{3}\frac{v^2\gamma^3}{(\Delta G_{\text{nuc}})^2} \qquad \text{(eq. S55)}$$

where $v$ is an effective molecular volume, $\gamma$ is an effective interfacial tension between the nucleating droplet and the surrounding phase, and:

$$\Delta G_{\text{nuc}} = k_B T \log\left(\frac{c_0}{c_{\text{sat}}}\right) = \ k_B T S \qquad \text{(eq. S56)}$$

represents the thermodynamic driving force for nucleation. Substituting this expression yields:

$$J(S) = K\ \exp\left[-\left(\frac{S^*}{S}\right)^2\right] \qquad \text{(eq. S57)}$$

with the dimensionless nucleation parameter:

$$S^* = \sqrt{\frac{16\pi}{3} v \left(\frac{\gamma}{k_B T}\right)^{3/2}} \qquad \text{(eq. S58)}$$

This expression captures the crossover between slow activated nucleation at weak supersaturation ($S \ll S^*$) and rapid nucleation at strong supersaturation. In the context of multicomponent coacervates, the parameter $S^*$ should be interpreted as an effective nucleation quantity characterizing the free-energy cost of forming a critical droplet.

In our experiments, the nucleation rate was measured as a function of the total concentration $c_0$ at different green laser levels. The data were fitted using the functional form:

$$J = K \, \exp\left[-\left(\frac{S^*}{\log\left(\frac{c_0}{c_{\text{sat}}}\right)}\right)^2\right] + a_0 \qquad \text{(eq. S59)}$$

where $a_0$ accounts for a small background contribution arising from fluorescence detection.

Experimentally, fits in fig. S11 yields a low dimensionless barrier $S^*$ = 1.36 ± 0.05 (mean ± s.e.) consistent with the weak interfacial tensions typically reported for complex coacervates and biomolecular condensates (*55*).

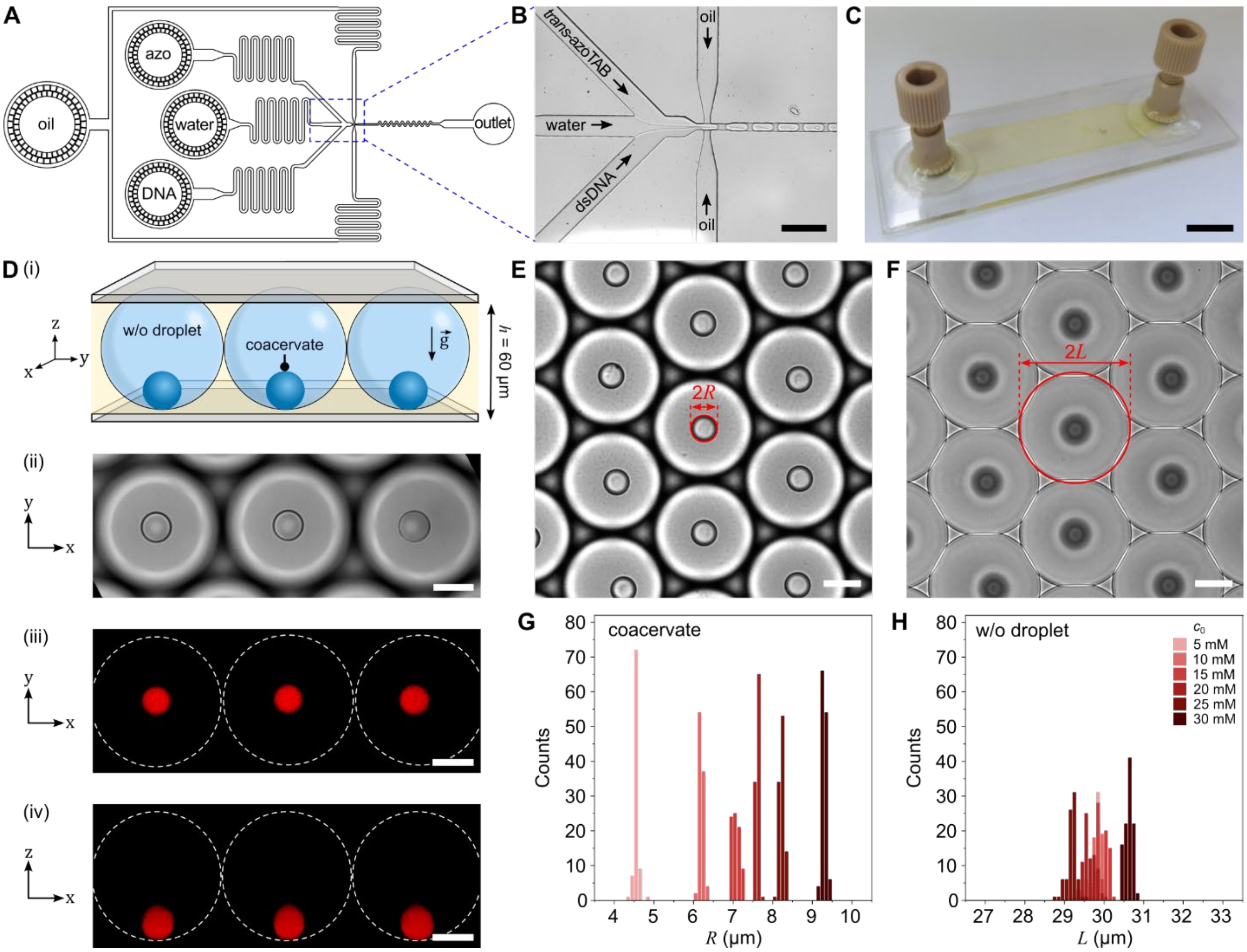

**Fig. S1. Microfluidic set-up and size distributions.** (**A**) Schematic of the microfluidic set-up used to produce coacervates within water-in-oil (w/o) emulsion droplets. (**B**) Optical microscopy image of the chip at the co-flowing junction, showing the two aqueous streams of *trans*-azoTAB and dsDNA (supplemented with NaCl, 10–100 mM), together with an additional stream of water (also supplemented with NaCl, 10–100 mM) co-flowed between them to prevent coacervation prior to droplet formation. Scale bar, 200 µm. (**C**) Photograph of the observation chamber containing coacervate-encapsulating w/o emulsion droplets. Scale bar, 1 cm. (**D**) (i) 3D schematic of the 60 µm-high observation chamber containing w/o emulsion droplets encapsulating coacervates. (ii) Optical microscopy image of w/o droplets encapsulating coacervates ($c_0$ = 15 mM) viewed through a 63× objective in the (*x,y*) plane. (iii) Corresponding confocal fluorescence microscopy image in the equatorial plane of the coacervates using TAMRA-ssDNA (1.5 µM) as a fluorescent reported. (iv) 2D projection on the (*x,z*) plane of 3D reconstructions of the droplets. Scale bars in (ii-iv), 20 µm. White dotted lines highlight w/o droplets. Coacervates set to the bottom of the droplets due to their higher density, without wetting the w/o interface. (**E**,**F**) Bright-field microscopy images of samples prepared at $c_0$ = 15 mM and focused on the inner coacervate (E) or outer w/o emulsion (F). The coacervate radius $R$ is determined via segmentation (binary threshold), and the w/o emulsion droplet radius $L$ is determined manually, as highlighted for one droplet by red circles. Scale bars, 20 µm. (**G**,**H**) Representative radius distribution of coacervate droplets (G) and the corresponding w/o emulsion droplets (H) produced at varying concentrations

of coacervate components $c_0$ (values listed in H). Measurements were performed on > 70 droplets sampled from different positions within the same preparation.

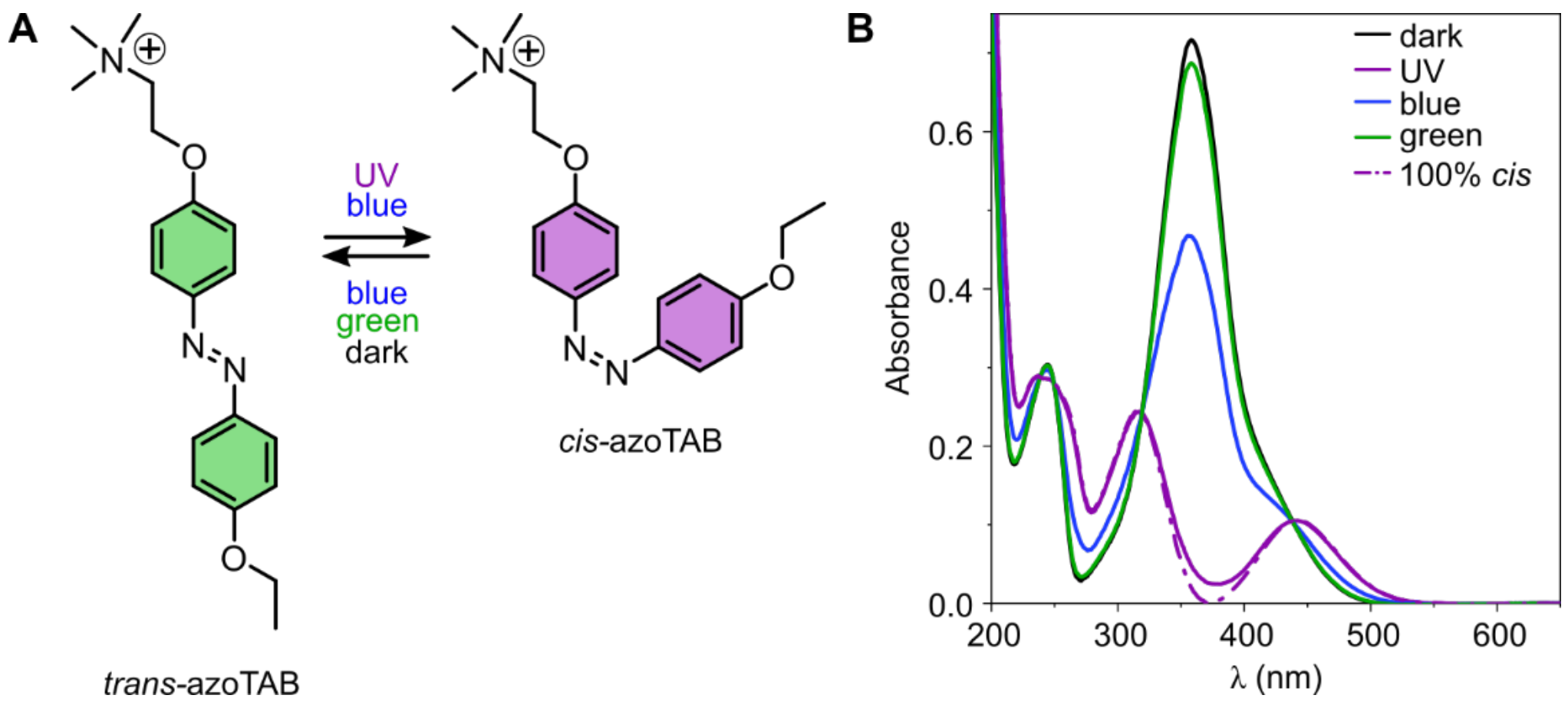

**Fig. S2. Equilibrium UV/vis spectra of azoTAB.** (**A**) Schematic of azoTAB isomerisation. UV and green light produce predominantly *cis*- and *trans*-azoTAB, respectively; in contrast, blue light produces a mixture of *trans*- and *cis*-azoTAB with an excess of the *trans* isomer. Dark promotes spontaneous relaxation to *trans*-azoTAB. (**B**) Equilibrium UV/vis spectra of a 30 µM azoTAB solution stored in the dark or irradiated with UV (375 nm), blue (460 nm), or green (560 nm) light. The spectrum acquired in the dark corresponds to 100% *trans*-azoTAB. The spectrum corresponding to 100% *cis*-azoTAB is derived from the spectra acquired under UV and blue light irradiation (Note S1). These light-adapted spectra are used to determine the *trans*:*cis*-azoTAB composition at the different photostationary states (table S1).

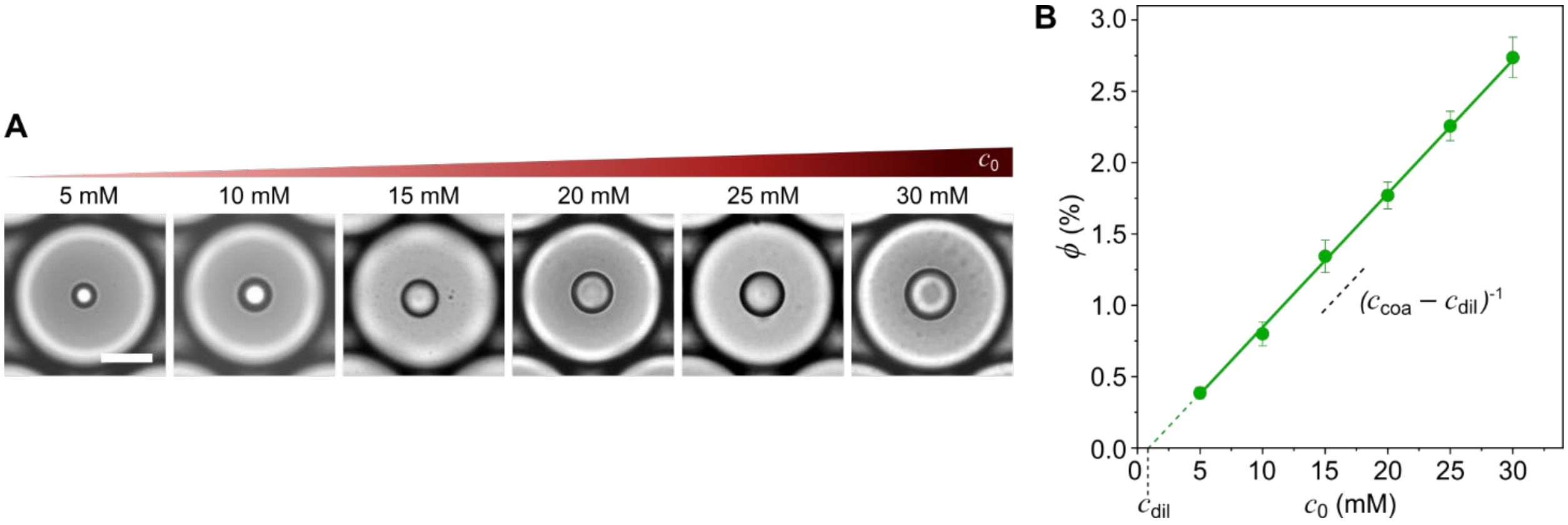

**Fig. S3. Equilibrium composition derived from coacervate volume fractions.** (**A**) Bright-field microscopy images of w/o emulsion droplets containing a single coacervate droplet prepared at varying components concentrations $c_0$ in the dark (100% *trans*-azoTAB). Scale bar, 20 µm. (**B**) Corresponding coacervate volume fraction $\phi$ as a function of $c_0$. The solid line shows a linear fit. Following mass conservation, the concentrations in the dilute ($c_{dil}$) and condensed ($c_{coa}$) phases are obtained from the *x*-intercept and the inverse slope, respectively.

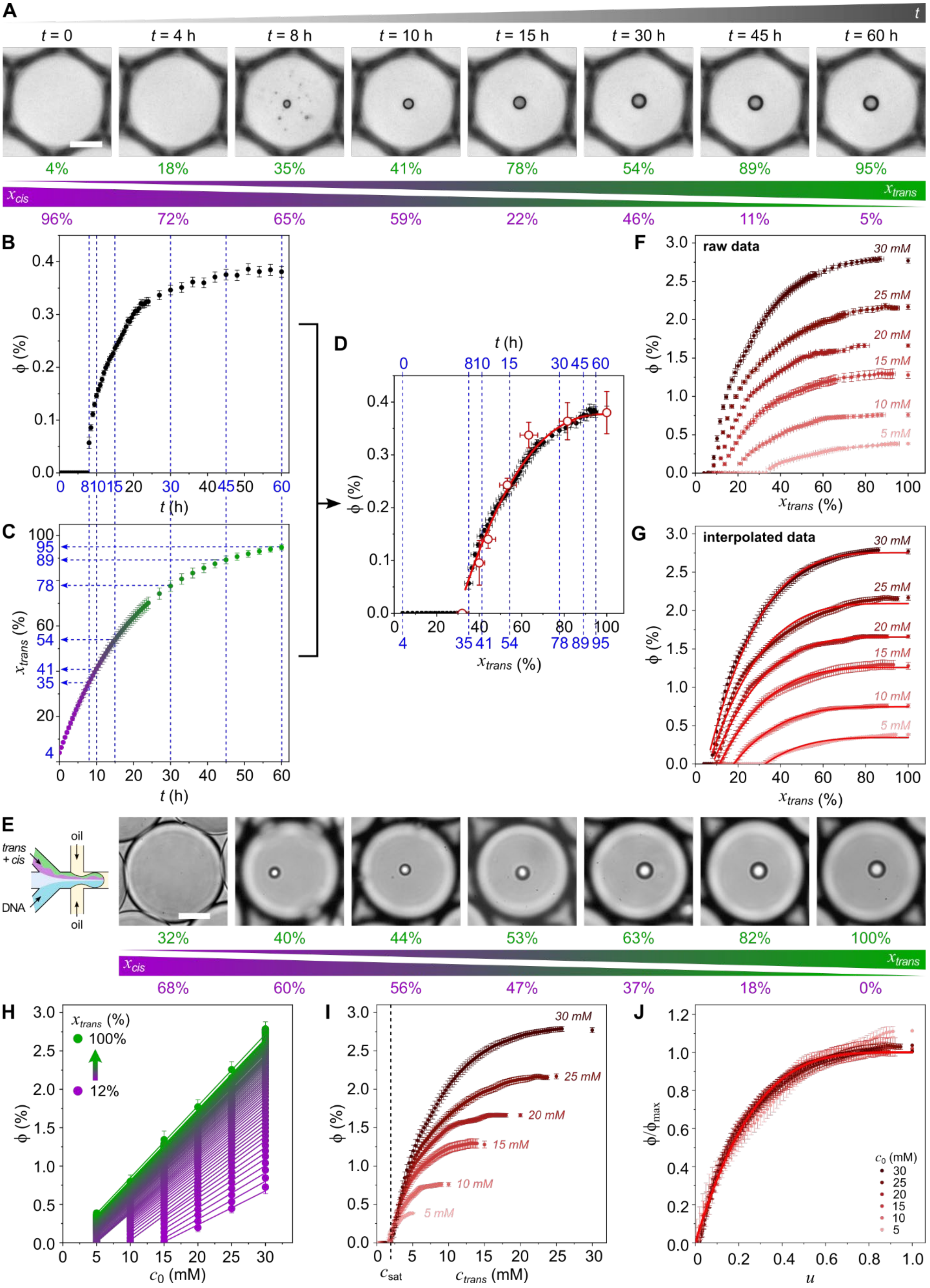
A
t = 0
t = 4 h
t = 8 h
t = 10 h
t = 15 h
t = 30 h
t = 45 h
t = 60 h
4%
18%
35%
41%
78%
54%
89%
95%
x_cis
x_trans
96%
72%
65%
59%
22%
46%
11%
5%
B
ϕ (%)
t (h)
C
x_trans (%)
D
t (h)
ϕ (%)
x_trans (%)
E
trans + cis
oil
DNA
oil
32%
40%
44%
53%
63%
82%
100%
68%
60%
56%
47%
37%
18%
0%
F
raw data
30 mM
25 mM
20 mM
15 mM
10 mM
5 mM
G
interpolated data
H
x_trans (%)
100%
12%
c_0 (mM)
I
c_sat
c_trans (mM)
J
ϕ/ϕ_max
u
c_0 (mM)
30
25
20
15
10
5

**Fig. S4. Relaxation-based coacervate growth.** (**A**) Time-lapse bright-field microscopy images of a w/o emulsion droplet prepared at $c_0$ = 5 mM, pre-irradiated with UV light and subsequently incubated in the dark at 26 ± 1 °C, showing gradual growth of a single coacervate droplet driven by *cis*→*trans*-azoTAB thermal relaxation. Scale bar, 20 µm. (**B**) Time-dependent evolution of the coacervate volume fraction $\phi$ in samples prepared at $c_0$ = 5 mM, pre-irradiated with UV light and subsequently incubated in the dark at 26 ± 1 °C. Data are shown as mean ± s.d. for *n* > 120 (typically ~150) droplets from the same sample. (**C**) Time-dependent evolution of the *trans* fraction $x_{trans}$ at 26 ± 1 °C derived from UV/vis measurements (Note S3). Data are shown as mean ± s.d., obtained by Monte Carlo propagation of the temperature uncertainty (5000 samples). This trace is used to convert the time axis into $x_{trans}$. (**D**) Coacervate volume fraction $\phi$ as a function of $x_{trans}$ derived from the two previous graphs. Data are shown as mean ± s.d. for *n* > 120 (typically ~150) droplets from the same sample; error bars on $x_{trans}$ arise from Monte Carlo propagation of the ± 1 °C temperature uncertainty in (B). Open red circles show equilibrium-prepared samples at predefined *trans*/*cis*-azoTAB ratios ($c_0$ = 5 mM, see panel (E)). Data are shown as mean ± s.d. for *n* = 25 droplets; error bars on $x_{trans}$ reflect uncertainty in the effective *trans*-azoTAB fraction at the time of measurement, calculated by propagating a ± 1 °C temperature uncertainty during coacervate formation. The red line shows a fit to the coacervate growth model (Note S3). (**E**) Bright-field microscopy images of a single w/o microfluidic droplet containing an equimolar charge solution of azoTAB and dsDNA prepared at $c_0$ = 5 mM and at varying *trans*-azoTAB fractions. The samples were produced using an inlet azoTAB stock solution containing pre-mixed *trans*- and *cis*-azoTAB at a given composition, with a total final concentration of azoTAB in the w/o droplet fixed at 5 mM. Scale bar, 20 µm. Note that a single coacervate droplet forms under all conditions, including for the same composition as obtained at equilibrium under blue light (≈ 65% *trans*). (**F**,**G**) Raw (F) and interpolated (G) coacervate volume fraction $\phi$ as a function of $x_{trans}$ in samples prepared at varying $c_0$, pre-irradiated with UV light and subsequently incubated in the dark at fixed temperature. Data in (F) are shown as mean ± s.d. for *n* > 120 (typically ~150) droplets from the same sample; error bars on $x_{trans}$ arise from Monte Carlo propagation of the ± 1 °C temperature uncertainty. Data in (G) are interpolated onto a common $x_{trans}$ grid (1% increments), and uncertainties in $x_{trans}$ are propagated into $\phi$ via Monte Carlo sampling. Shown error bars represent the combined standard deviation arising from temperature-induced uncertainty in $x_{trans}$ and experimental measurement uncertainty in $\phi$ (added in quadrature). Red lines in (G) show fits to the diffusion-limited growth model (Note S3). (**H**) Coacervate volume fraction $\phi$ as a function of $c_0$ for varying fractions of *trans*-azoTAB, $x_{trans}$. Lines indicate linear fits. Following mass conservation, the concentrations in the dilute phase, $c_{\mathrm{dil}}$, is obtained from the *x*-intercept of the fits and $c_{\mathrm{coa}} - c_{\mathrm{dil}}$ from the inverse slope, from which the concentration in the coacervate phase ($c_{\mathrm{coa}}$) is calculated. (I) Coacervate volume fraction $\phi$ as a function of the absolute *trans*-azoTAB concentration $c_{trans} = x_{trans} c_0$ for varying $c_0$. All datasets exhibit a common onset at a fixed threshold concentration $c_{\mathrm{sat}}$, indicating that nucleation is controlled by the absolute *trans* concentration. (J) Master-curve collapse obtained by plotting the normalized volume fraction $\phi/\phi_{\mathrm{max}}$ as a function of the rescaled *trans* fraction $u = (x_{trans} - x^*_{trans})/(1 - x^*_{trans})$. The red line shows the universal scaling predicted by the diffusion-limited growth model (Note S3).

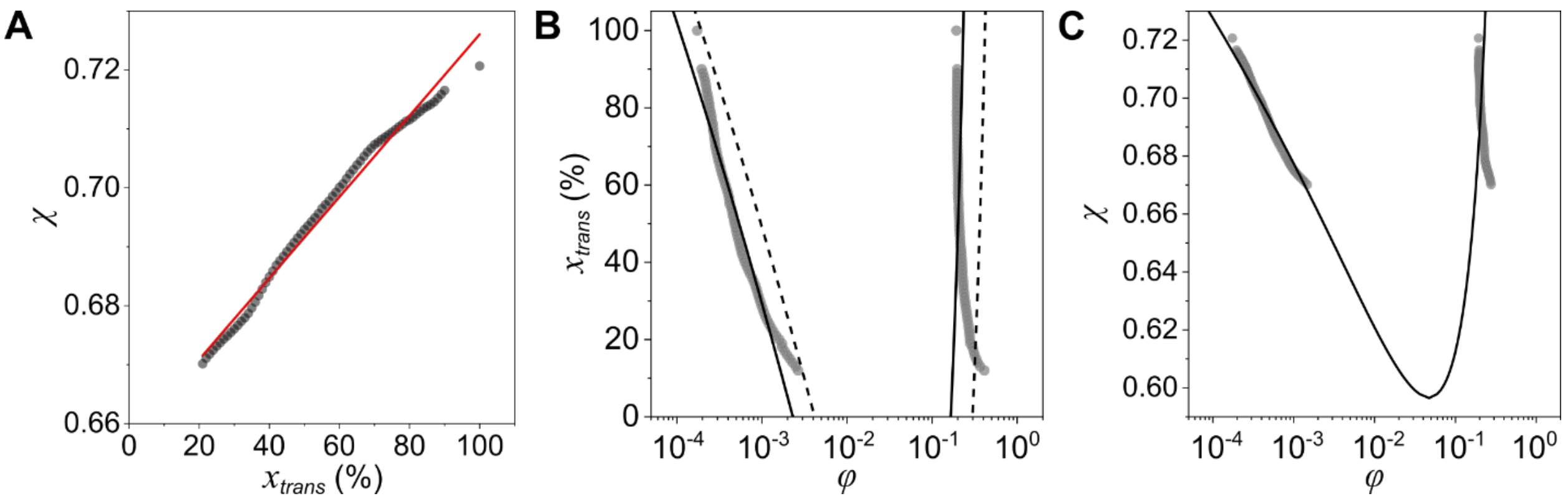

**Fig. S5. Flory-Huggins modeling of photoswitchable coacervation.** (**A**) Plot of the Flory parameter $\chi$ vs the fraction of *trans*-azoTAB $x_{trans}$, calculated from the experimental values of the DNA volume fractions in the dilute ($\varphi_{\mathrm{dil}}$) and condensed ($\varphi_{\mathrm{coa}}$) phases (black dots), and linear fit (red line). (**B**) Phase diagram of DNA/azoTAB coacervation in the ($x_{trans}$, $\varphi$) space showing values from microfluidics experiments (grey dots), and computed Flory-Huggins binodal before (dashed black line) and after (solid black line) rescaling by a factor $m$ = 1.8. (C) Binodal in the ($\chi$, $\varphi$) representation, combining experimental $\chi$ values from microfluidics (grey dots) and the theoretical curve (black line).

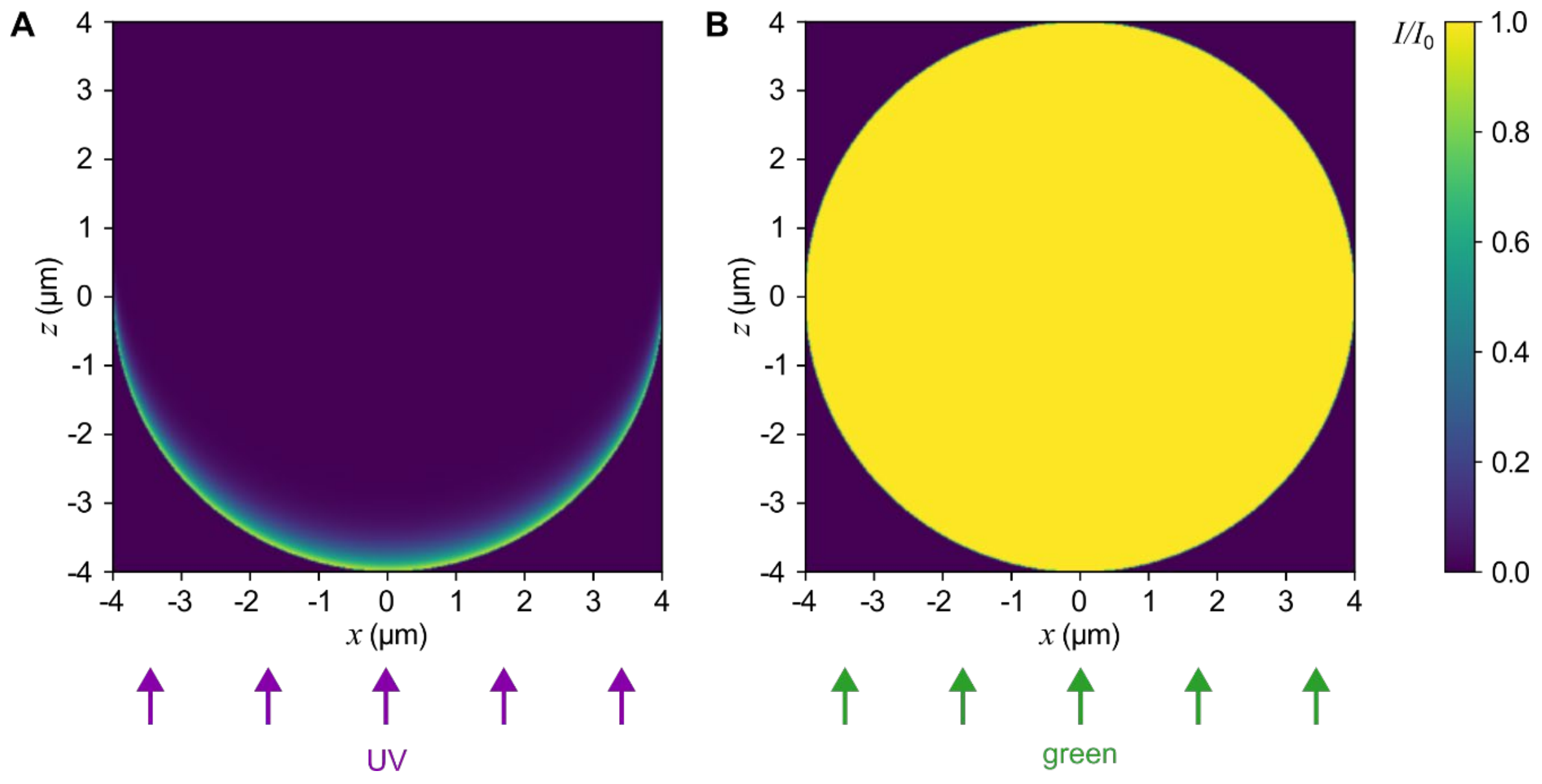

**Fig. S6. Spatial distribution of light intensity within coacervates.** (**A**,**B**) Calculated normalized light intensity profiles $I/I_0$ inside a spherical coacervate droplet (radius $R$ = 4 μm, *trans*-azoTAB concentration of 1 mol·L$^{-1}$) under collimated UV or green illumination along the $z$-axis (irradiation from the bottom), using a Beer-Lambert attenuation model. For *trans*-rich coacervates, UV illumination (high absorption) shows strong attenuation, confining light to a shallow interfacial layer (A). Green illumination (no absorption) exhibits weak attenuation and near-uniform intensity across the coacervate (B). All intensity maps are normalized to the incident intensity $I_0$; non-absorbing regions outside the spherical domain are masked; and refraction at the droplet interface is neglected due to small refractive index contrast between the two phases.

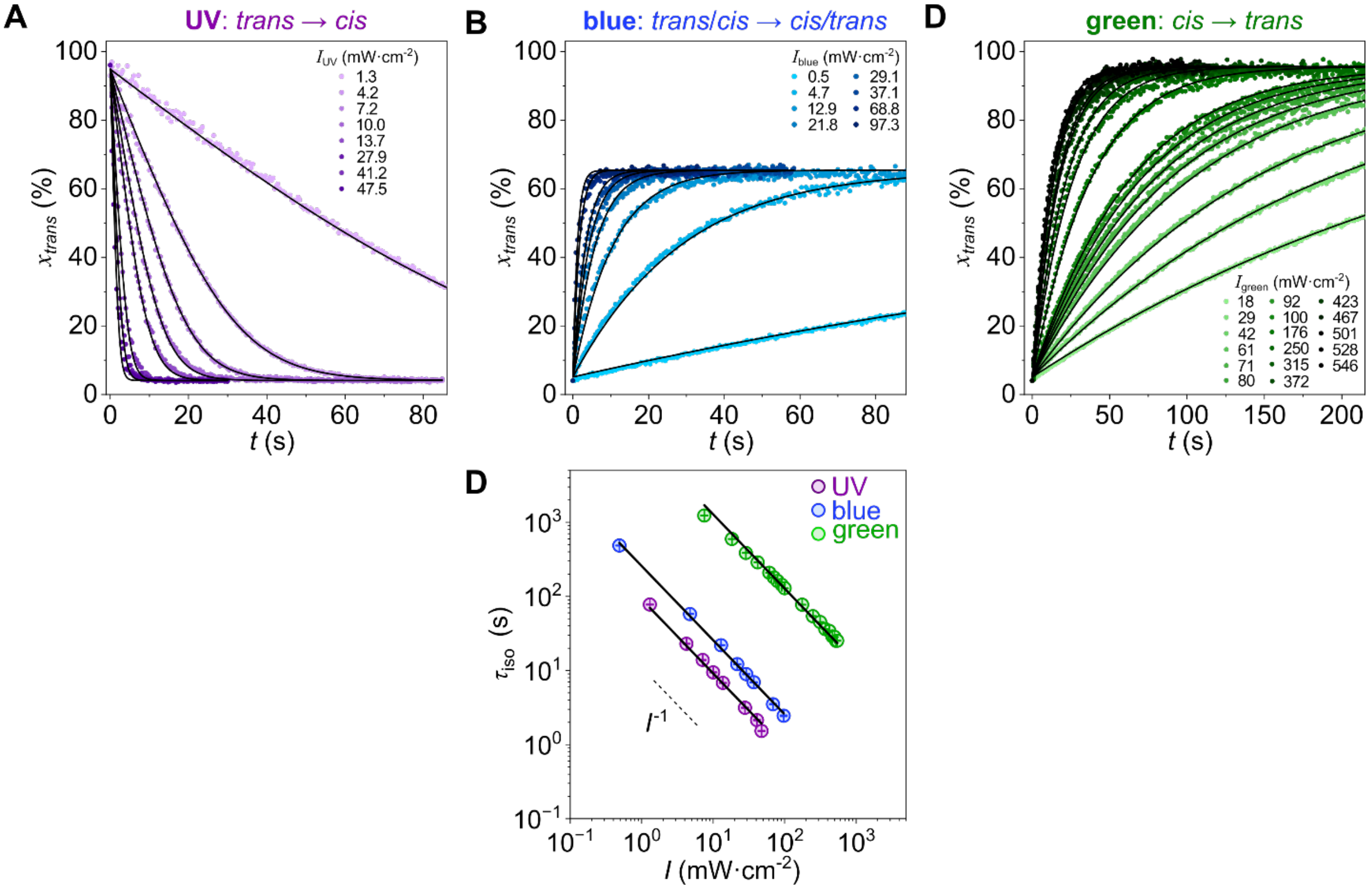

**Fig. S7. Isomerization kinetics of azoTAB.** (**A-C**) Time-dependent evolution of the fraction of *trans*-azoTAB $x_{trans}$ in a 50 μM solution irradiated with UV (A), blue (B) and green (C) light, respectively. Black lines are global fits using the model developed in Note S5, from with the characteristic isomerisation times, $\tau_{\text{iso}}$, are determined. (**D**) Plot of $\tau_{\text{iso}}$ under UV, blue or green light irradiation as a function of the light intensity, $I$. Data is shown as mean ± s.e. of the fits in (A-C). Black solid lines show power law fits using a scaling exponent of -1, confirming that $\tau_{\text{iso}} \propto I^{-1}$.

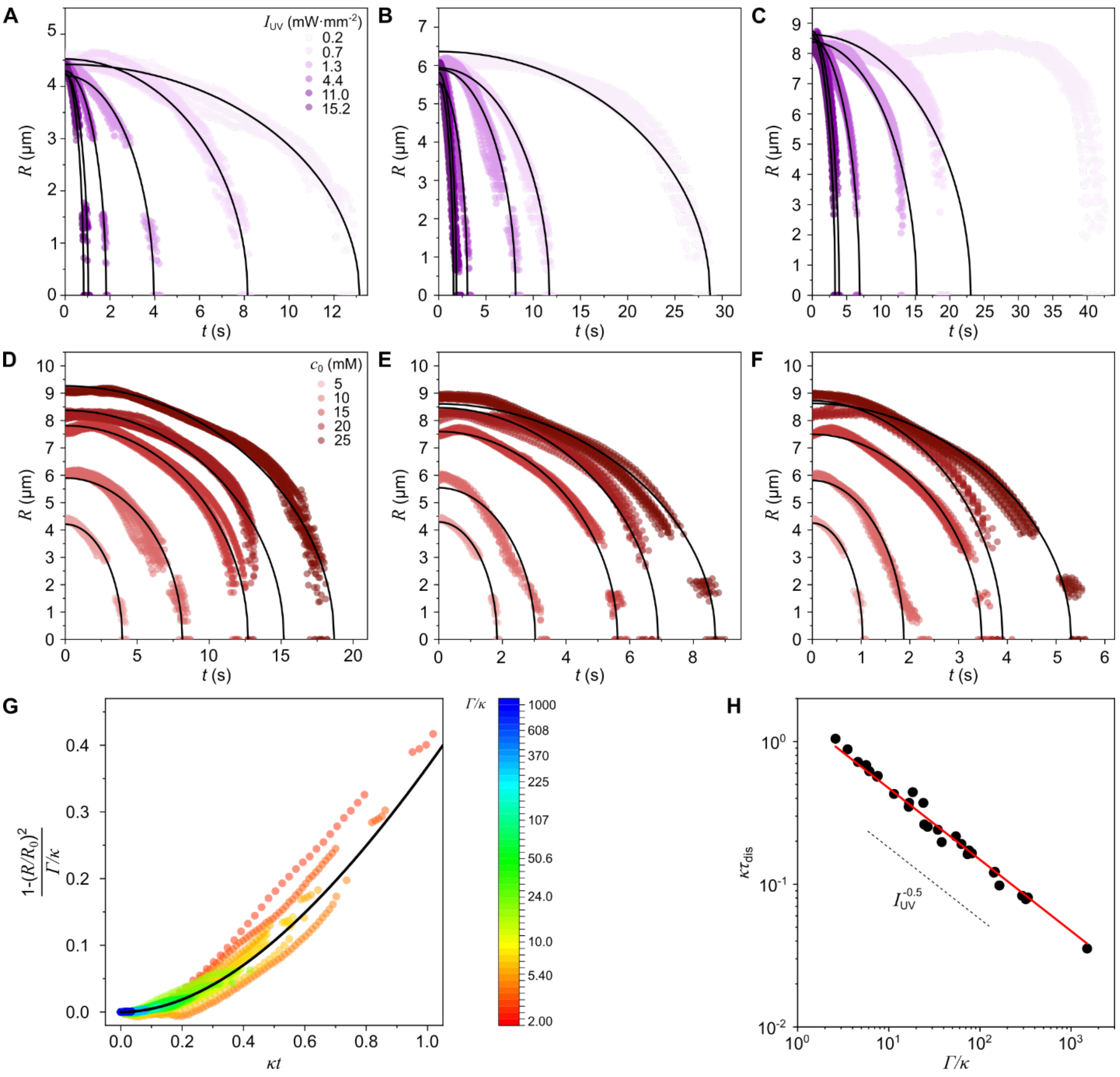

**Fig. S8. Coacervate dissolution dynamics.** (**A-F**) Representative time evolution of the coacervate radius $R(t)$ for coacervates prepared at varying initial concentrations $c_0$ and irradiated at different UV light intensities $I_{UV}$. Panels (A-C) show datasets at fixed $c_0$ (A: 5 mM, B: 10 mM, C: 20 mM), while panels (D-F) correspond to fixed $I_{UV}$ (D: 1.3 mW·mm$^{-2}$, E: 4.4 mW·mm$^{-2}$, F: 11.0 mW·mm$^{-2}$, k: 15.2 mW·mm$^{-2}$). Symbols represent trajectories for 10 individual coacervates for each condition; solid lines indicate global fits to the dissolution model (Note S6). At low UV intensities and high $c_0$, deviations from the model are observed (see in particular panel (C), $c_0$ = 20 mM and $I_{UV}$ = 0.2 mW·mm$^{-2}$). A transient increase in $R(t)$ prior to collapse may be observed, which prevents fitting with the dissolution model. More generally, initial plateaus or slight swelling are apparent under several conditions, becoming more pronounced at low $I_{UV}$ and high $c_0$. These effects likely reflect osmotically driven water influx and are not captured by the homogeneous dissolution model. (**G**) Rescaled dissolution trajectories plotted as $(1-(R/R_0)^2)/(\Gamma/\kappa)$ versus $\kappa t$, with photochemical rate $\kappa$ and diffusive rate $\Gamma$ from the dissolution model (Note S6). Each dataset corresponds to a single coacervate trajectory (one droplet per experimental condition).

Datasets are color-coded by $\Gamma/\kappa$ (log scale, 2-1000, rainbow colormap). All trajectories collapse onto a single master curve described by $e^{-\kappa t} + \kappa t - 1$ (solid black line). (**H**) Dimensionless dissolution time plotted as $\kappa\tau_{\mathrm{dis}}$ versus $\Gamma/\kappa$ for all experimental conditions. Each point corresponds to the average over 10 droplets for a given condition. A power-law dependence $\kappa\tau_{\mathrm{dis}} \propto (\Gamma/\kappa)^{-0.5}$ is observed over the explored range (red solid line). This scaling is consistent with the intermediate regime of dissolution, where $\kappa\tau_{\mathrm{dis}} \sim \mathcal{O}(1)$ and dissolution results from the coupled action of photochemical conversion and diffusive transport (Note S6).

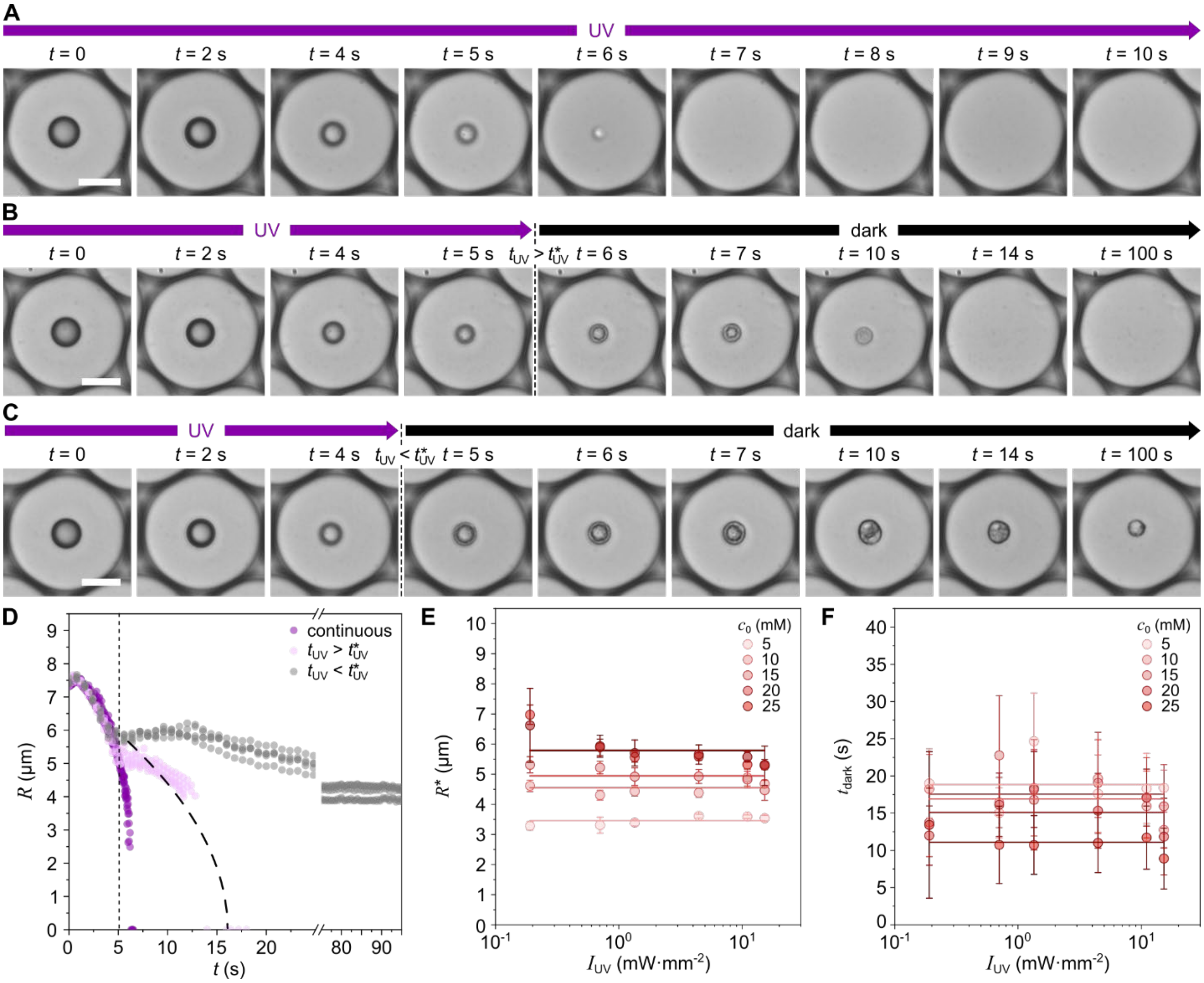

**Fig. S9. Irradiation-interrupted dissolution experiments.** (**A-C**) Time-lapse bright-field microscopy images of a w/o emulsion droplet prepared at $c_0$ = 15 mM and irradiated with UV light ($I_{UV}$ = 4.4 mW·mm$^{-2}$) either continuously (A) or for a varying time $t_{UV} > t^*_{UV}$ (B) resulting in coacervate dissolution in the dark; and $t_{UV} < t^*_{UV}$ (C) resulting to incomplete coacervate dissolution in the dark. Scale bars, 20 µm. (**D**) Corresponding time evolution of the coacervate radius $R$. The vertical dotted line identifies $t^*_{UV}$. Symbols show trajectories for 5 individual coacervate droplets at each condition. The dashed line shows a fit to a diffusion-like model (for $t_{UV} > t^*_{UV}$). (**E**) Critical coacervate radius $R^*$ measured at $t_{UV} = t^*_{UV}$ as a function of UV light intensity at different $c_0$. The corresponding critical coacervate volume fraction is used to estimate the critical fraction of *trans*-azoTAB below which coacervate spontaneously dissolve when light is turned off, reported on the phase diagram in Fig. 2F (purple triangles). (**F**) Dissolution time in the dark $t_{dark}$, measured after irradiation for $t_{UV} = t^*_{UV}$ as a function of UV light intensity at different $c_0$. The light-independency confirms that dissolution proceeds purely via diffusion once light is turned off.

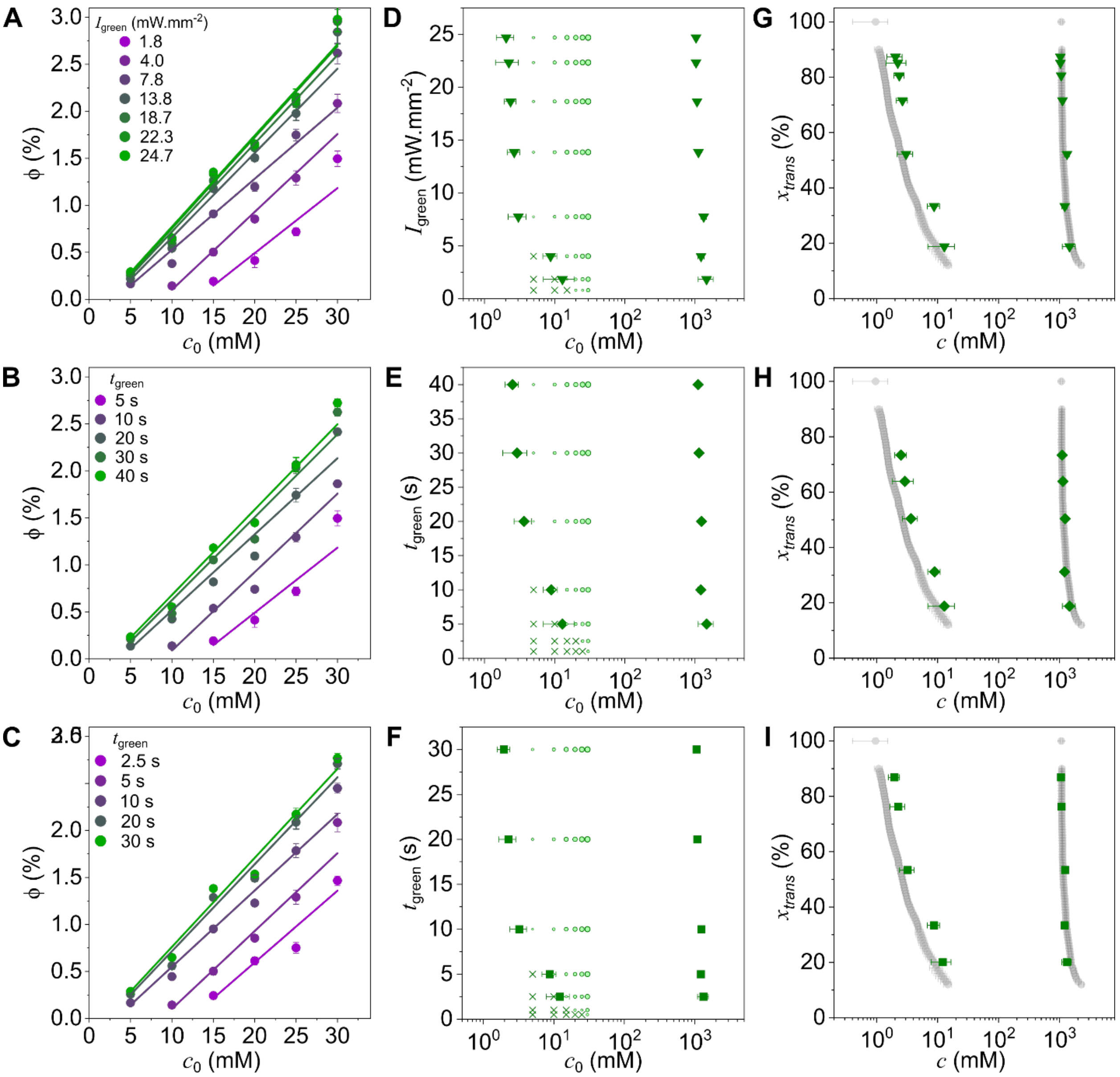

**Fig. S10. Reconstruction of the phase diagram under controlled green light illumination.** (**A-C**) Coacervate volume fraction $\phi$ as a function of $c_0$ measured 30 min after 5 s of green light irradiation at different light intensities (A), and after varying duration of green light irradiation at a fixed intensity of 1.8 mW·mm$^{-2}$ (B) and 4.0 mW·mm$^{-2}$ (C). Solid lines indicate linear fits used to extract binodal concentrations. (**D-F**) Phase diagram of azoTAB-dsDNA coacervation as a function of green light intensity for 5 s irradiation (D), and irradiation duration at fixed light intensity of 1.8 mW·mm$^{-2}$ (E), 4.0 mW·mm$^{-2}$ (F), showing the dilute-phase concentration $c_{\mathrm{dil}}$ (left branch) and the condensed-phase concentration $c_{\mathrm{coa}}$ (right branch) as solid dark green symbols. Error bars on concentration represent the standard error propagated from linear fits of $\phi$ versus $c_0$ at fixed $I_{\mathrm{green}}$ (for D) and fixed $t_{\mathrm{green}}$ (for E,F). Light green circles indicate conditions where coacervates form, with symbol size proportional to $\phi$, and crosses indicate conditions where no coacervates are observed. (**G-I**) Overlaid phase diagrams combining relaxation-based data (grey symbols) with data obtained after 5 s of green light irradiation at varying intensities (G, green symbols), or varying irradiation duration at a fixed light intensity of 1.8 mW·mm$^{-2}$ (H, green

symbols) or 4.0 mW·mm$^{-2}$ (I, green symbols). For green light data, $x_{trans}$ are calculated using eq. S52 in Note S7 by taking into account the given irradiation duration and light intensity, and neglecting the thermal relaxation during incubation in the dark.

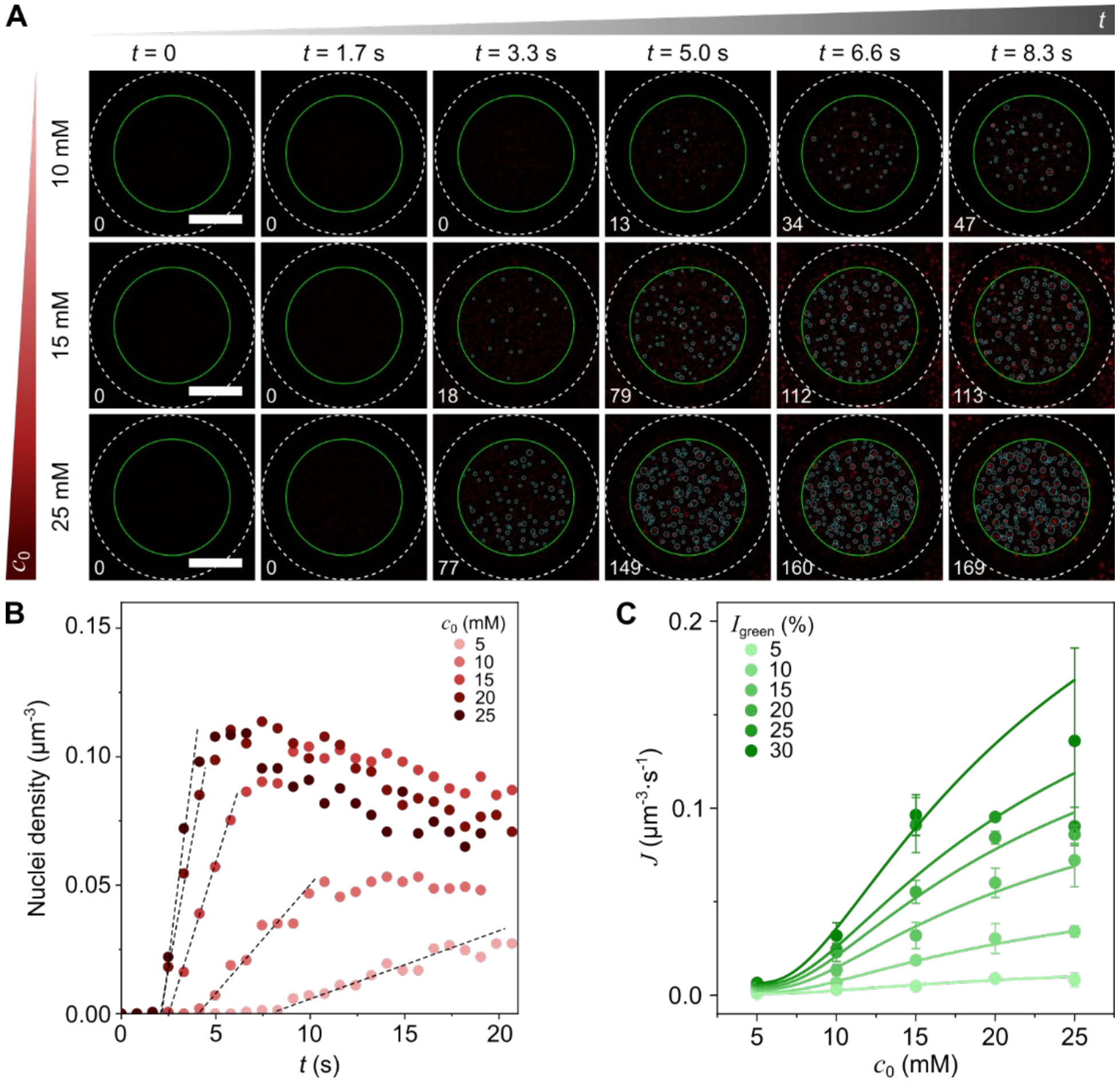

**Fig. S11. Nucleation rate under green light illumination.** (**A**) Representative time-lapse confocal fluorescence microscopy images of a w/o emulsion droplet prepared at varying $c_0$ and fixed green laser level (10%). Coacervates were doped with TAMRA-ssDNA (ssDNA : dsDNA = 1 : $10^4$ molar ratio). White dotted circles highlight the boundary of w/o droplets. Coacervate nuclei are detected via thresholding, individually circled and counted within the green circle boundary (number of nuclei listed on the bottom left corner of each image) to determine the nucleation rate. Samples were first equilibrated with UV light to induce coacervate dissolution prior to green light irradiation. Scale bars, 20 µm. (**B**) Representative plots of nuclei density vs. time at varying $c_0$ and fixed green laser level (10%). The doted lines show the slopes used to determine the nucleation rate $J$. (C) Nucleation rate $J$ as a function of $c_0$ for samples irradiated at varying green light intensities. Data are shown as mean ± s.d. ($n$ = 3 droplets). Solid lines show fits with the model predicted by classical nucleation theory (Note S8).

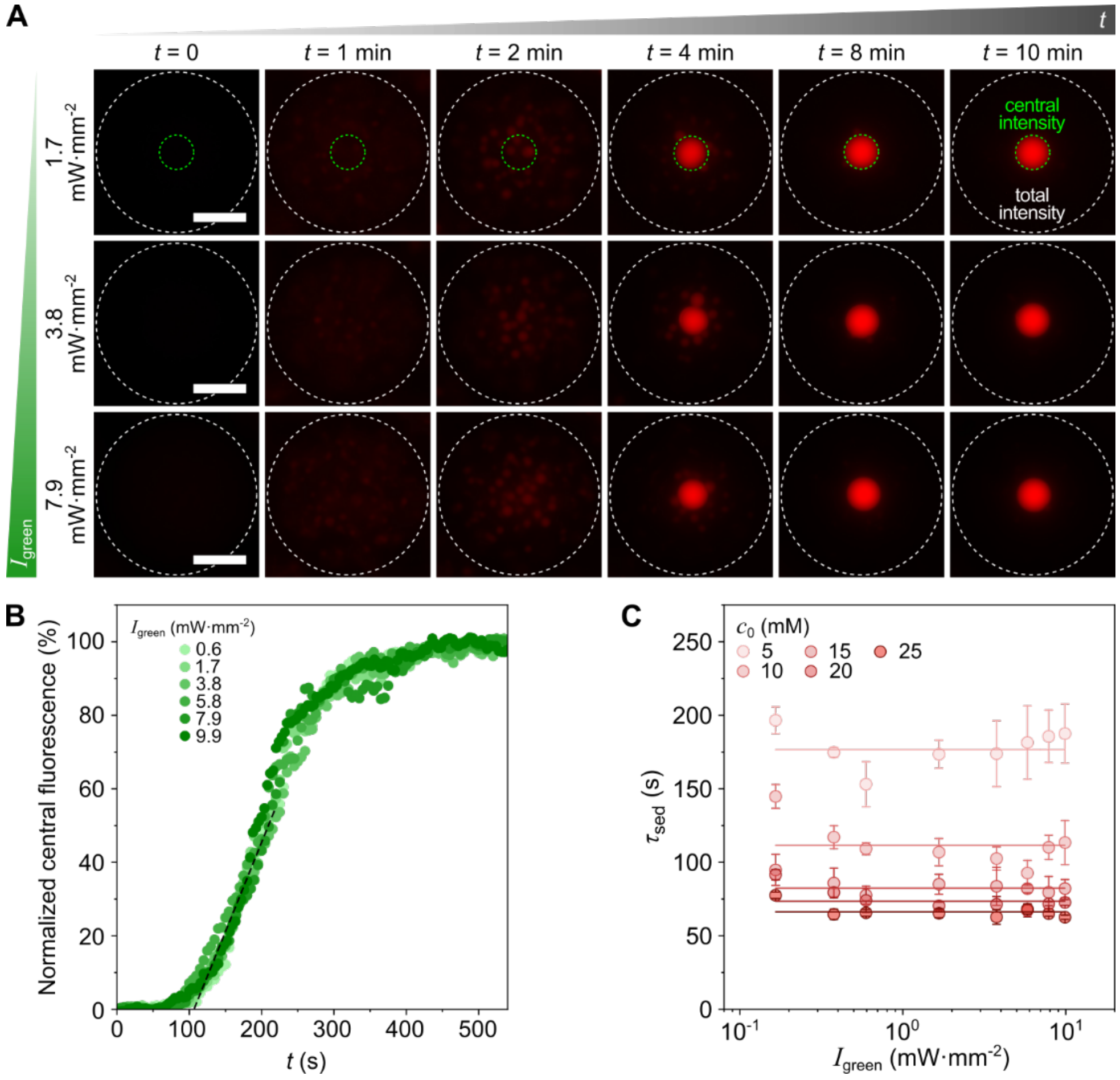

**Fig. S12. Coacervate sedimentation after nucleation.** (**A**) Representative time-lapse epifluorescence microscopy images of a w/o emulsion droplet prepared at fixed $c_0$ = 10 mM and irradiated continuously with green light of varying intensity. Coacervates were doped with TAMRA-ssDNA (ssDNA : dsDNA = 1 : 2.5 × $10^4$ molar ratio). White dotted circles highlight the boundary of w/o droplets. Green dotted circles in a identify the region used to calculate the central fluorescence intensity fraction (used in (C)). Coacervate sedimentation, observed by the gradual concentration of red fluorescence at the centre of the w/o droplets, occurs qualitatively at the same speed regardless of $I_{green}$. Samples were first equilibrated with UV light to induce coacervate dissolution prior to green light irradiation. Scale bars, 20 µm. (**B**) Representative time-dependent plots of the normalized central fluorescence intensity fraction of UV-equilibrated w/o droplets prepared at $c_0$ = 10 mM and irradiated at varying green light intensity. The onset of increase in the fluorescence intensity (identified by a black dotted line) was used to determine the sedimentation time. (C) Sedimentation time, $\tau_{sed}$, of coacervate droplets prepared at varying $c_0$ as a function of green light intensity, $I_{green}$. Data are shown as mean ± s.d. ($n$ = 3 droplets). Solid line are linear fits with zero slope.

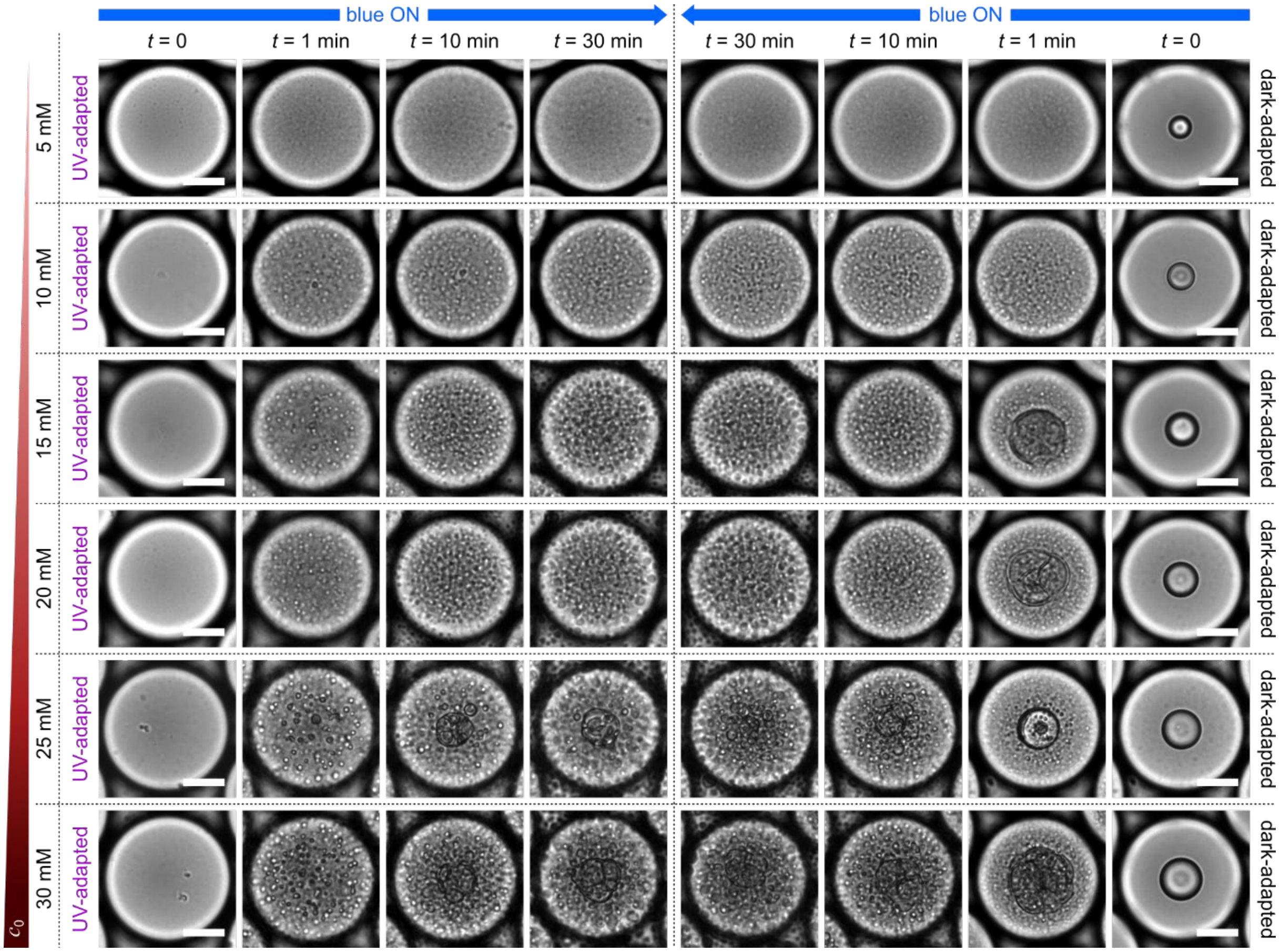

**Fig. S13. Blue-light induced arrested coarsening.** Representative time-lapse bright-field microscopy images of a w/o emulsion droplet prepared at varying $c_0$, either in the dark or pre-equilibrated with UV light, and irradiated continuously with blue light ($I_{blue}$ = 60 mW·mm$^{-2}$). For $c_0$ > 20 mM, residual structures remain after 30 min of irradiation in both the UV- and dark-adapted samples, indicating that coarsening or fragmentation is incomplete at these higher concentrations. These residual features are heterogeneous, often appearing as fused or vacuolated assemblies, distinct from the homogeneously dispersed coacervates observed at lower $c_0$. Scale bars, 20 μm.

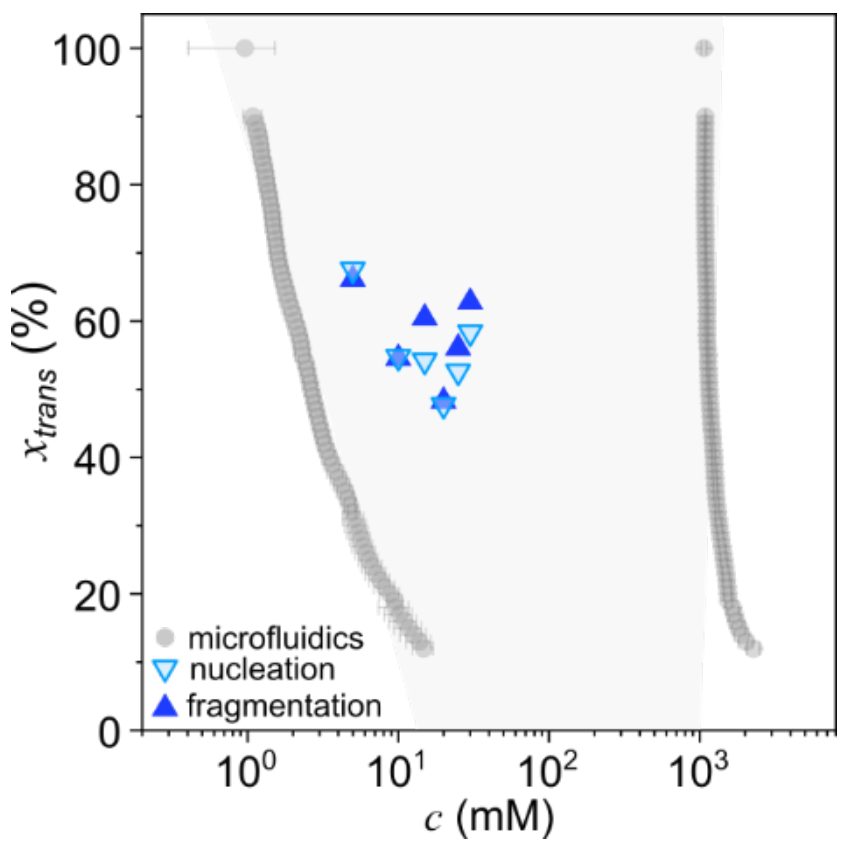

**Fig. S14. Average azoTAB composition in blue-equilibrated samples.** Fraction of *trans*-azoTAB $x_{trans}$ estimated from the coacervate volume fraction at varying $c_0$ after blue-light induced nucleation (from UV-adapted samples) or fragmentation (from dark-adapted samples), and after 30 min dark incubation, reported on the phase diagram.

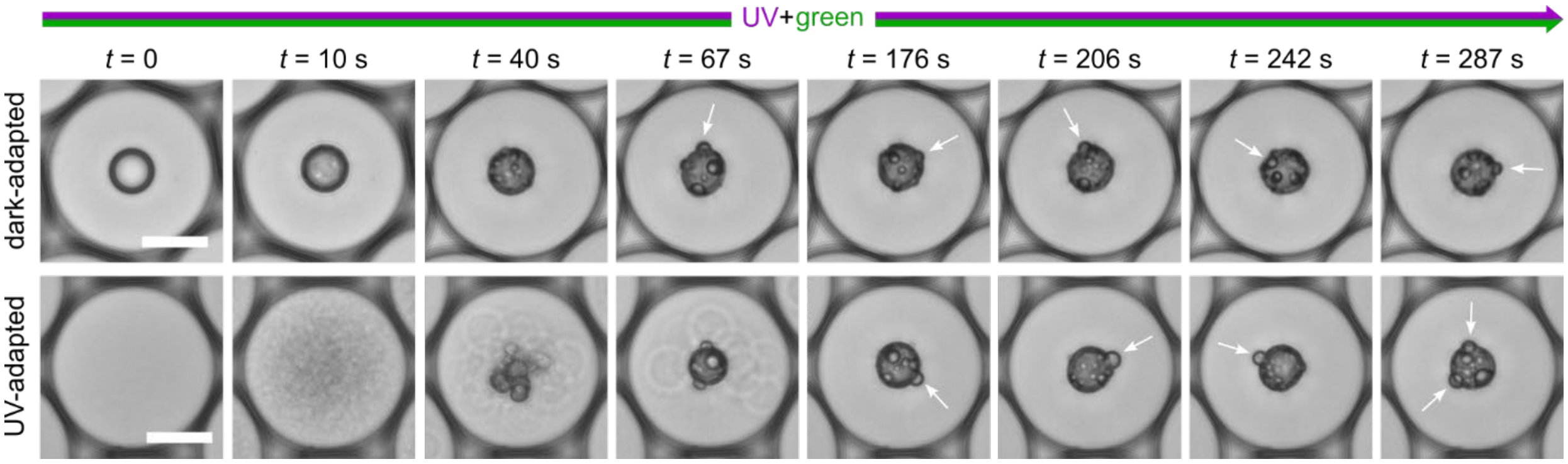

**Fig. S15. Coacervate budding from the dissolved state.** Representative time-lapse bright-field microscopy images of a w/o emulsion droplet prepared at $c_0$ = 20 mM in the dark (top) or initially equilibrated with UV light to induce coacervate dissolution (bottom), then irradiated simultaneously with UV ($I_{\mathrm{UV}}$ = 1.0 mW·mm$^{-2}$) and green ($I_{\mathrm{green}}$ = 20 mW·mm$^{-2}$) lights (budding regime). Images show gradual coacervate formation (at $t \leq 67$ s) and concomitant emergence of budding behaviour (at t ≥ 67s). White arrows highlight buds protruding from the coacervate surface. Scale bars, 20 µm.

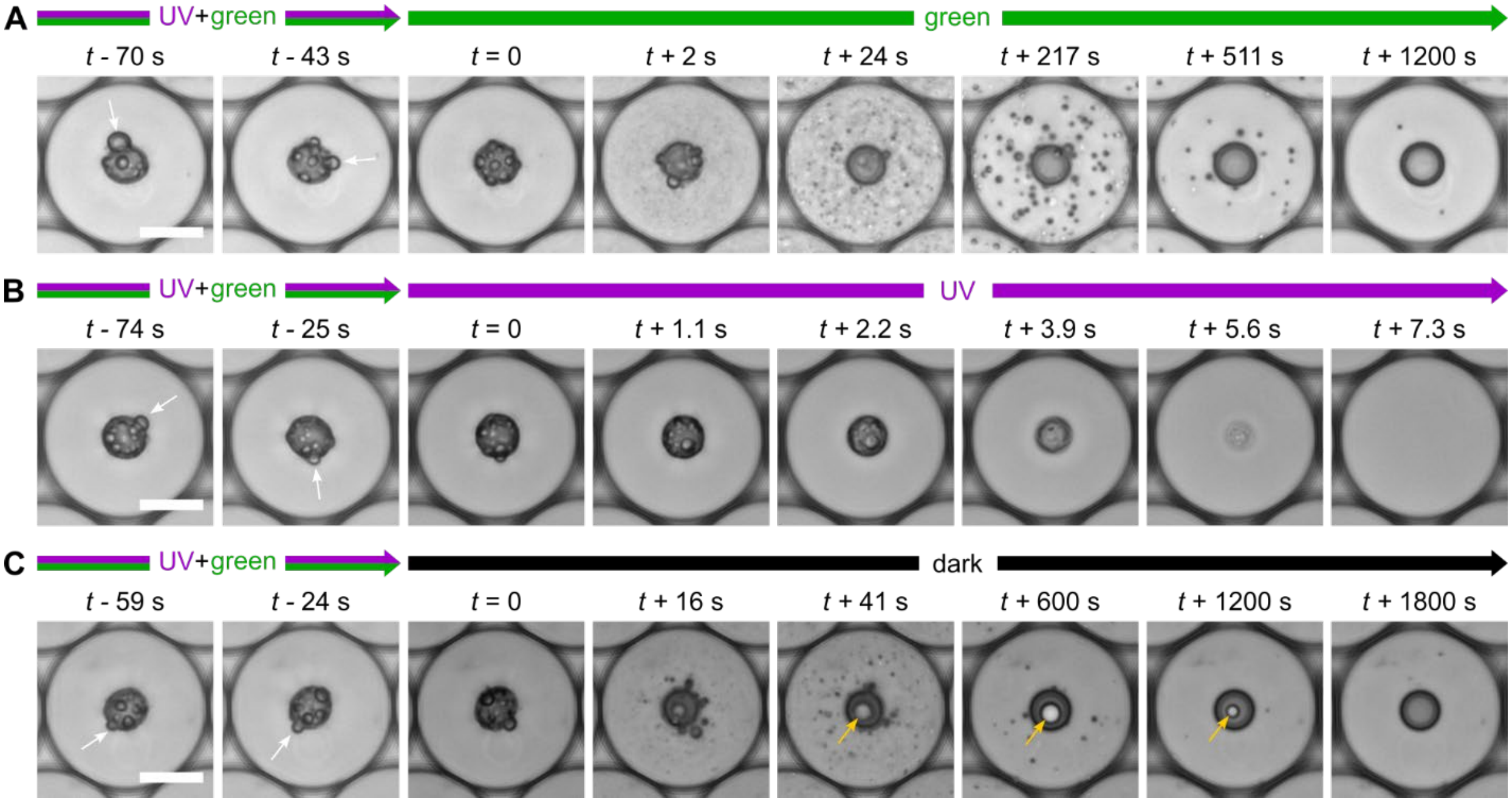

**Fig. S16. Coacervate evolution upon cessation of UV and/or green irradiation.** (**A-C**) Time-lapse bright-field microscopy images of a w/o emulsion droplet prepared at $c_0$ = 20 mM, initially irradiated simultaneously with UV ($I_{UV}$ = 1.0 mW·mm$^{-2}$) and green ($I_{green}$ = 20 mW·mm$^{-2}$) light for 5 minutes (budding regime), and subsequently monitored after switching off UV (A), green (B) or both lights sources (C). Irradiation is stopped at $t$ = 0. Switching off UV leads to nucleation and growth of smaller droplets in the dilute phase, followed by coalescence into the main droplet to produce a single stable coacervate (A). Switching off green light induces dissolution of the coacervate (B). When both light sources are turned off, droplets nucleate and grow in the dilute phase, and a transient vacuole (yellow arrow) forms within the parent droplet before relaxation to a single coacervate after 30 min (C). In all cases, budding ceases a few seconds after removal of either or both light sources.

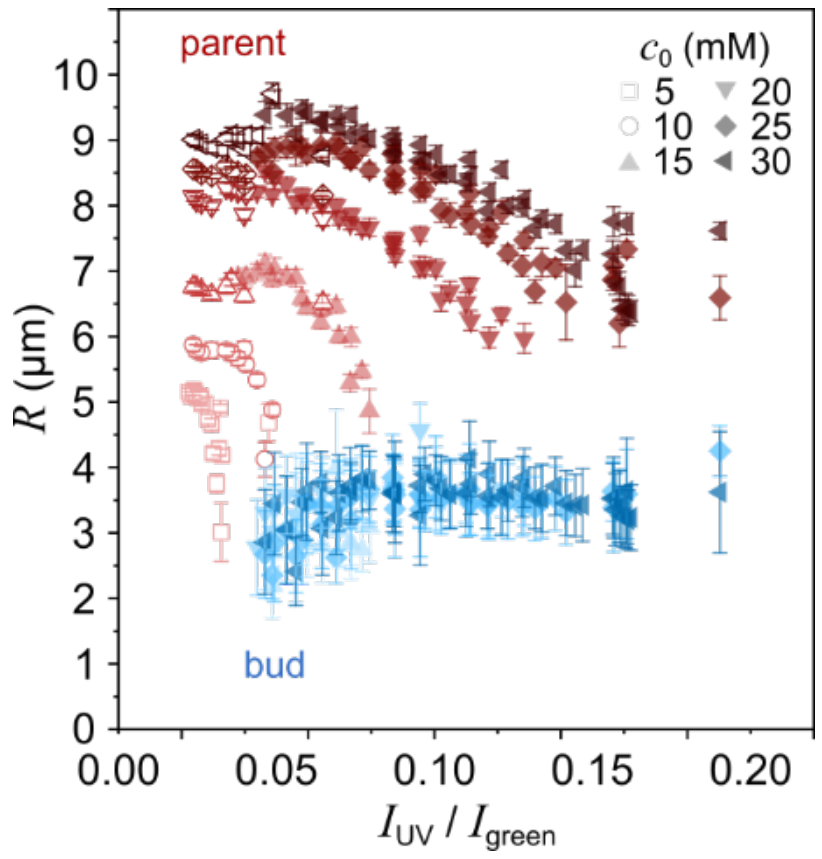

**Fig. S17. Intensity-dependent parent and bud size.** Non-normalized parent droplet (red) and bud (blue) radius as a function of the illumination ratio $I_{UV}/I_{green}$ for varying $c_0$: 5 mM (squares), 10 mM (circles), 15 mM (upward triangles), 20 mM (downward triangles), 25 mM (diamonds), 30 mM (leftward triangles). Darker hues correspond to higher $c_0$. Filled symbols indicate droplets exhibiting budding; open symbols correspond to droplets not developing buds. Data are shown as mean ± s.d. ($n \geq 10$ parent droplets and buds).

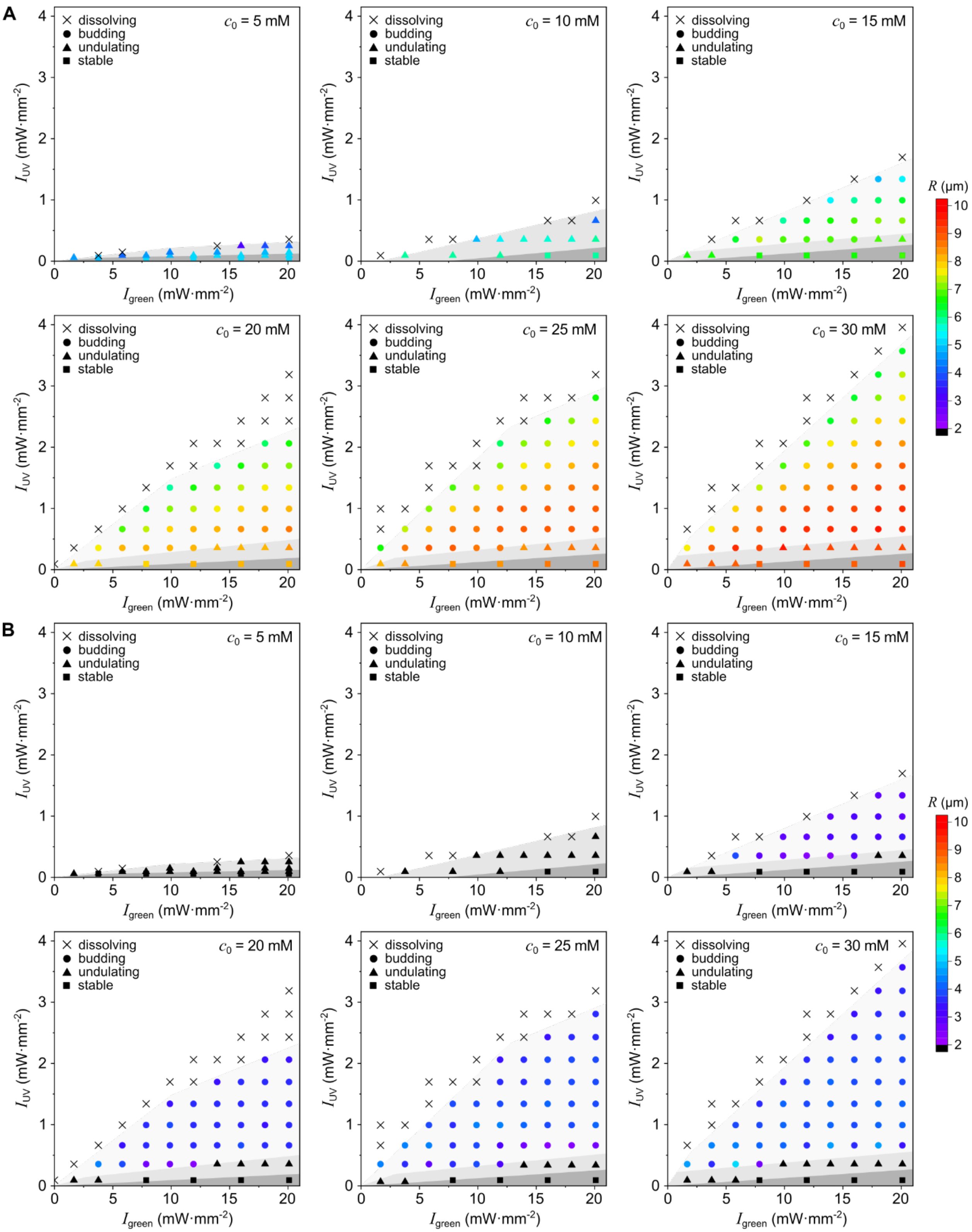

**Fig. S18. Phase response under dual UV/green illumination at varying concentration.** (**A**,**B**) Non-equilibrium response diagrams in the ($I_{UV}, I_{green}$) parameter space showing dissolving (crosses), budding (circles), undulating (triangles) and stable (squares) regimes for 100 mM NaCl concentration and varying $c_0$: 5, 10, 15, 20, 25 and 30 mM. Symbol color encodes the mean parent

droplet (A) and bud (B) radius measured during the final minute of dual illumination (rainbow color scale: 2-10 μm). Grey shading indicates the observed regimes: dark grey, stable; grey, undulating; light grey, budding.

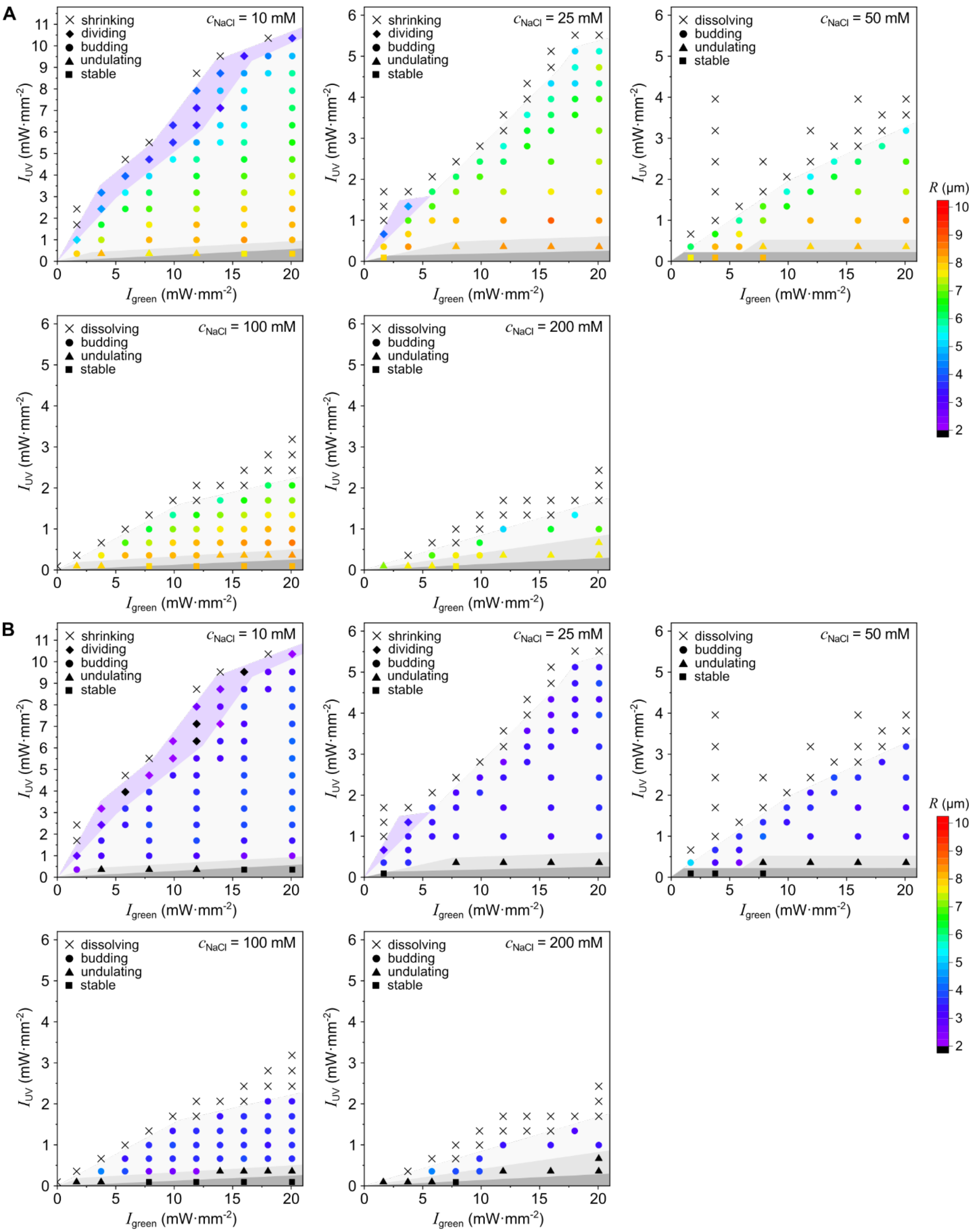

**Fig. S19. Phase response under dual UV/green illumination at varying ionic strength.** (**A**,**B**) Non-equilibrium response diagrams in the ($I_{\mathrm{UV}}, I_{\mathrm{green}}$) parameter space showing dissolving or shrinking (crosses), budding (circles), undulating (triangles) and stable (squares) regimes for $c_0$ = 20 mM and varying salt concentrations $c_{\mathrm{NaCl}}$: 10, 25, 50, 100 and 200 mM. Symbol color encodes

the mean parent droplet (A) and bud (B) radius measured during the final minute of dual illumination. In dividing conditions, the color corresponds to the radius of the detached daughter droplet immediately after fission (rainbow color scale: 2-10 μm). Black squares and triangles correspond to conditions where no budding is detected. Grey and purple shading indicate the observed regimes: dark grey, stable; grey, undulating; light grey, budding; purple, dividing. Note that the $I_{\mathrm{UV}}$ axis range is higher at $c_{\mathrm{NaCl}}$ = 10 mM to match the experimentally accessible regime of phase behavior at low salt concentration.

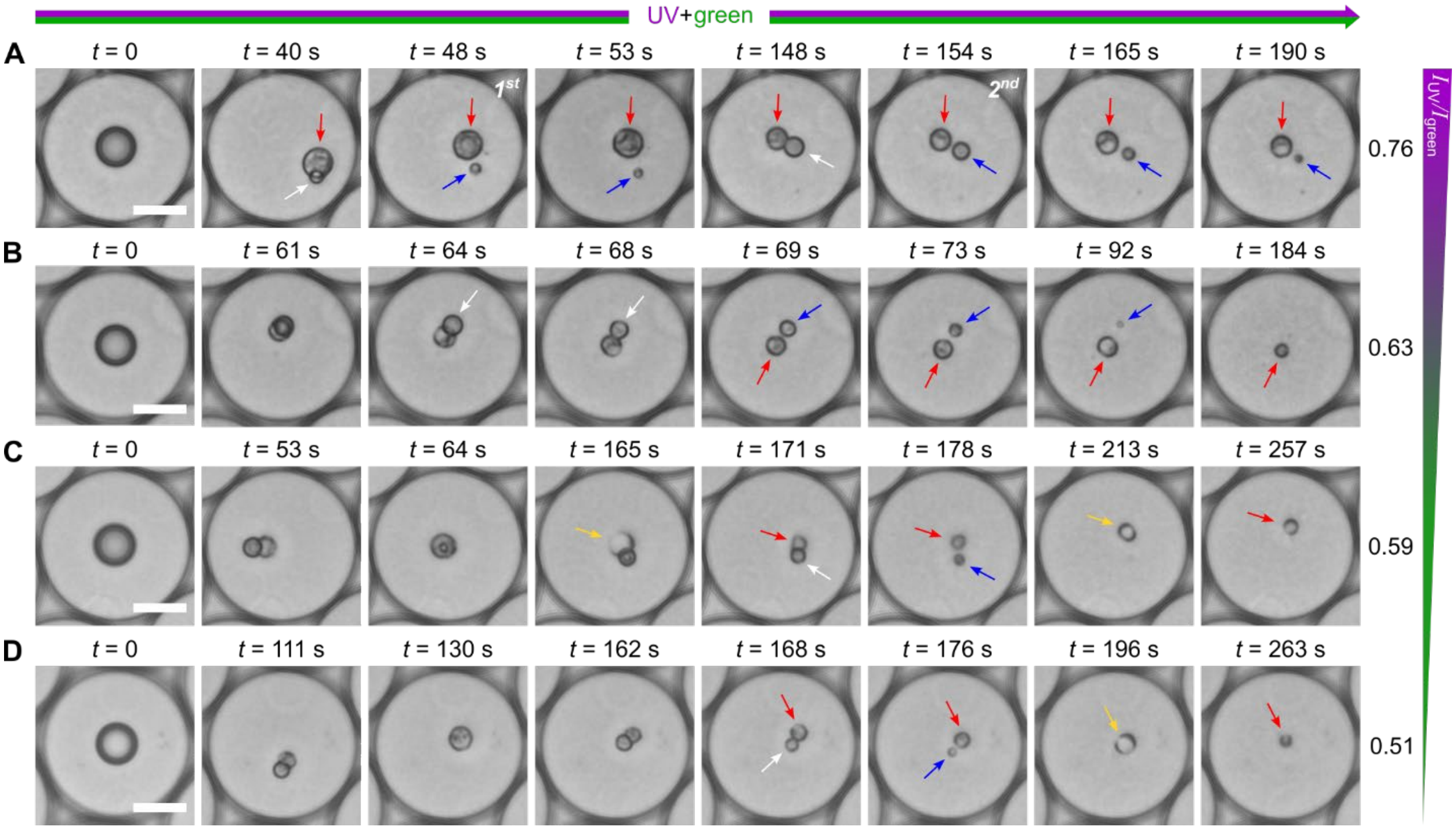

**Fig. S20. Division examples across illumination ratios.** (**A–D**) Time-lapse bright-field microscopy images of a w/o emulsion droplet prepared at $c_0$ = 20 mM in 10 mM NaCl and co-irradiated continuously with UV and green light at varying intensities (A: $I_{\mathrm{UV}}$ = 1.3 mW·mm$^{-2}$, $I_{\mathrm{green}}$ = 1.7 mW·mm$^{-2}$; B: $I_{\mathrm{UV}}$ = 2.4 mW·mm$^{-2}$, $I_{\mathrm{green}}$ = 3.8 mW·mm$^{-2}$; C: $I_{\mathrm{UV}}$ = 4.7 mW·mm$^{-2}$, $I_{\mathrm{green}}$ = 7.9 mW·mm$^{-2}$; D: $I_{\mathrm{UV}}$ = 7.1 mW·mm$^{-2}$, $I_{\mathrm{green}}$ = 14 mW·mm$^{-2}$). Corresponding $I_{\mathrm{UV}}/I_{\mathrm{green}}$ ratios are listed on the right. In (A), two sequential division events are observed, although droplets continue shrinking over time so that division eventually stops. White arrows indicate buds protruding from the coacervate surface; red and blue arrows denote the remnant and daughter droplets, respectively; yellow arrows in (C–D) indicate vacuolated droplets, identified by a high-contrast shell surrounding a lower-contrast interior. Scale bars, 20 µm.

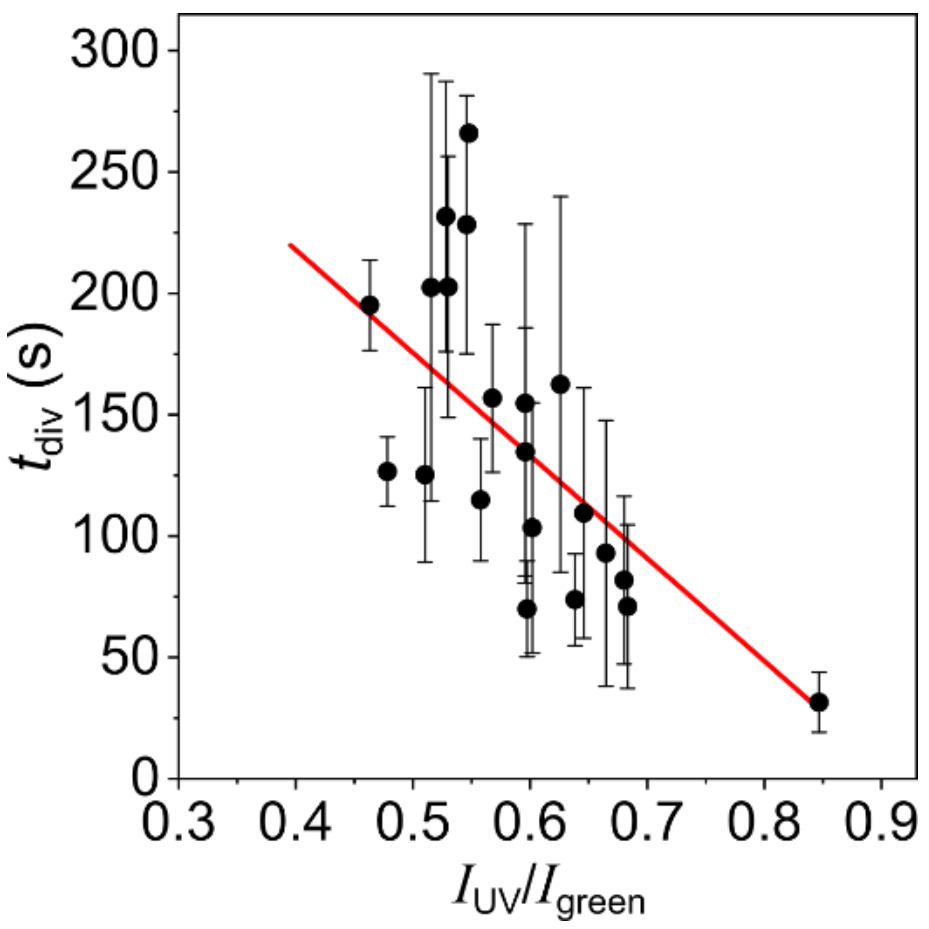

**Fig. S21. Division time.** Time to first division $t_{\mathrm{div}}$ as a function of the illumination ratio $I_{\mathrm{UV}}/I_{\mathrm{green}}$ for $c_0$ = 20 mM in 10 mM NaCl. The solid line serves as a guide to the eye. Data are shown as mean ± s.d. ($n \geq 4$).

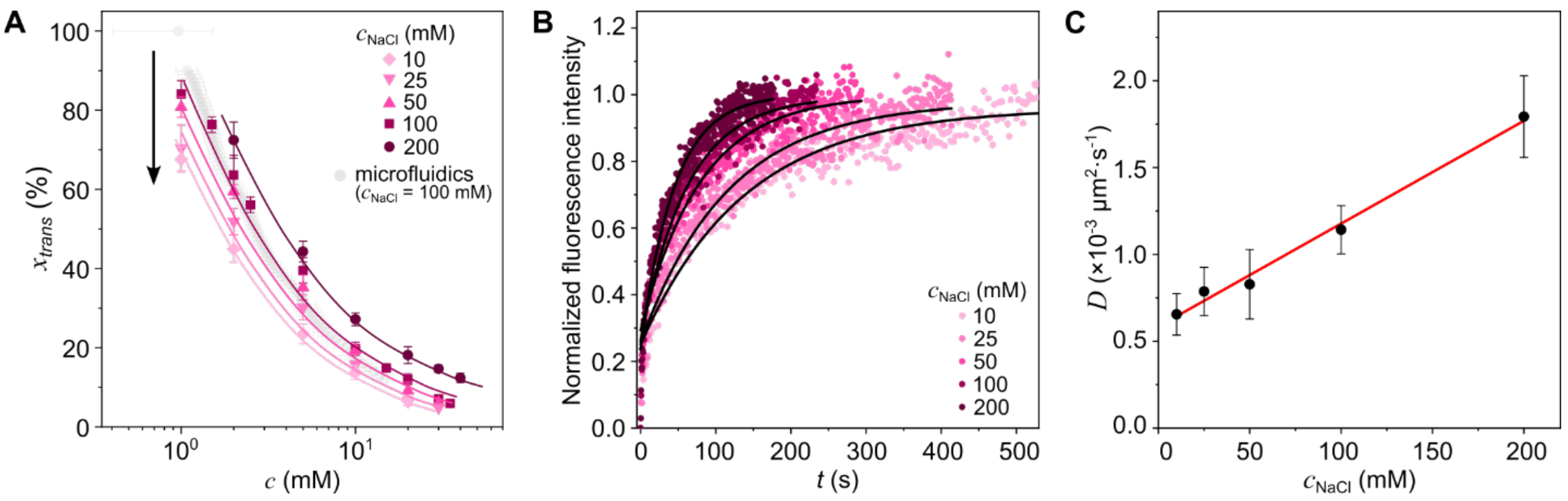

**Fig. S22. Phase stabilization at lower ionic strength.** (**A**) Dilute phase boundary determined from turbidity recovery assays for samples prepared at varying concentration and salt concentration. A shift of the phase boundary to lower fractions of *trans*-azoTAB is observed as the ionic strength decreases, pointing to the stabilization of the condensed phase. Data are shown as mean ± s.d. ($n$ = 3 independent samples per condition). Solid lines serve as guide to the eye. (**B**) Time-dependent normalized fluorescence recovery after photobleaching intensity of Cy5-ssDNA within coacervates prepared at $c_0$ = 20 mM and varying salt concentrations. Solid lines indicate fit to a mono-exponential growth. (C) Diffusion coefficient of Cy5-ssDNA within coacervates as a function of salt concentration, showing a slower molecular diffusion at the ionic strength decreases, consistent with stronger attractive interactions within the coacervate. Data are shown as mean ± s.d. ($n$ = 6). The red line shows a linear fit.

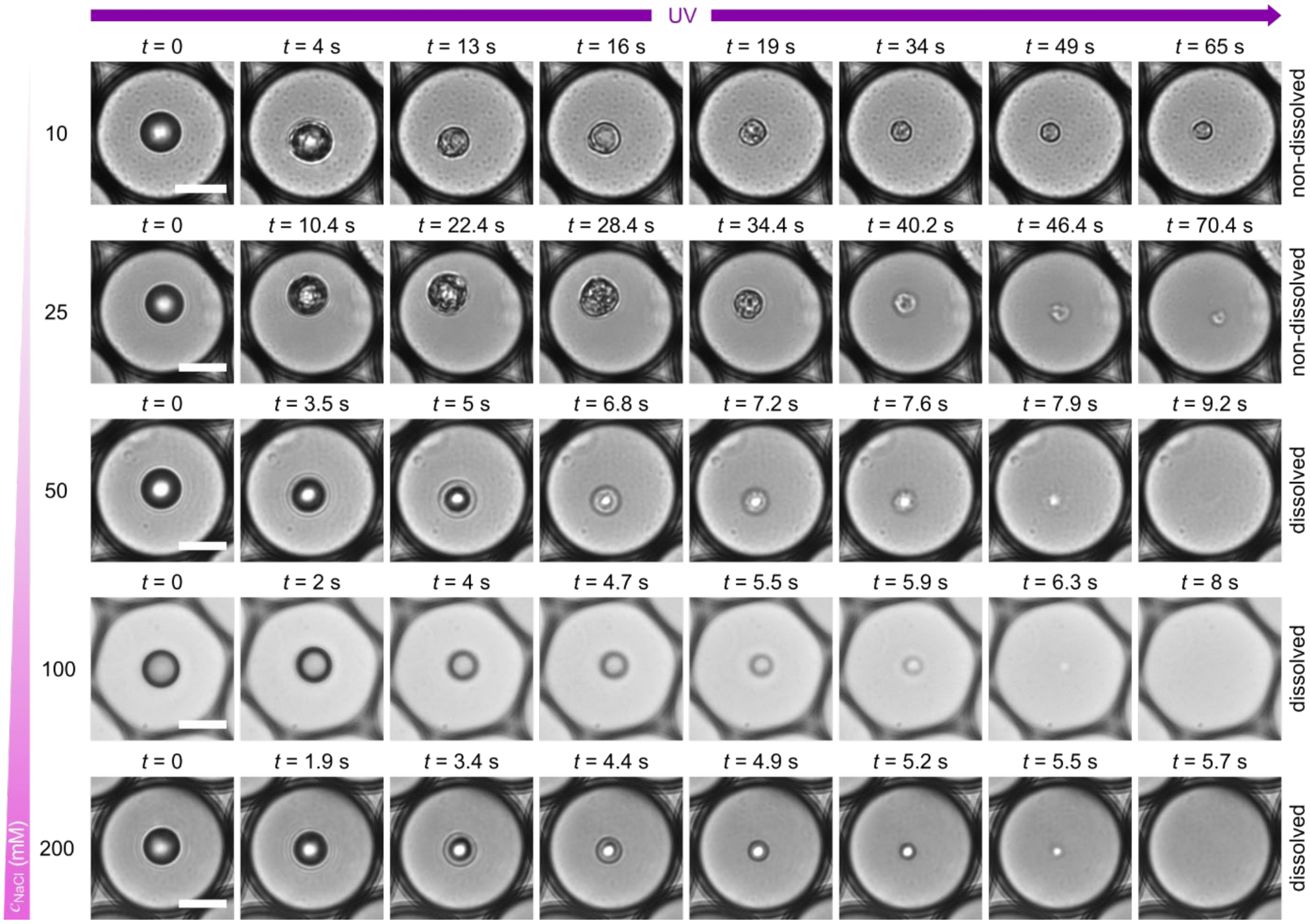

**Fig. S23. Complete coacervate dissolution is suppressed at lower ionic strength.** Time-lapse bright-field microscopy images of a w/o emulsion droplet prepared at $c_0$ = 20 mM and varying salt concentration irradiated continuously with UV light ($I_{UV}$ = 4.4 mW·mm$^{-2}$), showing the shrinkage of coacervates without complete dissolution at $c_{NaCl} \leq$ 25mM. Scale bars, 20 μm.

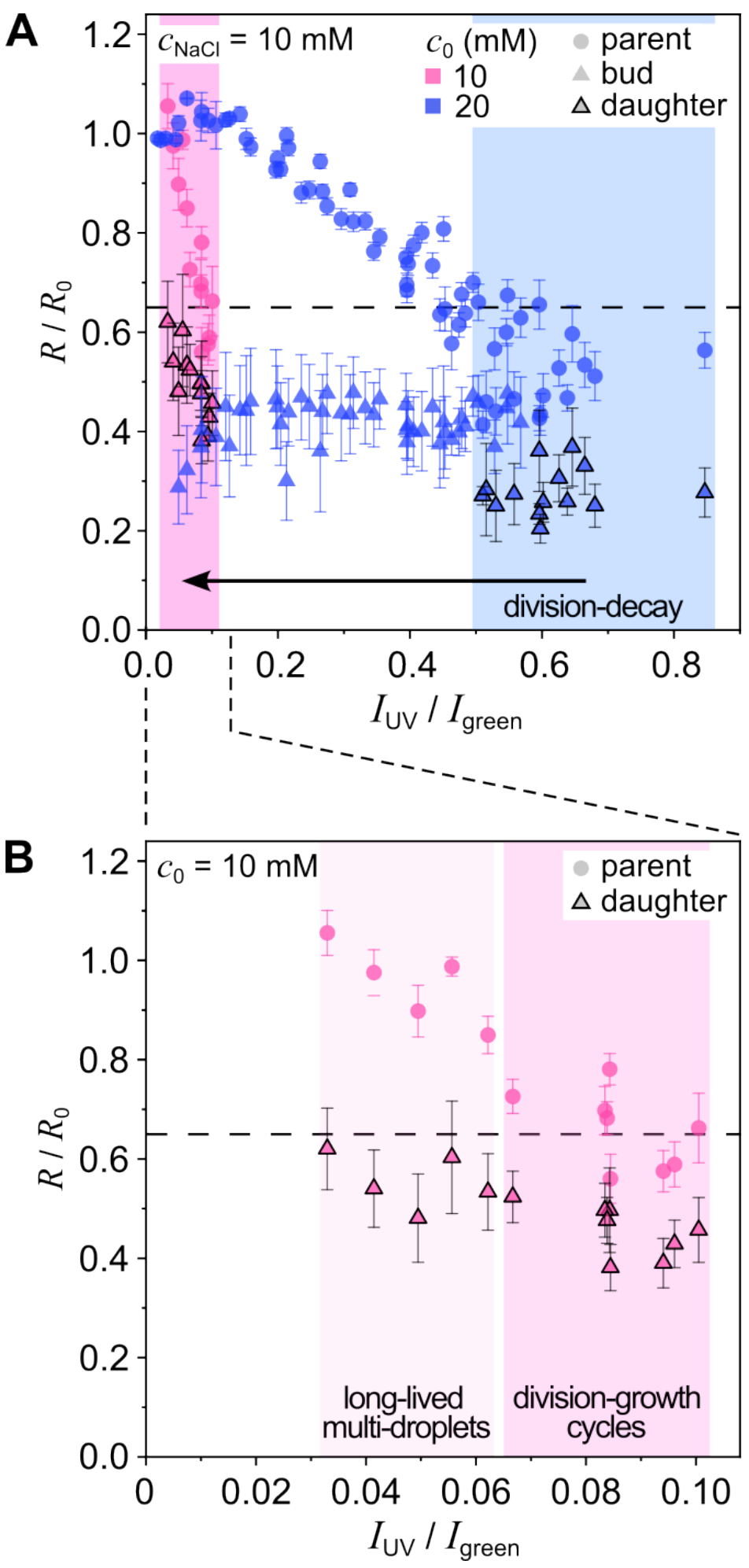

**Fig. S24. Division regime for different coacervate sizes.** (**A**) Normalized parent droplet (circles) and bud (triangles) or daughter droplet (edged triangles) radius as a function of the illumination ratio $I_{UV}/I_{green}$ for $c_0$ = 10 and 20 mM at 10 mM NaCl concentrations. The dotted line marks $R/R_0$ = 0.65. Data are shown as mean ± s.d. ($n \geq 4$ parent droplets, buds and daughter droplets). The division regime is accessed at ca. 10 times lower $I_{UV}/I_{green}$ for $c_0$ = 10 mM (pink area) compared to $c_0$ = 20 mM (blue area). (**B**) Normalized parent droplet (circles) and daughter droplet (edged triangles) radius as a function of the illumination ratio $I_{UV}/I_{green}$ for $c_0$ = 10 mM at $c_{NaCl}$ = 10 mM. The dotted lines mark $R/R_0$ = 0.65. Pink shaded areas identify the long-lived multi-droplets (dark pink) and division/growth cycles (light pink) regimes. Data are shown as mean ± s.d. ($n \geq 6$ parent and daughter droplets).

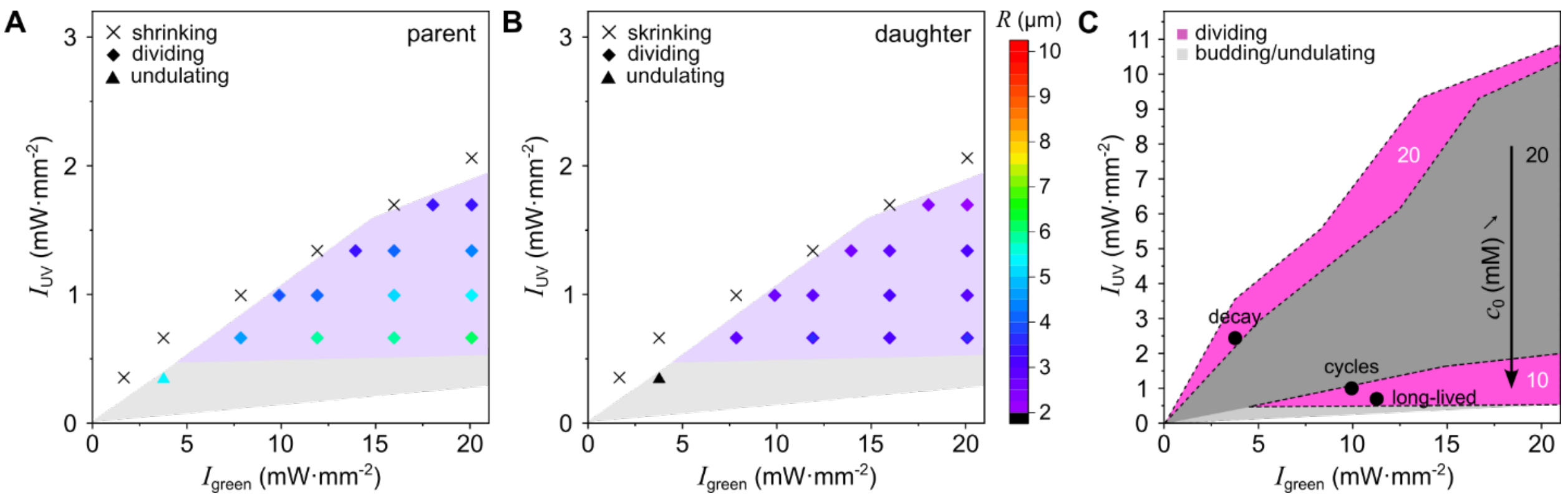

**Fig. S25. Phase response of smaller droplets under dual UV/green illumination.** (**A**,**B**) Non-equilibrium response diagrams in the ($I_{UV}$, $I_{green}$) parameter space showing shrinking (crosses), dividing (diamonds) and undulating (triangles) for $c_0$ = 10 mM and $c_{NaCl}$ = 10 mM. Grey and purple shading indicate the observed regimes: grey, undulating; purple, dividing. Symbol color encodes the mean parent (A) and daughter (B) droplet radius measured during ≥1 cycle for $n$=6 droplets (rainbow color scale: 2-10 μm). (**C**) Comparison of the nonequilibrium phase boundaries in the ($I_{UV}$, $I_{green}$) parameter space at $c_{NaCl}$ = 10 mM and varying $c_0$, as indicated, showing undulating/budding regimes (grey) and dividing regimes (pink, dotted boundary). Conditions for the three division regimes are marked as black dots. The overall extent of the instability regime in the ($I_{UV}$, $I_{green}$) phase space is lower at $c_0$ = 10 mM than 20 mM, but the dominant non-equilibrium outcome shifts from budding to droplet division.

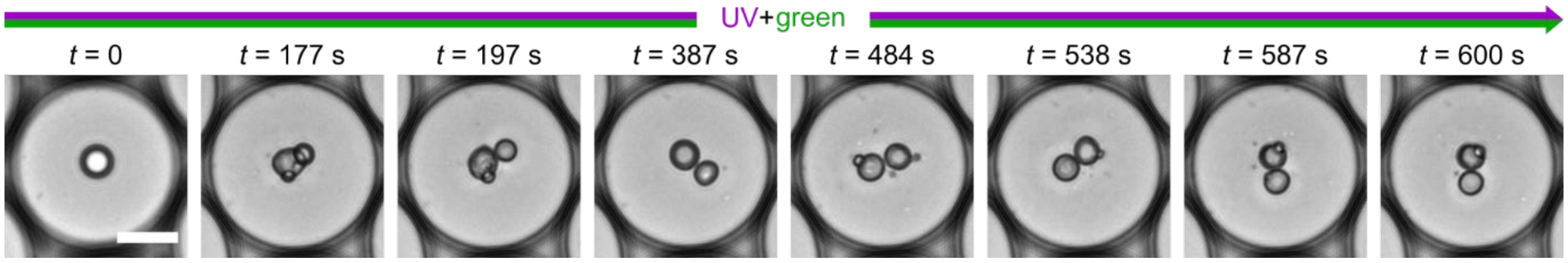

**Fig. S26. Long-lived divided droplets.** Representative time-lapse bright-field microscopy images in the long-lived multi-droplet regime ($c_0$ = 10 mM, $c_{\mathrm{NaCl}}$ = 10 mM, $I_{\mathrm{UV}}$ = 0.7 mW·mm$^{-2}$, $I_{\mathrm{green}}$ = 11.9 mW·mm$^{-2}$), where two same-size daughter droplets persist for tens of seconds after division. Despite touching each other repeatedly, the two droplets resist coalescence. Scale bar, 20 μm.

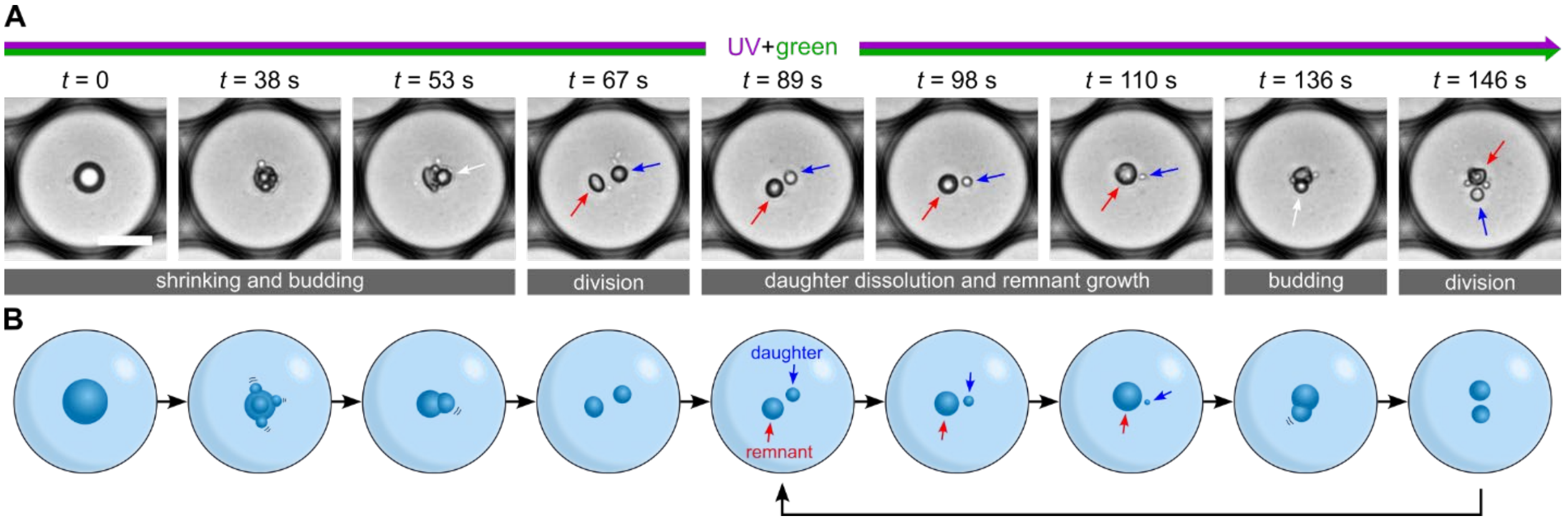

**Fig. S27. Representative division-growth-division cycle under continuous illumination.** (**A**) Time-lapse bright-field microscopy images showing a complete growth-division cycle of a coacervate droplet under continuous UV and green ($c_0$ = 10 mM, $c_{\mathrm{NaCl}}$ = 10 mM, $I_{\mathrm{UV}}$ = 1.0 mW·mm$^{-2}$, $I_{\mathrm{green}}$ = 9.9 mW·mm$^{-2}$; see also movie S8). The sequence shows the onset of interfacial instability and budding, droplet division, subsequent dissolution of the daughter droplet accompanied by regrowth of the remnant, and a second division event. Scale bar, 20 µm. (**B**) Schematic representation of the division/growth cycle.

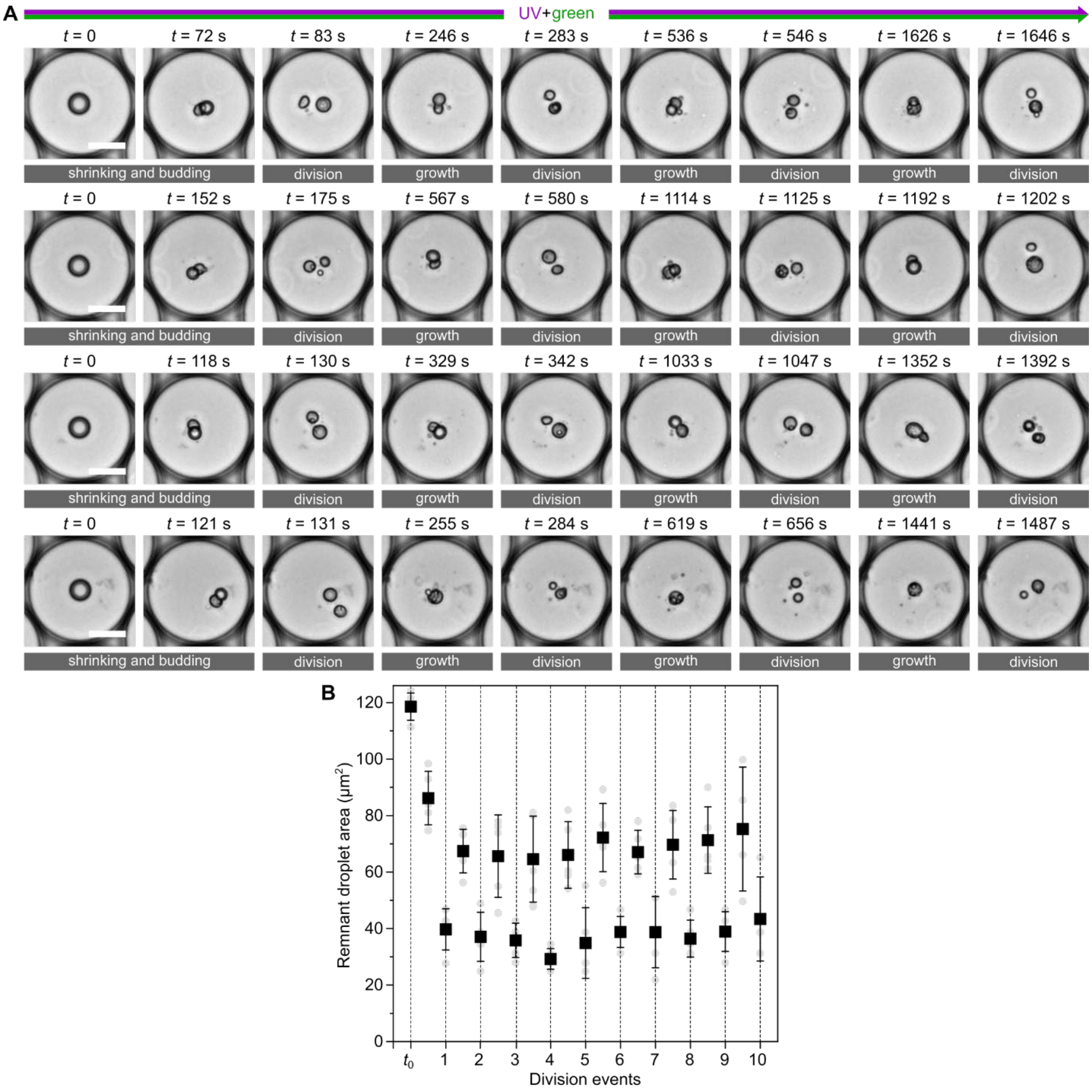

**Fig. S28. Sustained growth-division cycles under continuous illumination.** (**A**) Representative time-lapse bright-field microscopy images of growth-division cycles ($c_0$ = 10 mM, $c_{\mathrm{NaCl}}$ = 10 mM, $I_{\mathrm{UV}}$ = 1.0 mW·mm$^{-2}$, $I_{\mathrm{green}}$ = 9.9 mW·mm$^{-2}$; see also movie S8). Four examples are shown. Scale bar, 20 μm. (**B**) Remnant droplet area measured before and after successive division events, following the initial shrinking from $t_0$. Grey symbols show individual measurements from five droplets, and black symbols show the mean ± s.d. Up to ten successive growth-division events were observed in individual droplets.

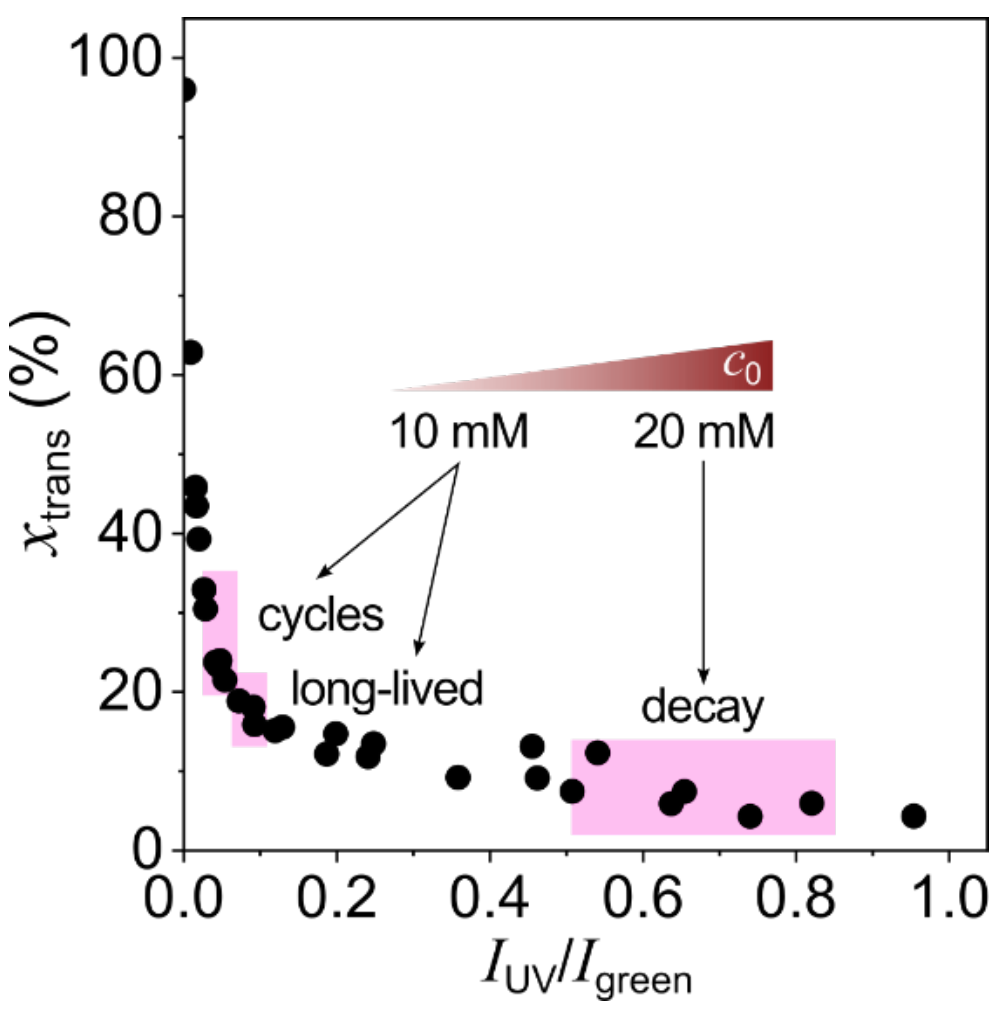

**Fig. S29. Division regimes depends on photostationary *trans* fraction.** Fraction of *trans*-azoTAB $x_{trans}$ at the photostationary state as a function of UV-to-green light illumination ratio $I_{UV}/I_{green}$. The fraction was determined by UV/vis spectroscopy during co-irradiation of a 50 µM azoTAB solution with UV and green light at varying light intensities. Pink shaded areas highlight the $I_{UV}/I_{green}$ range where division is observed depending on $c_0$, together with the division regime depending on $c_0$: $c_0$ = 20 mM, 5% ≲ $x_{trans}$ ≲ 15%: division and decay; $c_0$ = 10 mM, 15% ≲ $x_{trans}$ ≲ 20%: long-lived multi-droplets, 20% ≲ $x_{trans}$ ≲ 35% division/growth cycles.

**Table S1. Photostationary states of azoTAB under different irradiation conditions.** The fractions were calculated from equilibrium UV/vis spectra (fig. S2 and Note S1).

| **Condition** | **λ (nm)** | ***trans*-azoTAB (%)** | ***cis*-azoTAB (%)** |
|---|---|---|---|
| dark | - | 100 | 0 |
| UV | 375 | 4 | 96 |
| blue | 460 | 64 | 36 |
| green | 555 | 96 | 4 |

**Table S2. Extinction coefficients of azoTAB.** Values were obtained from concentration-dependent equilibrium UV/vis spectra by linear fitting of absorbance vs. concentration traces at the given wavelengths, and are given as mean ± s.e of the fits.

| Condition | λ (nm) | $\varepsilon_{trans}$ (mol$^{-1}$·L·cm$^{-1}$) | $\varepsilon_{cis}$ (mol$^{-1}$·L·cm$^{-1}$) |
|---|---|---|---|
| UV | 375 | 20,500 ± 110 | ~0 |
| blue | 460 | 1,660 ± 9 | 3,050 ± 46 |
| green | 555 | 0 | 28 ± 1 |

**Table S3. Sequence of fluorescent oligonucleotides.**

| Abbreviation | Sequence (5', 3') | Length (nt) | $M_w$ ($g \cdot mol^{-1}$) |
|---|---|---|---|
| TAMRA-ssDNA | GTT AGC AGC CGG ATC TCA GTG GT/36TAMSp/ | 23 | 8104.5 |
| Cy5-ssDNA | GTT AGC AGC CGG ATC TCA GTG GT/3Cy5Sp/ | 23 | 7766.3 |

**Movie 1.** Time-lapse bright-field microscopy images of coacervates in w/o emulsion droplets ($c_0$ = 15 mM) irradiated with UV light ($I_{\mathrm{UV}}$ = 1.3 mW·mm$^{-2}$), showing homogeneous coacervate dissolution over a few seconds. The movie is shown at 10 frames per seconds and real-time speed. Scale bar, 20 µm.

**Movie S2.** Time-lapse bright-field microscopy images of coacervates in w/o emulsion droplets ($c_0$ = 15 mM) initially equilibrated with UV light (no coacervates) then irradiated with green light ($I_{\mathrm{green}}$ = 16.4 mW·mm$^{-2}$); showing coacervate nucleation and coarsening over > 30 minutes. The movie is shown at 10 frames per seconds and ×100 real-time speed. Scale bar, 20 µm.

**Movie S3.** Time-lapse bright-field microscopy images of coacervates in w/o emulsion droplets ($c_0$ = 15 mM) initially equilibrated with UV light (no coacervates) then irradiated with blue light ($I_{\mathrm{blue}}$ = 60.3 mW·mm$^{-2}$), showing coacervate nucleation and arrested growth. The movie is shown at 6.6 frames per seconds and ×200 real-time speed. Scale bar, 20 µm.

**Movie S4.** Time-lapse bright-field microscopy images of coacervates in w/o emulsion droplets ($c_0$ = 15 mM) initially equilibrated in the dark then irradiated with blue light ($I_{\mathrm{blue}}$ = 60.3 mW·mm$^{-2}$), showing coacervate fragmentation into finite size. The movie is shown at 6.6 frames per seconds and ×200 real-time speed. Scale bar, 20 µm.

**Movie S5.** Time-lapse bright-field microscopy images of coacervates in w/o emulsion droplets ($c_0$ = 20 mM) co-irradiated continuously with UV ($I_{\mathrm{UV}}$ = 0.09 mW·mm$^{-2}$) and green ($I_{\mathrm{green}}$ = 1.7 mW·mm$^{-2}$) light, showing undulation at the coacervate's surface. The movie is shown at 10.7 frames per seconds and ×6 real-time speed. Scale bar, 20 µm.

**Movie S6.** Time-lapse bright-field microscopy images of coacervates in w/o emulsion droplets ($c_0$ = 20 mM) co-irradiated continuously with UV ($I_{\mathrm{UV}}$ = 1.3 mW·mm$^{-2}$) and green ($I_{\mathrm{green}}$ = 20 mW·mm$^{-2}$) light, showing coacervate budding. The movie is shown at 10.7 frames per seconds and ×6 real-time speed. Scale bar, 20 µm.

**Movie S7.** Time-lapse bright-field microscopy images of a coacervate a in w/o emulsion droplet ($c_0$ = 20 mM) prepared at 10 mM NaCl concentration and co-irradiated continuously with UV ($I_{\mathrm{UV}}$ = 2.4 mW·mm$^{-2}$) and green ($I_{\mathrm{green}}$ = 3.8 mW·mm$^{-2}$) light for 5 min, showing a coacervate division event before decay. The movie is shown at 21.4 frames per seconds and ×12 real-time speed. Scale bar, 10 µm.

**Movie S8.** Time-lapse bright-field microscopy images of a coacervate in a w/o emulsion droplet ($c_0$ = 10 mM) prepared at 10 mM NaCl concentration and co-irradiated continuously with UV ($I_{\mathrm{UV}}$ = 1.0 mW·mm$^{-2}$) and green ($I_{\mathrm{green}}$ = 9.9 mW·mm$^{-2}$) light for 10 min, showing cyclic division events. The movie is shown at 21.4 frames per seconds and ×12 real-time speed. Scale bar, 10 µm.